\documentclass[twocolumn,twoside]{IEEEtran}      
\usepackage{subfigure} 
\usepackage{bm} 
\usepackage{graphicx,graphics} 
\usepackage{amsmath, amsthm, amsfonts, amssymb, amsbsy}
\usepackage{setspace} 
\usepackage{pstricks,arydshln}
\usepackage{multirow,multicol} 
\usepackage{booktabs}
\usepackage{algorithmic,amscd,textcomp,amssymb,epsfig,graphics,epsf,color,amsmath,balance,algorithm,cite}
\usepackage{caption} 
\usepackage{cite} 
\usepackage{tabularx}
\usepackage{makecell}

\begin{document}
\title{Orbital Angular Momentum Waves: Generation, Detection and Emerging Applications}

\author{Rui Chen,~\IEEEmembership{Member,~IEEE,} Hong Zhou, Marco Moretti,~\IEEEmembership{Member,~IEEE}, Xiaodong Wang,~\IEEEmembership{Fellow,~IEEE,} and Jiandong Li,~\IEEEmembership{Senior Member,~IEEE}
\thanks{This work was supported in part by the Fundamental Research Funds for the Central Universities (JB180109), the 111 Project (B08038) and the University of Pisa under the PRA 2018-2019 Research Project CONCEPT.}
\thanks{Rui Chen, Hong Zhou and Jiandong Li are with the State Key Laboratory of ISN, Xidian University, Shaanxi 710071, China (e-mail: rchen@xidian.edu.cn, hzhou\_1@stu.xidian.edu.cn, jdli@mail.xidian.edu.cn). Rui Chen is also with the Science and Technology on Communication Networks Laboratory.} %
\thanks{Marco Moretti is with the University of Pisa, Dipartimento di Ingegneria dell'Informazione, Italy (e-mail: marco.moretti@iet.unipi.it).}
\thanks{Xiaodong Wang is with the Electrical Engineering Department, Columbia University, New York, NY 10027 USA (e-mail: wangx@ee.columbia.edu).}
}

\maketitle

\thispagestyle{empty}

\begin{abstract}
Orbital angular momentum (OAM) has aroused a widespread interest in many fields, especially in telecommunications due to its potential for unleashing new capacity in the severely congested spectrum of commercial communication systems.  Beams carrying OAM have a helical phase front and a field strength with a singularity along the axial center, which can be used for information transmission, imaging and particle manipulation. The number of orthogonal OAM modes in a single beam is theoretically infinite and each mode is an element of a complete orthogonal basis that can be employed for multiplexing  different signals,  thus greatly improving the spectrum efficiency. In this paper, we comprehensively  summarize and compare the methods for generation and detection  of optical OAM, radio OAM and acoustic OAM. Then, we represent the applications and technical challenges of OAM in communications, including free-space optical communications, optical fiber communications, radio communications and acoustic communications. To complete our survey, we also discuss the state of art of particle manipulation and target imaging with OAM beams.
\end{abstract}


\section{Introduction}

\IEEEPARstart{E}lectromagnetic waves carry both energy and momentum, where momentum comprises linear momentum $\mathbf{P}$ and angular momentum $\mathbf{L}$. In particular, angular momentum has an additive component linked to polarization, \emph{spin angular momentum} (SAM), and another one associated with spatial distribution, which is called \emph{orbital angular momentum} (OAM).
The relationship between SAM and OAM can be explained by referring to the model of electron rotation around the nucleus: the momentum generated by the circular motion of electrons around the nucleus is equivalent to OAM, and the momentum generated by the spin of electrons is equivalent to SAM.

In 1992, Allen et al. first combined the concept of OAM with the idea of \emph{optical vortex} \cite{Allen1992Orbital}:  in an optical vortex the planes of constant phase of the electric and magnetic vector fields form a corkscrew or helicoid running in the direction of propagation.  The vortex is characterized by a number, called the \emph{topological charge}, which indicates the number of twists the light does in one wavelength. The larger the number of  twists, the faster the light is rotating around the axis. Accordingly, the OAM carried by the optical vortex theoretically has an infinite number of eigenstates and is defined in an infinite-dimensional Hilbert vector space. Because of this, the application potential of OAM in the field of communications are enormous, even if there are still some problems to be solved before a full deployment. If the OAM dimension of photons can be fully utilized for information modulation or multiplexing, the information capacity of a single photon can be significantly improved, thereby leading to an increase of the transmission capacity of single-wavelength and single-mode fibers. In addition, since the vortex beam has a helical wavefront, its axial center field in the direction of propagation is null, creating the  potential for applications in particle manipulation and imaging.

The potential of employing OAM  for communications are not restricted to  electromagnetic waves at light frequencies. In 2007, Thid\'e et al.  \cite{Thid2007Utilization} proposed to apply  the concept of optical vortex to the field of wireless communications, i.e., in a range of lower radio frequencies than light.
Moreover, a different line of research has showed that, unlike electromagnetic waves, sound waves do not have polarization or spin effects and cannot carry SAM but only carry OAM. Although the concept of acoustic vortex was first proposed in 1979 \cite{Broadbent1979Acoustic}, twenty years later, Hefner and Marston \cite{Hefner1999An} derived the relationship between sound pressure and angular momentum, perfecting the theory of the \emph{acoustic vortex}. The application of OAM to wireless communications and acoustic communications, especially underwater communications, is expected to open new fields of research and possibly break the limits of existing communication systems.

\subsection{Contribution with respect to existing literature}

\newcommand{\tabincell}[2]{\begin{tabular}{@{}#1@{}}#2\end{tabular}}
This paper is devoted to reviewing the research progress on the use of OAM waves for communications  in the last three decades. Our survey  includes optical, radio and acoustic OAM generation, detection and applications in communications,  as well as the applications of OAM waves in particle manipulation and imaging.  Main technical challenges are addressed  and  potential solutions are analyzed.  The key contributions of this survey can be summarized as
\begin{itemize}
\item The most common methods for generating and detecting optical, radio and acoustic OAM waves are introduced and compared from multiple perspectives.
\item Recent advances in OAM applications in free-space optical communications, optical fiber communications, radio communications and acoustic communications are carefully reviewed, so that the main technical problems and their potential solutions are  discussed in details. Indeed, integrating OAM into existing communication systems presents some very serious challenges such as OAM beam divergence, misalignment between transmitter and receiver, atmospheric turbulence effects in free-space optical links, mode coupling in fiber links, and multipath effects in radio communication links. These effects need to be carefully considered in link design.
\item Since OAM has been considered as a potential solution for precise particle manipulation and target imaging, the application prospects of OAM in these fields are also briefly discussed.
\end{itemize}

This is not the first survey on OAM waves, since there are other works on the subject, such as \cite{Yao2011Orbital}, \cite{Willner2017Recent}, and \cite{Willner2015Optical}. However, \cite{Yao2011Orbital} is more inclined to discuss the physical properties of OAM waves, and \cite{Willner2017Recent} addresses only the application of OAM multiplexing in free-space optical and radio communications.  The work in \cite{Willner2015Optical} surveys both generation/detection and application of OAM beams but mainly for optical communications and there is no mention of the potential of OAM in acoustics.
Compared with these surveys, our work has the merit of comprehensively and consistently  discussing the application of OAM waves for optical, radio and acoustic communications. For each system we present  a complete and accurate  review of  the generation/detection methods, their application and the most pressing technical challenges.

\subsection{Paper Organization}
The rest of the paper is organized as follows. Section II describes the main principles of OAM,  introducing the basic characteristics of OAM waves and their application potential. Subsequently, the generation and detection methods for optical, radio and acoustic OAM communication systems are summarized and compared in Section III and Section IV, respectively. Section V focuses on  the most recent advances in OAM applications to free-space optical, optical fiber, radio and acoustic communications including the corresponding technical challenges and potential solutions. Atmospheric turbulence effects in free-space optical links, mode coupling in fiber links, and multipath effects in radio communication links are all significant practical obstacles to the implementation of OAM-based communication systems. In addition, OAM beam divergence and misalignment also need to be considered in link design. Section VI discusses the application of OAM in other areas, including particle manipulation and imaging. Conclusions and perspectives then follow in Section VII.

\section{OAM Waves and their Potential}
The angular momentum of the electromagnetic field associated with a volume $V$ can be expressed as \cite{Mohammadi2010Orbital}
\begin{equation} \label{angmomentun}
\mathbf{J}=\int \varepsilon_0 \mathbf{r} \times \textrm{Re}\{\mathbf{E}\times \mathbf{B}^*\}dV,
\end{equation}
where $\varepsilon_0$ is the dielectric constant in vacuum, $\mathbf{r}$ is the radius vector of a point in the electric field, $\mathbf{E}$ and $\mathbf{B}$ are the electric field intensity and magnetic flux density vector, respectively, $\textrm{Re}\{\cdot\}$ represents the real part operation, and $(\cdot)^*$ represents the conjugate operator. The angular momentum $\mathbf{J}$ can be decomposed into a component $\mathbf{S}$ (SAM) associated with polarization and a component $\mathbf{L}$ (OAM) related to the spatial distribution of electromagnetic waves, i.e.
\begin{equation} \label{decomposition1}
\mathbf{J}=\mathbf{S}+\mathbf{L},
\end{equation}
where
\begin{equation} \label{decomposition2}
\begin{aligned}
&\mathbf{S}=\varepsilon_0 \int \textrm{Re}\left\{ \mathbf{E}^*\times \mathbf{A}\right\}dV,\\
&\mathbf{L}=\varepsilon_0 \int \textrm{Re}\left\{i \mathbf{E}^*\left(\hat{\mathbf{L}}\cdot \mathbf{A}\right)\right\}dV.\\
\end{aligned}
\end{equation}
In (\ref{decomposition2})$, \hat{\mathbf{L}}=-i\left(\mathbf{r}\times \triangledown \right)$ is the OAM operator, and $\mathbf{A}$ is the vector potential.

\begin{figure}[t]
\centering
\subfigure[]
{\includegraphics[scale=0.34]{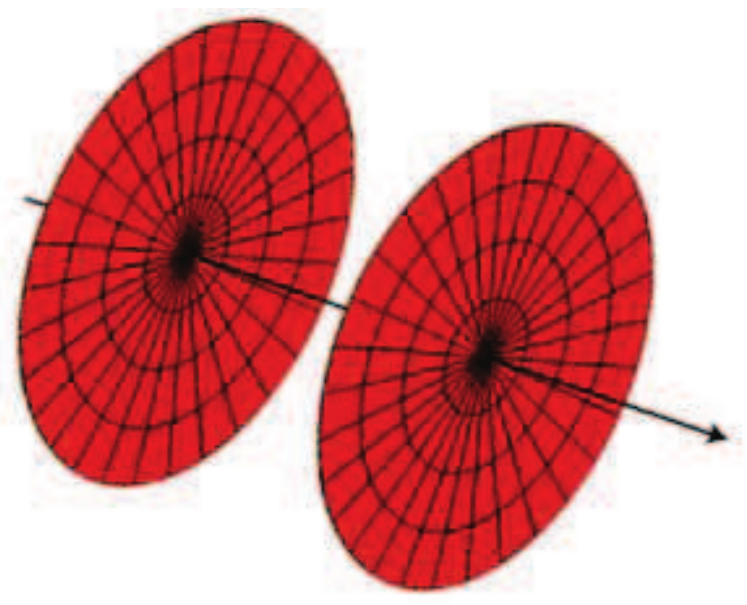}}
\subfigure[]
{\includegraphics[scale=0.35]{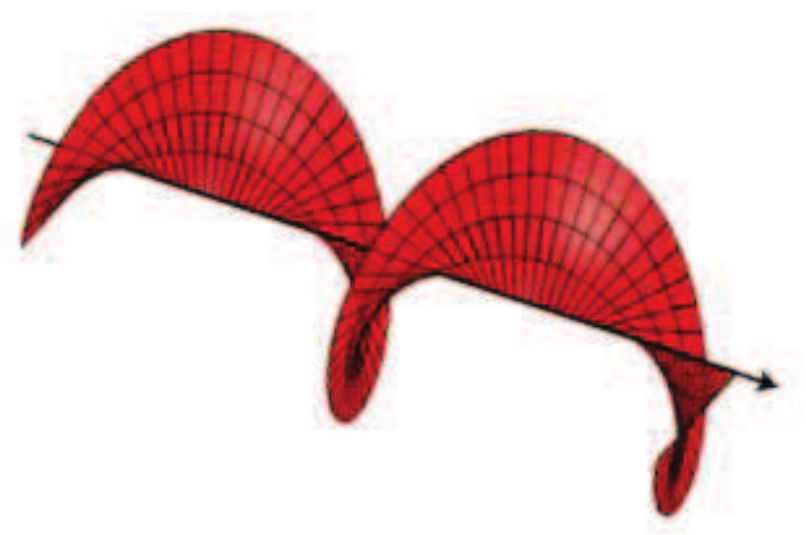}}
\subfigure[]
{\includegraphics[scale=0.35]{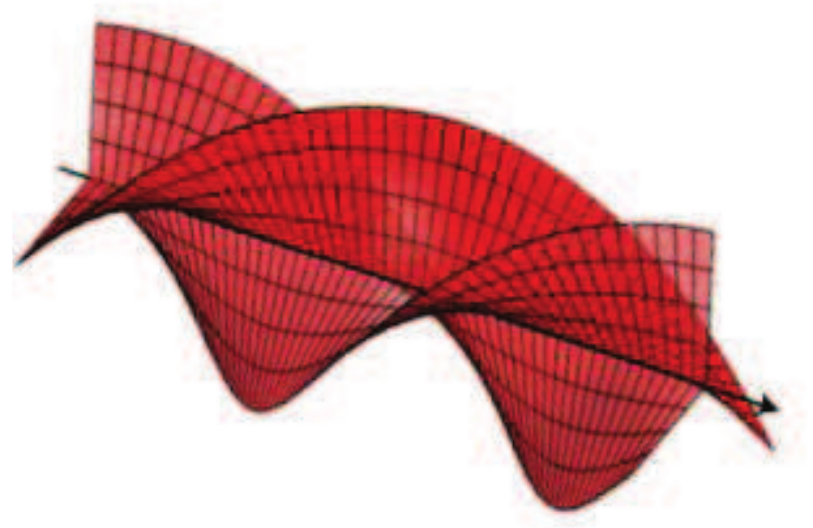}}
\caption{The helical phase fronts of OAM waves: (a) $\ell=0$, (b) $\ell=1$, and (c) $\ell=3$.}
\label{Fig0}
\end{figure}

It has been demonstrated that vortex waves having helical phase structure can carry OAM \cite{Allen1992Orbital}. Due to the phase singularity, the wavefront of electromagnetic waves will be twisted during the propagation, forming vortex waves. The vortex waves are characterized by helical phase fronts, as shown in Fig. \ref{Fig0}. The amplitude of the vortex wave field is null along the axis center so that there is a ``dark'' core at the location of the phase singularity. The helical phase structure is described by the phase term $\exp(i\ell\theta)$, where $\theta$ is the transverse azimuthal angle and $\ell$ is known as the \emph{topological charge} or \emph{OAM mode} and can be any real number. The magnitude and sign of topological charge $\ell$ determine the number and chirality of the wave torsions in one wavelength, respectively. When an OAM beam has integer desired topological charges the OAM modes are called \emph{pure}. In practice, it might happen that some of the energy of one pure mode leaks into other modes. Mode purity is measured as the ratio of the  power of the desired mode and the overall OAM power.

There are several special OAM waves that can be well described by their radial intensity distributions, among which Laguerre-Gaussian (LG) beams are probably the most widely known.  LG beams are paraxial solutions of the wave equation in cylindrical coordinates and homogeneous media (e.g., free space).
In cylindrical coordinates, the complex amplitude distribution of a LG beam propagating along the z-axis is given by \cite{Allen1992Orbital,Yao2011Orbital}
\begin{align} \label{LGbeam}
LG_{p,\ell}(r,\theta,z)=&\sqrt{\frac{2p!}{\pi w^2(z)(p+|\ell|)!}} \left[\frac{r\sqrt{2}}{w(z)}\right]^{|\ell|} L_p^{|\ell|} \left(\frac{2r^2}{w^2(z)}\right) \nonumber \\
&\exp\left[\frac{-r^2}{w^2(z)}\right] \exp\left[\frac{ikr^2z}{2(z^2+z_R^2)}\right] \exp(i\ell\theta) \nonumber \\
&\exp\left[-i(2p+|\ell|+1)\tan^{-1}\left(\frac{z}{z_R}\right)\right],
\end{align}
where $w(z)=w_0[(z^2+z_R^2)/z_R^2]^{1/2}$ is the radius of the beam, $w_0$ is the waist radius, $z_R$ is the Rayleigh range, $k$ is the wave number, $L_p^{|\ell|}(x)$ is the associated Laguerre polynomial and $(2p+|\ell|+1)\tan^{-1}\left(z/z_R\right)$ is the Gouy phase.
It can be seen from (\ref{LGbeam}) that LG beams are characterized by two indexes: the azimuthal index $\ell$ and the radial index $p$. The former is the topological charge, which characterizes the beam OAM, the latter is related to the number of radial nodes on the cross section of the beam intensity. Therefore, the azimuth and radial wavefronts of LG beams can be described by the $\ell$ and $p$ indexes.

The complex amplitude distribution of Bessel beams, another special OAM wave described by their radial intensity distributions,  is
\begin{equation} \label{Besselbeam}
B(r,\theta,z)=J_\ell(k_r r)\exp(ik_z z)\exp(i\ell\theta),
\end{equation}
where $J_\ell(k_r r)$ is the $\ell$-order Bessel function of the first kind, $k_r$ and $k_z$ are radial and axial wave numbers, respectively.

OAM beams have many attractive properties but probably the most important one is the orthogonality between modes. Considering two OAM waves with topological charges $\ell_1$ and $\ell_2$, it can be shown  that the two OAM modes are orthogonal through the inner product
\begin{equation} \label{inner}
\int_0^{2\pi} e^{i\ell_1 \theta}\left(e^{i\ell_2 \theta}\right)^* d\theta=\left\{
\begin{aligned}
&0 & & \ell_1 \neq \ell_2\\
&2\pi & & \ell_1 = \ell_2\\
\end{aligned}
.\right.
\end{equation}
Thanks to the inherent orthogonality between OAM modes, OAM waves with different $\ell$ can be used as a separate set of data transmission channels adding a new dimension independent of time, frequency, and polarization.   Accordingly, this new multiplexing dimension  can be combined with other existing multiplexing strategies to improve system capacity.

Moreover, OAM has certain application potential in particle manipulation and imaging. Due to the presence of the ``dark'' core, the pattern of OAM beams has a specific  \emph{doughnut shape}. This particular shape  allows OAM beams to capture particles or drive the particles to rotate around the beam axis, with great potential in biological and medical applications. Besides, this ring beam with helical phase structure is expected to obtain resolutions beyond the Rayleigh limit for imaging systems, and  provide new instruments for accurate radar targets imaging.
\begin{figure}[t]
\centering
\includegraphics[scale=0.7]{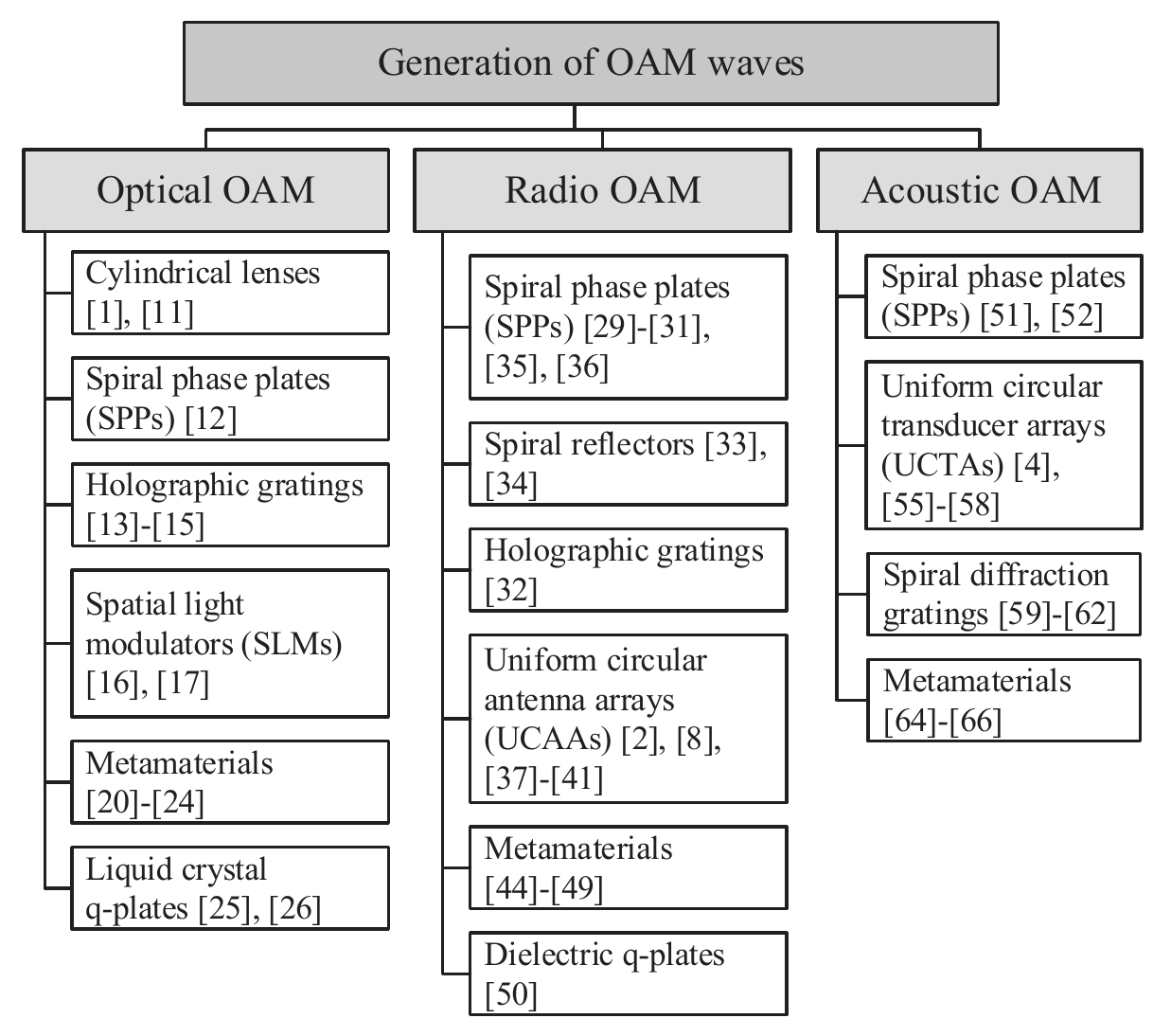}
\caption{Taxonomy diagram of generation of OAM waves.}
\label{Fig31}
\end{figure}

\section{Generation of OAM Waves}
Since Allen et al. in 1992 proved that the optical vortex with a spiral wavefront carries OAM, the research on OAM has gradually deepened and broadened. In this section, we discuss various state-of-the-art methods for generating OAM in the fields of light, radio, and sound waves.  Because of the large number of methods discussed, Fig. \ref{Fig31} shows a diagram that summarizes a  taxonomy of the techniques presented in this section.

\subsection{Optical OAM}
In optics, a vortex beam can  either be obtained  as the  direct output of a laser cavity, or can be generated by feeding a Gaussian beam into a  converter, such as a cylindrical lens, a spiral phase plate, a phase hologram, a metamaterial or a $q$-plate.

Even before Allen showed that the vortex beam carries OAM, research has been focused on vortex beams and on how to generate them. In \cite{Vaughan1983Temporal}, the resonance phenomenon is exploited to generate  a vortex from an optical cavity. The laser cavity normally  produces a mixture of multiple modes including the basic mode with $\ell=0$. By placing a component inside the laser cavity, such as a spot-defect mirror \cite{Kano2011Generation}, the laser cavity can be forced to resonate on a specific OAM mode.

\begin{figure}[t]
\centering
\subfigure[]
{\includegraphics[scale=0.55]{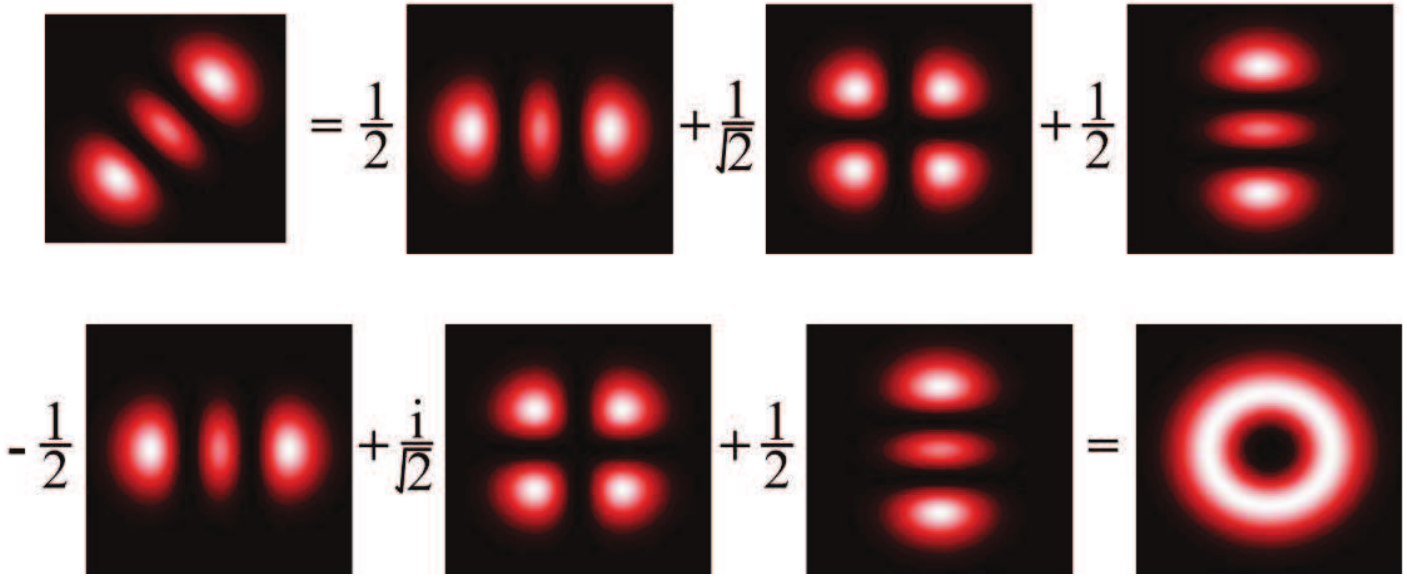}}

\subfigure[]
{\includegraphics[scale=0.55]{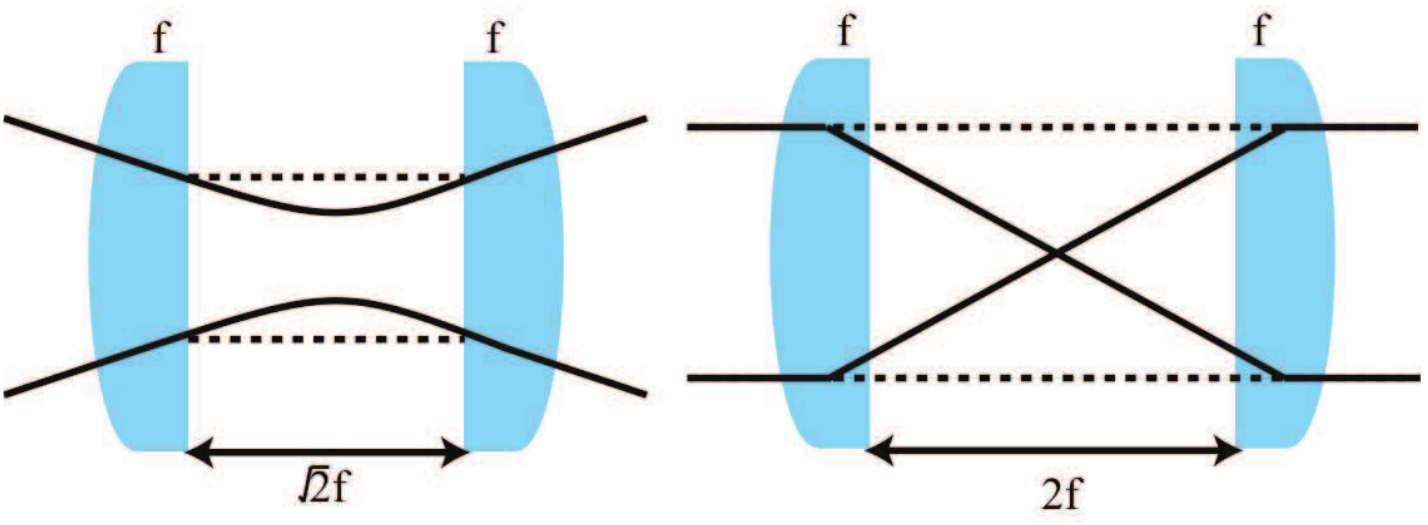}}
\caption{(a) An example of 2nd order HG mode decomposition and LG mode synthesis: HG$_{02}$ at $45^{\circ} = \frac{1}{2}\textrm{HG}_{02}+\frac{1}{\sqrt{2}}\textrm{HG}_{11}+\frac{1}{2}\textrm{HG}_{20}$ and $ -\frac{1}{2}\textrm{HG}_{02}+\frac{i}{\sqrt{2}}\textrm{HG}_{11}+\frac{1}{2}\textrm{HG}_{20}=\textrm{LG}_{02}$. (b) Schematic diagram of $\pi/2$-converter and $\pi$-converter. Both converters consist of two identical cylindrical lenses of focal length f \cite{Yao2011Orbital}.}
\label{Fig1}
\end{figure}

In \cite{Allen1992Orbital,Beijersbergen1993Astigmatic}, \emph{cylindrical lenses} are used to transform a Hermite-Gaussian (HG) laser beam  into a helically-phased LG beam.
An example of a 2nd order HG mode decomposition and LG mode synthesis  is shown in Fig. \ref{Fig1} (a). The 45$^{\circ}$ angle HG mode can be decomposed into a series of HG modes, and this series of HG modes can be rephased to obtain the LG mode. This rephasing can be achieved by changing the Gouy phase shift in the HG mode. Two cylindrical lens mode converters, the $\pi/2$-converter and the $\pi$-converter, are proposed in \cite{Beijersbergen1993Astigmatic}, as shown in Fig. \ref{Fig1} (b). The functions of the mode converters are similar to the polarization of birefringence $\lambda/4$ and $\lambda/2$ plates. The $\pi/2$-converter can convert a HG mode with indices $m, n$, oriented at 45$^{\circ}$ angle to the lens axis, into a LG mode  with topological charge $\ell=m-n$ and number of radial nodes in the intensity distribution $p=\min(m, n)$. The $\pi$-converter changes the index of the input mode, that is, HG$_{m,n}$ turns into HG$_{n,m}$ or LG$_{m,n}$ turns into LG$_{n,m}$, which has an azimuthal dependence of the opposite sign. Cylindrical lens mode converters have high conversion efficiency and generate OAM with high purity, but they require high construction precision, and, at the same time, have poor flexibility because  they require a very precise incident field angle.

\begin{figure}[t]
\centering
\includegraphics[scale=0.4]{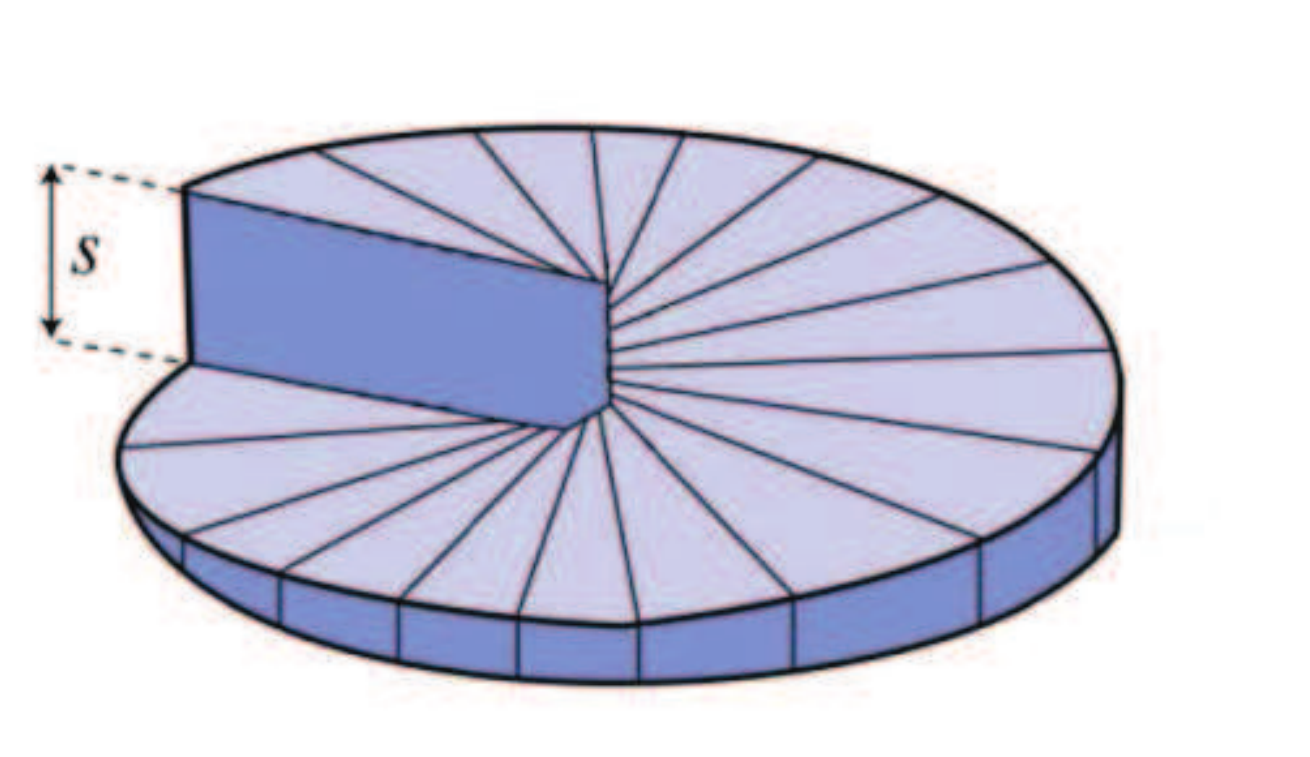}
\caption{Schematic diagram of SPP with step height $s$ \cite{Yao2011Orbital}.}
\label{Fig2}
\end{figure}

The \emph{spiral phase plate} (SPP) is another way to implement the vortex beam \cite{Beijersbergen1994Helical}. As shown in Fig. \ref{Fig2},  the SPP is designed like a rotating step so that its thickness increases as the azimuth angle increases but is uniform radially. The step height is expressed as $s=\ell\lambda\theta/[2\pi(n-n_0)]$, where $n$ is the refractive index of the  transparent dielectric material  of the plate, $n_0$ is the refractive index of the external medium, $\ell$ is the topological charge, $\lambda$ is the wavelength of the incident light, and $\theta$ is the spatial azimuth. When a plane wave of Gaussian light passes through the phase plate, the  light beam experiences  a different phase in the azimuth direction due to the spiral thickness of the phase plate and is converted into a helically-phased beam with topological charge $\ell$. The SPP has high conversion efficiency and can be used for high power laser beams but has also the  limit that it can generate a single mode only and it  has very tight precision requirements.

\begin{figure}[t]
\centering
\includegraphics[scale=0.6]{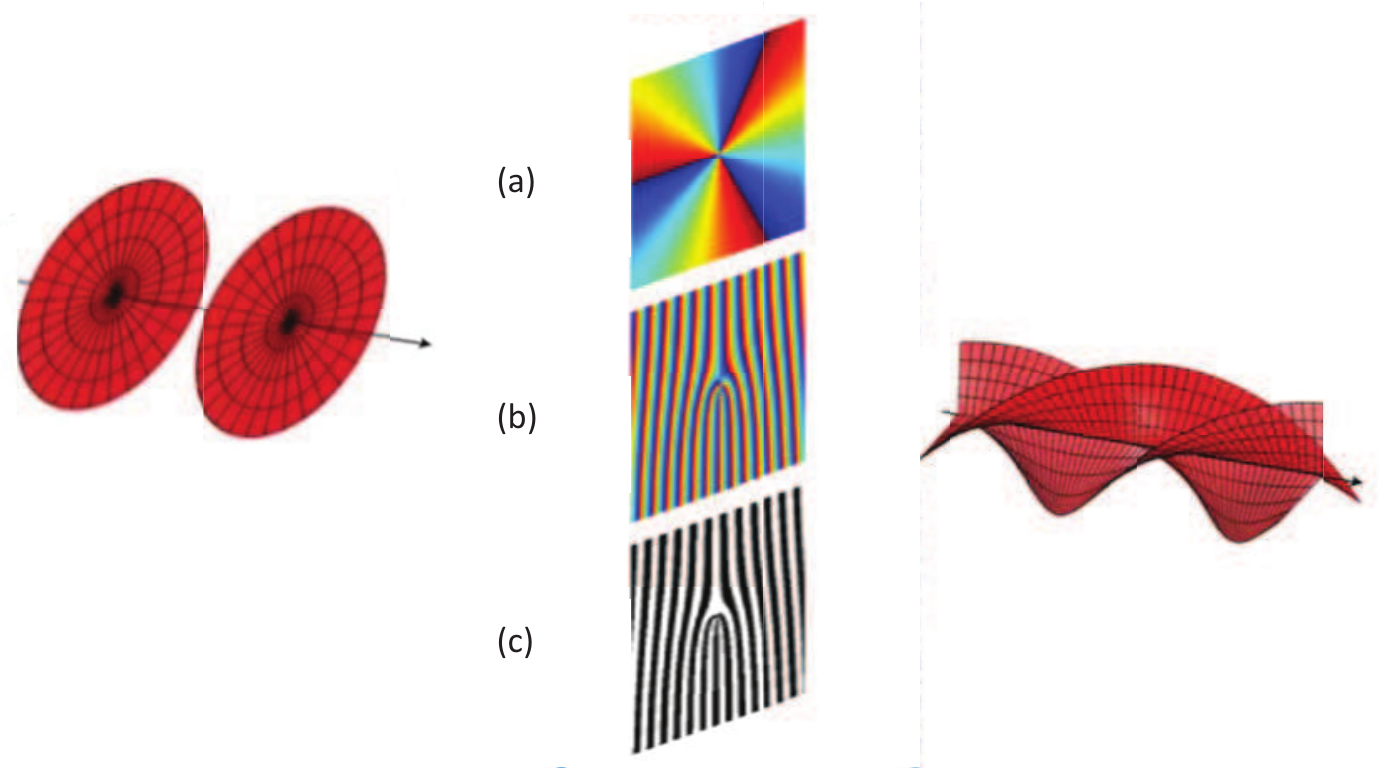}
\caption{Phase holograms with $\ell=3$ in the form of (a) a spiral phase hologram, (b) an $\ell$-fold forked hologram, and (c) a binarized $\ell$-fold forked grating.}

\label{Fig3}
\end{figure}

Instead of producing a complex refractive optical element to generate a vortex wave, one can employ a computer-generated hologram to design a diffractive optical element whose transmittance function is related to the helical phase $\exp(i\ell\theta)$. \emph{Holographic gratings}, such as Fresnel spiral diffraction gratings and $\ell$-fold forked diffraction gratings \cite{Yu1990Laser,Bazhenov1992Screw,Heckenberg1992Generation}, are generated by using a photolithographic process to record an interference pattern on a photorestist-coated substrate.  When Gaussian light is incident on the $\ell$-fold forked diffraction grating, a vortex beam with the topological charge of $n\ell$ can be obtained at the $n$th diffraction order. Holographic diffraction gratings are relatively simple and fast to produce and provide good wavefront flatness and high efficiency for a single polarization plane. However, as the number of diffraction orders increases, the quality of the vortex beam is seriously degraded and this method  is generally only used to generate low-order vortex beams.

Unlike the SPP and the diffraction grating described above, the \emph{spatial light modulator} (SLM) can generate vortex beams with different topological charges \cite{Curtis2002Dynamic,Maurer2007Tailoring}. A SLM is a pixelated liquid crystal device whose liquid crystal molecules can be programmed to dynamically change the incident beam parameters, including the beam phase in the transverse plane, to create a vortex beam. In detail,  a phase hologram with a transmittance function of $\exp(i\ell\theta)$ is digitally generated and loaded on the SLM so that the SLM will determine the phase of each point in the two-dimensional space according to the value of each pixel of the input phase hologram.
In general, to generate a vortex beam the phase hologram loaded onto the SLM  is in the form of: a) a spiral phase hologram, b) a $\ell$-fold forked hologram, and c) a binarized $\ell$-fold forked grating (shown in Figs. \ref{Fig3}(a), \ref{Fig3}(b), and \ref{Fig3}(c), respectively). The spiral phase hologram produces a vortex beam, which exits in the direction perpendicular to the hologram plane, with only a single topological charge. The forked hologram can control the exit direction of the vortex beam, while the binarized forked grating has multiple diffraction orders, which produce different topological charges at different diffraction orders. Liquid crystal SLMs are very flexible since they can control various parameters of the vortex beam but they are relatively expensive  and have an energy threshold that makes their use impossible with high power laser beams. Recently, in addition to the liquid crystal SLM, a diffractive optical element using a digital micro-mirror device (DMD) has been introduced \cite{Mirhosseini2013Rapid,Mitchell2016High}. In contrast to liquid crystal devices, DMDs cost less and are faster but their diffraction efficiency is lower.

\begin{figure}[t]
\centering
\includegraphics[scale=0.7]{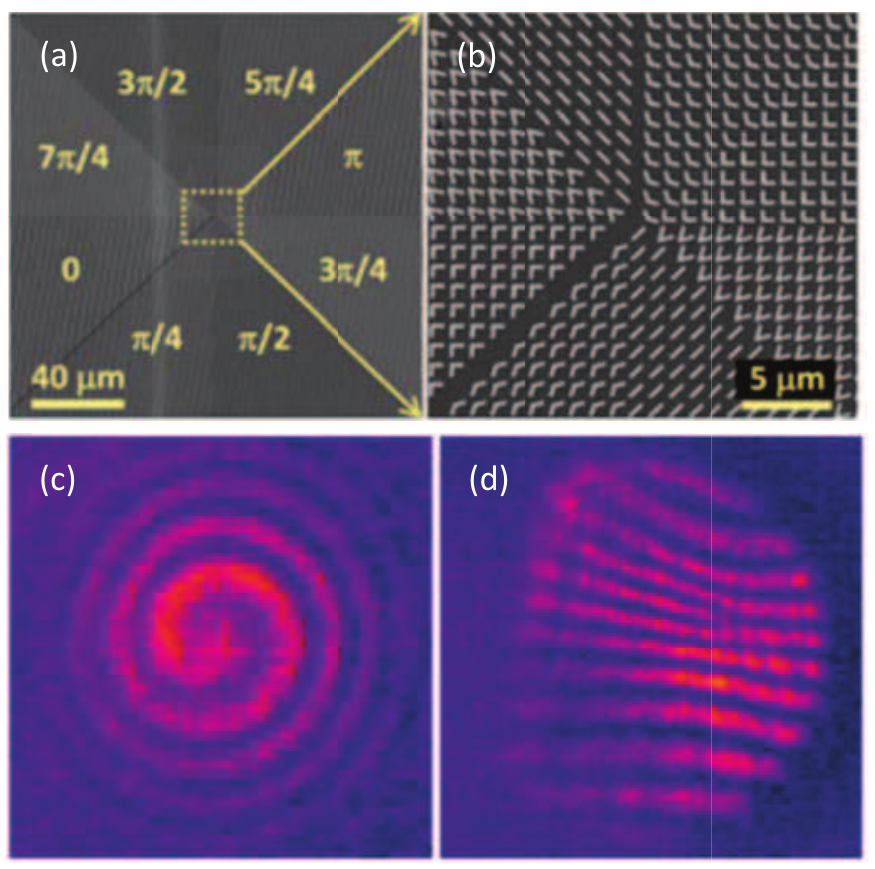}
\caption{(a) A metamaterial composed by a  sub-wavelength V-shaped antenna array. The entire structure consists of eight regions and the phase response difference between  adjacent regions is $\pi/4$. (b) Partially enlarged view of the center part of (a). (c) Measured pattern with a spiral fringe created by the interference of the vortex beam with $\ell=1$ and a co-propagating Gaussian beam. (d) Measured pattern with a dislocated fringe created by the interference of the vortex beam with $\ell=1$ and a Gaussian beam when the two are tilted with respect to each other \cite{Yu2011Light}.}
\label{Fig4}
\end{figure}

Transformation optics is a recently developed  discipline that employs  complex artificial materials, called \emph{metamaterials}, to make  transformations in optical space. Generation of optical OAM is one of the fields where the findings of transformation optics can  be applied: the metamaterials  are  planar ultra-thin optical components, usually composed of sub-wavelength constitutional units, such as V-shaped antennas \cite{Yu2011Light,Genevet2012Ultra},  L-shaped antennas \cite{Karimi2014Generating}, rectangular apertures \cite{Zhao2013Metamaterials}, and rectangular split-ring resonators \cite{Wang2015Ultra}. Rather than relying on the continuous transitions of  conventional optics, these planar ultra-thin optical components  operate by forcing a sudden change of beam phase, amplitude or polarization at the interface. Thus, optical OAM is obtained by controlling the geometrical parameters (shape, size, direction, etc.) of the  metamaterial to manipulate the phases of different azimuths and change the spatial phase of the incident light. As shown in Fig. \ref{Fig4}, the metamaterial structure is composed by a sub-wavelength V-shaped antenna array, and the beam phase can be controlled by adjusting the angle between the two arms of the V-shaped antenna. The entire surface is divided into eight regions, and the phase response difference between the adjacent regions is $\pi/4$. This object generates a helical phase shift relative to the front of the incident beam producing a vortex with the topological charge of $\ell = 1$. The biggest advantage of the use of metamaterials is their small size that allows for easy integration. Unfortunately, only a vortex beam with a particular topological charge can be generated by a single device.

\begin{figure}[t]
\centering
\includegraphics[scale=0.5]{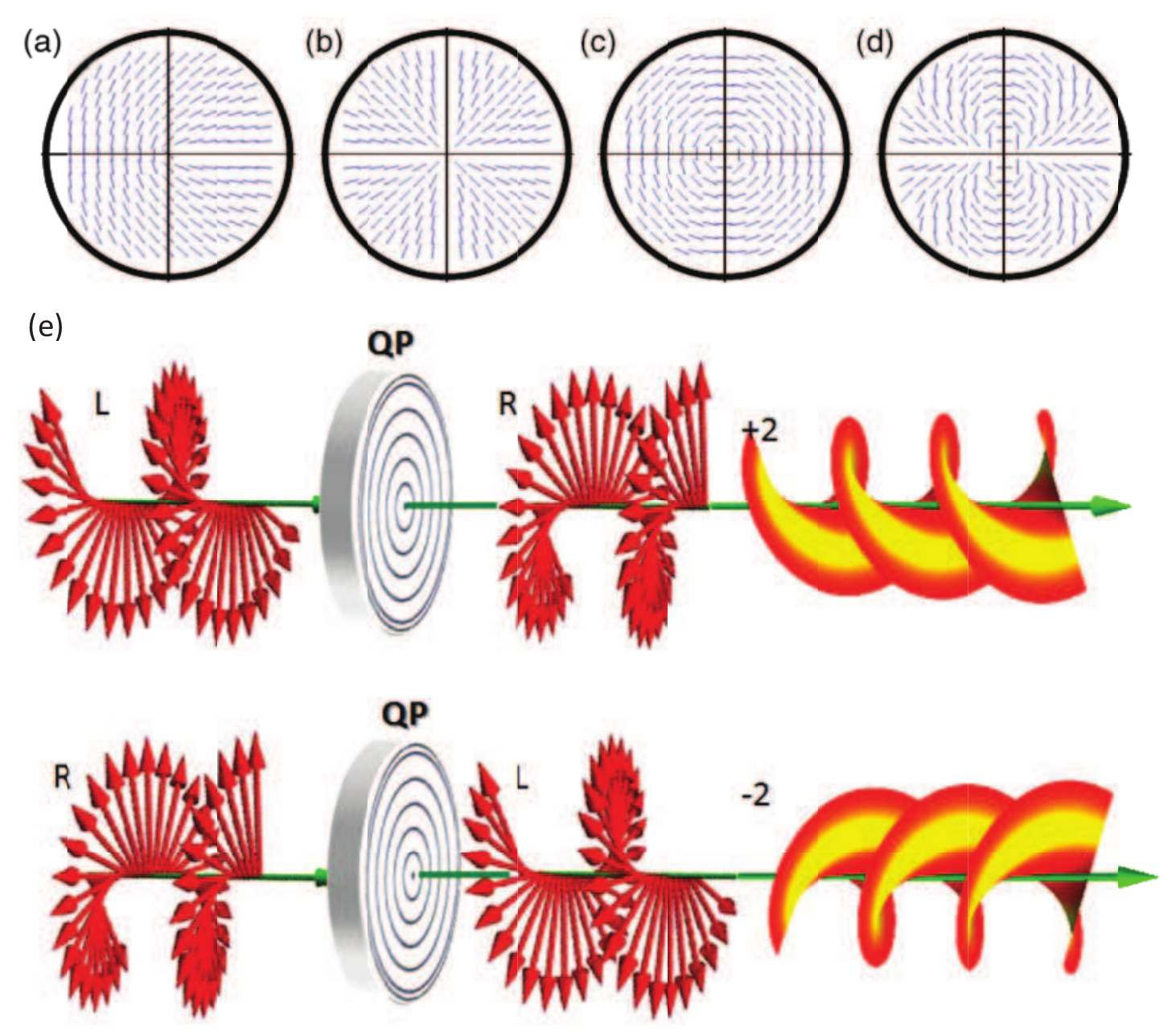}
\caption{Four examples of $q$-plate patterns with: (a) $(q,\alpha_0)=(1/2,0)$, (b) $(q,\alpha_0)=(1,0)$, (c) $(q,\alpha_0)=(1,\pi/2)$ and (d) $(q,\alpha_0)=(2,0)$. The segments indicate the optical axis orientation in the transverse plane. (e) Pictorial illustration of the optical action of a tuned $q$-plate with $q=1$ on an input circularly polarized plane beam. A left-circular (or right-circular) polarized Gaussian beam passing through a tuned $q$-plate with $q=1$ turns into a helically phased beam with $\ell=+2$ (or $\ell=-2$) and right-circular (or left-circular) polarization \cite{Marrucci2011Spin}.}
\label{Fig5}
\end{figure}

\begin{table*}[!bt]
\small
\centering
\caption{Comparison of Optical OAM Generation Methods.}
  \begin{tabular}{lllllll}
  \toprule
  \textbf{Features} &\textbf{Cylindrical lenses} &\textbf{SPPs} &\textbf{\tabincell{l}{Holographic \\ gratings}} &\textbf{SLMs} &\textbf{Metamaterials} &\textbf{\tabincell{l}{Liquid crystal \\ $q$-plates}} \\
  \midrule
  \specialrule{0em}{2pt}{2pt}
  \textbf{Cost} &Normal &Low &Low &High &Low &High \\
  \specialrule{0em}{2pt}{2pt}
  \textbf{Speed} &Normal &Normal &Fast &Normal &Normal &Fast \\
  \specialrule{0em}{2pt}{2pt}
  \textbf{Conversion efficiency} &High &High &Low &Normal &Relatively high &Relatively high \\
  \specialrule{0em}{2pt}{2pt}
  \textbf{OAM mode} &\tabincell{l}{Multiple modes; \\Pure mode}    &\tabincell{l}{Single mode; \\Non-pure mode}       &Multiple mode        &\tabincell{l}{Multiple modes; \\Composite mode} &Single mode  &\tabincell{l}{Single mode; \\Pure mode} \\
  \specialrule{0em}{2pt}{2pt}
  \textbf{Flexibility} &Low &Low &Low &High &Low &High \\
  \specialrule{0em}{2pt}{2pt}
  \textbf{Working frequency} &All &One point &All &All &A range &All \\
  \specialrule{0em}{2pt}{2pt}
  \textbf{Processing difficulty} &High &High &High &Low &High &Low \\
  \specialrule{0em}{2pt}{2pt}
  \textbf{System complexity} &High &Low &Low &Low &Low &Low \\
  \specialrule{0em}{2pt}{2pt}
  \textbf{Market readyness} &Yes &Yes &Yes &Yes &No &Yes \\
  \specialrule{0em}{2pt}{2pt}
  \bottomrule
  \label{Table1}
 \end{tabular}
\end{table*}

The \emph{$q$-plate} \cite{Marrucci2006Optical,Marrucci2011Spin} is a liquid crystal panel that has  a uniform birefringence phase retardation $\delta$ and a transverse optical axis pattern with a non-zero topological charge. The angle between the optical axis orientation and the $x$-axis in $xy$ plane can be expressed as $\alpha (r,\varphi)=q\varphi+\alpha_0$, where $\alpha_0$ is an initial optical axis orientation. $Q$-plate patterns with different values of $q$ and $\alpha_0$ are shown in Fig. \ref{Fig5}. When $q$-plate is optimally tuned, i.e. $\delta=\pi$, a light beam incident on it is modified to have a topological charge variation $\Delta\ell=\pm2q$. For example a circularly polarized Gaussian passing through a tuned $q$-plate with $q=1$ has a helical phase front with topological charges $\ell = \pm2$, where the sign depends on the chirality of the input polarized light, as shown in Fig. \ref{Fig5}(e). The characteristic of $q$-plates is that they generate OAM  as a result of optical spin-to-orbital angular momentum conversion, exploiting the law of conservation of angular momentum, and for $q=1$, the maximum topological charge that can be generated  is $\ell =2$. The $q$-plate is essentially a Pancharatnam-Berry phase optical component that controls the wavefront shape by transitioning the polarization state. In addition to the $q$-plate, the computer-generated sub-wavelength dielectric grating \cite{Hasman2002Space,Biener2002Formation} and the L-shaped antenna array metamaterial proposed in \cite{Karimi2014Generating} are also Pancharatnam-Berry optical devices.  The $q$-plate is capable of producing a pure OAM mode but, being made of liquid crystal, has  also an energy threshold and thus cannot be used in conjunction with high power laser beams.

Table \ref{Table1} compares the different  methods of generating  optical OAM  with respect to several parameters: \emph{cost}, \emph{speed}, \emph{conversion efficiency}, the ratio of outgoing power to incident power, \emph{OAM mode}, the capacity of  generating pure integer  modes, multiple modes, i.e. several different pure modes,  or composite modes OAM, \emph{flexibility}, the capacity of controlling the OAM wavefront parameters such as the beam direction or the size of OAM beam ring, \emph{working frequency}, \emph{processing difficulty}, the difficulty in manufacturing the generator,  \emph{system complexity}, the level of difficulty of integrating the generator in a transmission system, \emph{market readyness}, the availabilty on the market of the technology as a product.
As can be seen, the lesson learned is that each method has advantages and disadvantages. Taking the cylindrical lens as an example, it has high mode conversion efficiency and generates OAM modes with high purity, but processing difficulty and system complexity are high.
The OAM generation method should be chosen according to practical considerations. Flexibility is the biggest advantage of SLMs. If different OAM modes or superimposed modes are required, SLMs should be the first choice. Besides, SPPs perform well in cost, conversion efficiency and structural complexity. If mode purity and processing difficulty are not considered, SPPs are more recommended in a single OAM mode application due to the  low cost and the  simple structure.
\par
In optical OAM communication systems, SLMs loaded with phase holograms are widely used to generate OAM, and SPPs are second only to SLMs. In general, cylindrical lenses require a special incident field that limits their application. At the same time, low conversion efficiency makes the application of holographic gratings much less practical than SLMs and SPPs. Metamaterials are low cost, small size and high efficiency, so they are more suitable for small integrated vortex beam generators. The $q$-plate is a spin-orbital angular momentum conversion device and as such is not recommended yet for OAM communications because of the low value of the  maximum topological charge and  is more suitable for particle manipulation.

\begin{figure}[t]
\centering
\includegraphics[scale=0.6]{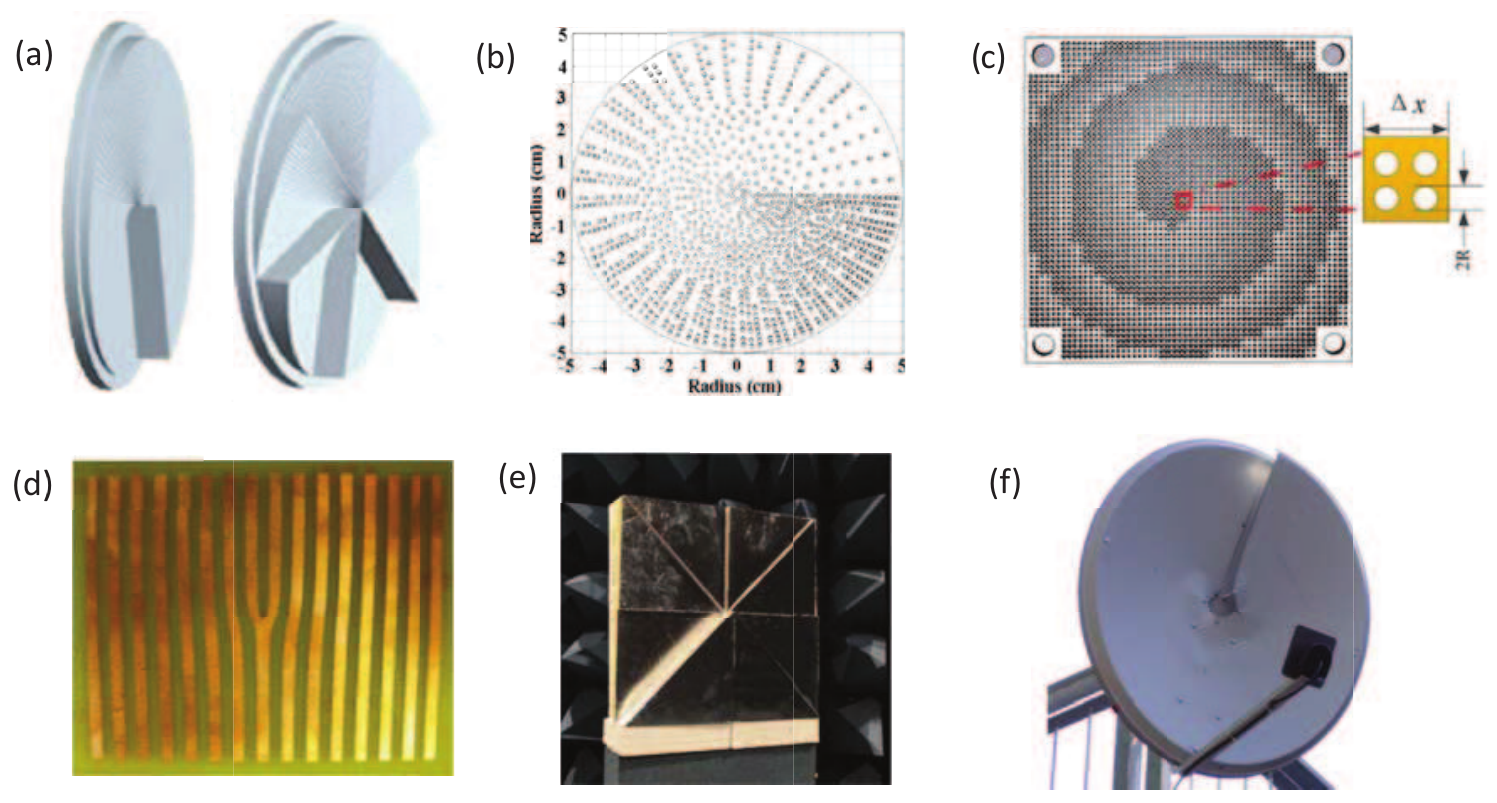}
\caption{(a) Single-stage and multi-stage SPPs \cite{Zhu2014Experimental}, planar SPPs with varying borehole densities (b) \cite{Bennis2013Flat} and radius (c) \cite{Cheng2014Generation}, (d) forked gratings  \cite{Mahmouli2012Orbital}, (e) stepped spiral reflecting surface \cite{Tamburini2011Experimental} and (f) spiral parabolic antenna \cite{Tamburini2012Encoding}.}
\label{Fig6}
\end{figure}

\subsection{Radio OAM}
OAM is a fundamental characteristic of electromagnetic waves at  all frequencies. It is not limited to the optical band and can be generated also in the radio band \cite{Thid2007Utilization}.

In analogy with the generation of optical OAM,  \emph{SPPs}, whose  thickness increases with the increase of the azimuth angle, have been the first method used to generate OAM \cite{Turnbull1996The,Yan2014Demonstration}  at high radio frequencies. 
To generate a high-order vortex beam, the spiral surface is usually replaced by a stepped surface to form a stepped SPP \cite{Zhu2014Experimental} as shown in Fig. \ref{Fig6} (a).
Recently, also planar SPPs \cite{Bennis2013Flat,Cheng2014Generation} have been proposed for generating radio OAM. Unlike conventional SPPs, which use thickness to change the beam phase, planar phase plates adjust the wavefront of the electromagnetic wave by controlling the change in dielectric constant using different borehole densities \cite{Bennis2013Flat} or radius \cite{Cheng2014Generation}. Although these devices have a planar structure, like ordinary phase plates, the requirements in terms of  structural precision is very high and the plate is difficult to manufacture.

\begin{table*}[!bt]
\small
\centering
\caption{Comparison of Radio OAM Generation Methods.}
  \begin{tabular}{lllllll}
  \toprule
  \textbf{Features} &\textbf{SPPs} &\textbf{\tabincell{l}{Holographic \\gratings}} &\textbf{Spiral reflectors} &\textbf{UCAAs} &\textbf{Metamaterials} &\textbf{\tabincell{l}{Dielectric \\$q$-plates}} \\
  \midrule
  \specialrule{0em}{2pt}{2pt}
  \textbf{Cost} &Low &Low &Low &High &Low &Low \\
  \specialrule{0em}{2pt}{2pt}
  \textbf{Speed} &Normal &Fast &Normal &Normal &Fast &Normal \\
  \specialrule{0em}{2pt}{2pt}
  \textbf{Conversion efficiency} &High &Low &Normal &Normal &Relatively high &/ (Not discussed)\\
  \specialrule{0em}{2pt}{2pt}
  \textbf{OAM mode} &\tabincell{l}{Single mode; \\Non-pure mode}   &Multiple mode   &\tabincell{l}{Single mode; \\Non-pure mode}   &\tabincell{l}{Multiple modes; \\Composite mode}  &Single mode  &\tabincell{l}{Single mode; \\Pure mode} \\
  \specialrule{0em}{2pt}{2pt}
  \textbf{Flexibility} &Low &Low &Low &High &Low &Low \\
  \specialrule{0em}{2pt}{2pt}
  \textbf{Working frequency} &One point  &All  &One point  &All  &One point  &One point \\
  \specialrule{0em}{2pt}{2pt}
  \textbf{Processing difficulty} &High &High &High &Low &High &High \\
  \specialrule{0em}{2pt}{2pt}
  \textbf{System complexity} &Low &Low &Low &High &Low &Low \\
  \specialrule{0em}{2pt}{2pt}
  \textbf{Market readyness} &No &No &No &Yes &No &No \\
  \specialrule{0em}{2pt}{2pt}
  \bottomrule
  \label{Table2}
 \end{tabular}
\end{table*}

To create OAM, the stepped spiral surface can be also built as a reflector illuminated by a conventional antenna \cite{Tamburini2011Experimental}. The \emph{spiral reflector} causes the reflected electromagnetic wave to have a wave path difference at different positions of the cross section, so that the desired electromagnetic vortex is created. As in Fig. \ref{Fig6} (e), the reflecting stepped spiral surface is divided into $N$ discrete regions, each of which introduces a  discontinuous phase difference $2\pi/N$. The topological charge of radio OAM of stepped surfaces is $\ell=2{\mu_0}(N+1)/(\lambda N)$, where $N$ is the number of regions into which the reflection surface is partitioned, $\mu_0$ is the total surface spacing, and $\lambda$ is the wavelength of the incident electromagnetic wave. When $N$ tends to infinity, the stepped spiral reflector becomes a continuous form, such as the spiral parabolic antenna used in the 2012 OAM wireless communication experiment \cite{Tamburini2012Encoding}. The spiral reflector is simple and easy to implement. However, it can only respond to specific frequencies and the OAM mode produced by this method is difficult to determine.

\begin{figure}[t]
\centering
\includegraphics[scale=0.8]{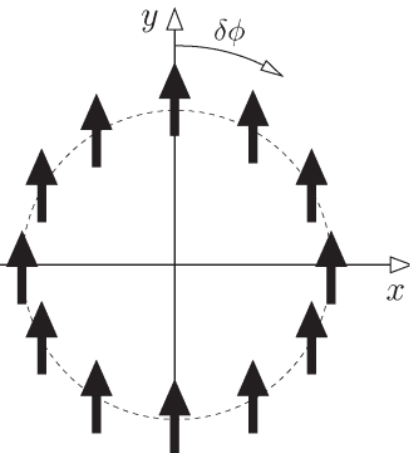}
\caption{Schematic diagram of a UCAA with $N$ elements uniformly distributed on the circumference, where the azimuth difference between adjacent elements is $\delta\phi=2\pi/N$ \cite{Mohammadi2010Orbital}.}
\label{Fig7}
\end{figure}

\emph{Holographic gratings} designed by means of computer-generated holograms are also a method of generating radio OAM in analogy with the generation method at optical frequencies \cite{Mahmouli2012Orbital}.

Nevertheless, a \emph{uniform circular antenna array} (UCAA),  as shown in Fig. \ref{Fig7}, is the most commonly used antenna for generating electromagnetic vortex \cite{Thid2007Utilization}, \cite{Mohammadi2010Orbital}. The UCAA is composed by $N$ elements, uniformly distributed on the circumference. Each array element is controlled by an input signal of the same amplitude but different phase and the phase difference between adjacent array elements is $\Delta\varphi=2\pi \ell/N$, where $\ell$ is the OAM mode, so that the phase difference over the whole array is $2\pi \ell$. Theoretically, a UCAA with $N$ array elements can produce a distortion-free vortex wave with $|\ell|<N/2$.
By precisely controlling the amplitude and the phase of the excitation signals, UCAAs have the potential to generate multiplexed radio OAM. In practice, UCAAs can be used in different forms of antennas, such as dipole antennas \cite{Thid2007Utilization, Mohammadi2010Orbital}, Vivaldi antennas \cite{Deng2013Generation}, horn antennas \cite{Bai2015Generation}, Yagi antennas \cite{Gaffoglio2016OAM}, and microstrip patch antennas \cite{Bai2013Generation,Sun2016The}. In addition to UCAAs, radio OAM generation with imperfect uniform circular arrays (IUCAs) has also be discussed in \cite{Lin2017Theoretical}.
In \cite{Tennant2013Generation} a circular time switched array (TSA) is employed  to generate vortex electromagnetic waves. This TSA method uses high-speed RF switches to activate the array elements sequentially, enabling the simultaneous generation of all OAM modes at the harmonic frequencies of the TSA switching cycle. The mode of the electromagnetic vortex generated by the UCAA can be controlled flexibly, but this usually requires a complicated feed network.

\begin{figure}[t]
\centering
\includegraphics[scale=0.4]{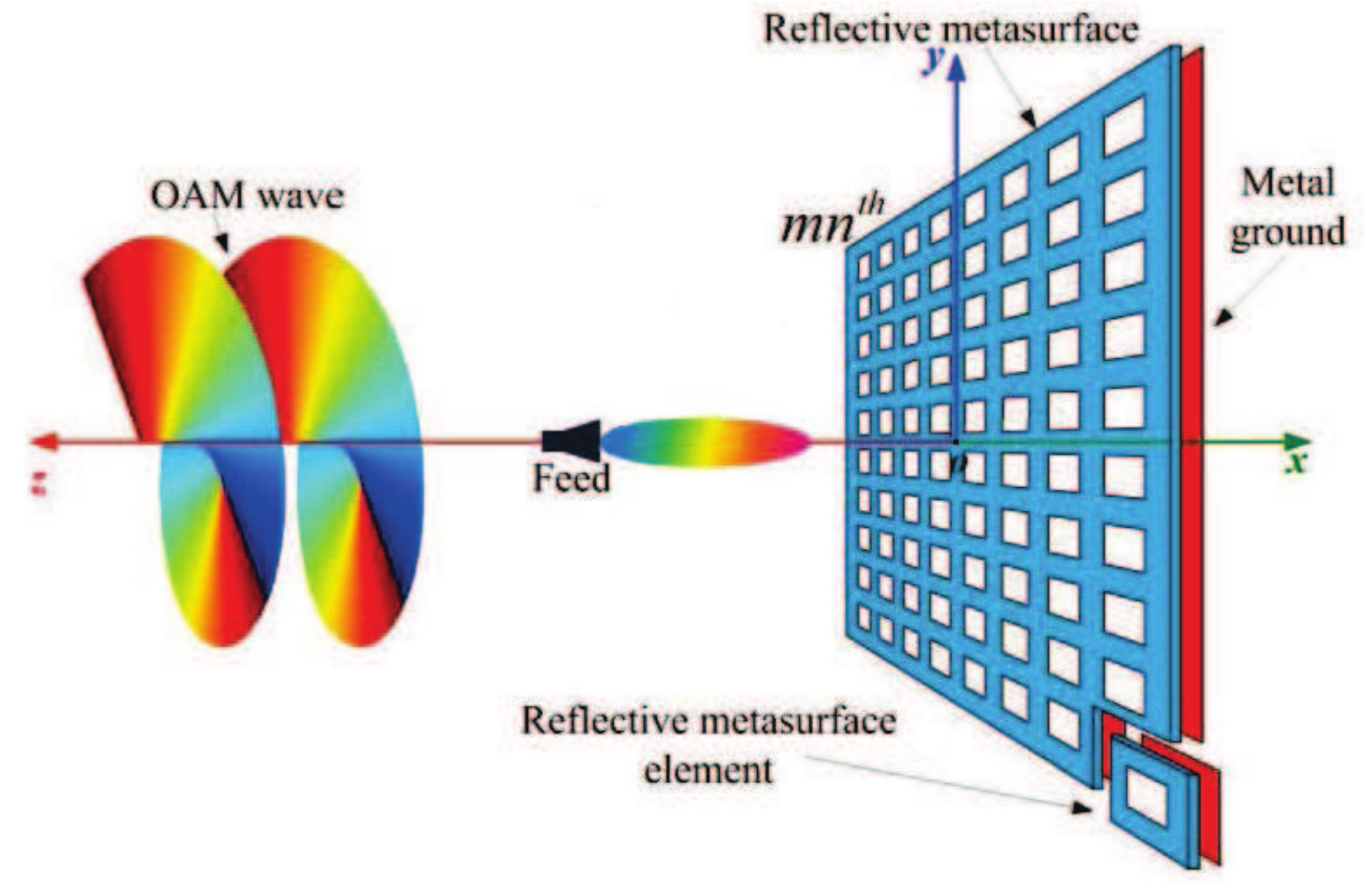}
\caption{Pictorial illustration of generating OAM with a reflective metamaterial \cite{Yu2016Design}.}
\label{Fig8}
\end{figure}

Similar to the generation method at optical frequencies, \emph{metamaterials}, which can be classified as reflective metamaterials \cite{Chen2016Artificial,Yu2016Design, Yu2016Generating,Yu2016Dual,Zhang2018Transforming} and transmissive metamaterials \cite{Kou2016Generation}, are also used to generate vortex electromagnetic waves. The schematic diagram of generating OAM using a reflective metamaterial is shown in Fig. \ref{Fig8}, where the incident wave at radio frequencies is reflected and converted  into a vortex  wave.

In addition to metamaterials, also \emph{$q$-plates}, made by carving a series of concentric rings on a dielectric  plate, can be used to generate a vortex wave at radio frequencies \cite{Maccalli2013q}.

Table \ref{Table2} compares the most common methods of generating vortex waves at radio frequencies. Normally, because their size depends on the signal wavelength, SPPs and holographic gratings are suitable for the generations of vortex waves at high frequencies such as millimeter-waves, while spiral reflectors and UCAAs are more suitable for OAM generation at lower frequencies. It is well known that massive multiple-input multiple-output (MIMO) technology is going to be one of the key technologies for $5$G mobile networks. Therefore, UCAAs have attracted a wide attention and it is currently under study the  introduction of OAM into $5$G or beyond $5$G networks. In addition, thanks to their capability of  controlling the wavefront, radar imaging systems frequently uses UCAAs to generate OAM waves. The low cost, high conversion efficiency and simple structure of SPPs make them recommended for radio OAM communication.  Since the spiral parabolic antenna can focus the OAM wave while it is being generated, it is more suitable for long distance transmissions. Moreover, SPPs, spiral parabolic antennas and other single transmit antennas are expected to be combined with MIMO technology to further increase capacity. Consistently with what happens with optical OAM, low conversion efficiency makes holographic gratings less useful, and metamaterials are more suitable for miniaturized integrated circuits due to the small size and  low cost. Also in this scenario, $q$-plates are not particularly recommended for radio OAM communications.

\subsection{Acoustic OAM}
OAM has found applications in audio bands as well \cite{Hefner1999An}.
Because of their simple structure and high conversion efficiency, SPPs are widely used to generate optical vortices and radio vortices. Similarly, \emph{SPPs} have also been employed to generate acoustic vortices. The acoustic SPP in \cite{Wunenburger2015Acoustic} has thickness $h$ that varies with azimuth, $h=\ell\theta/[2\pi f(c_0^{-1}-c^{-1})]$, where $c_0$ is the sound speed in the surrounding medium, $c$ is the sound speed in the phase plate, and $f$ is the frequency of the incident wave. Another absorbing SPP using optoacoustic technology was proposed in 2004 to generate OAM in the ultrasonic band \cite{Gspan2004Optoacoustic}. The difference is that the optoacoustic conversion efficiency of this SPP is very low, and this device requires high-energy short pulse excitation. In addition to passive SPPs, active sound sources with spiral thickness \cite{Hefner1998Acoustical, Ealo2011Airborne} are used to generate acoustic vortices. The dimensions of those spiral structures are generally dozens of wavelengths and their large volume limits the application in low frequency sound waves.

\begin{figure}[t]
\centering
\includegraphics[scale=0.8]{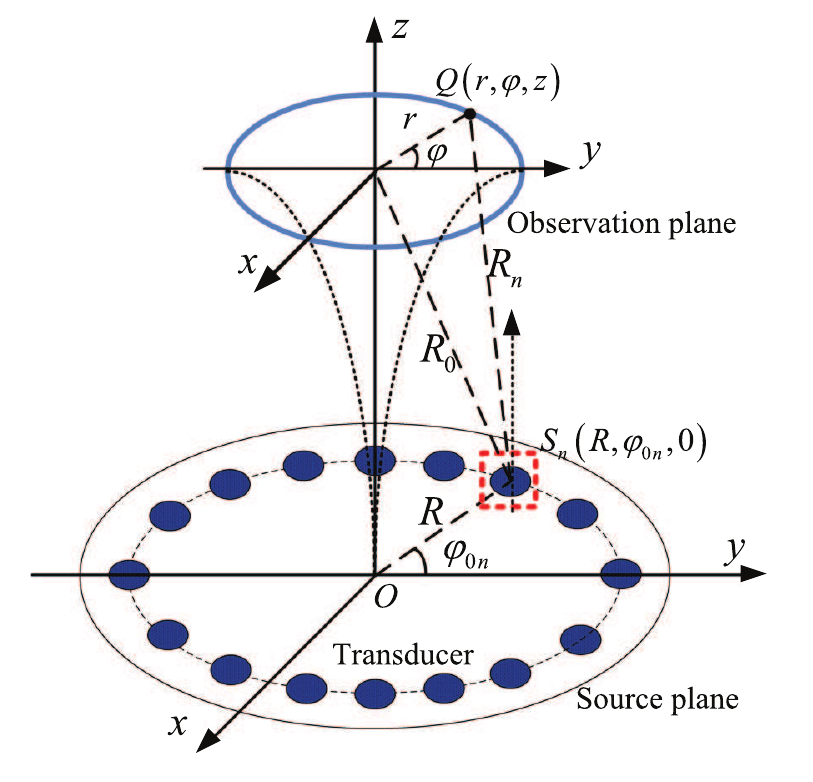}
\caption{Schematic of an acoustic vortex generation system using a UCTA composed by $N$ uniformly distributed elements. The positions of the $n$-th sound source $S_n$ and the observation point $Q$ are $(R, \varphi_{0n}, 0)$ and $(r, \varphi, z)$, respectively \cite{Li2017Deep}.}
\label{Fig9}
\end{figure}

In analogy with antenna arrays in radio OAM, transducer arrays especially \emph{uniform circular transducer arrays} (UCTAs) are also widely used to generate acoustic vortices carrying OAM. A simple four-panel transducer has first been proposed to produce acoustic vortices \cite{Hefner1999An}. Subsequently, hexagonal arrays \cite{Marchiano2005Synthesis} and UCTAs \cite{Volke2008Transfer, Yang2013Phase, Li2017Deep} have also been studied:  UCTAs use fewer transducers than hexagonal arrays, and, by adjusting the phase difference between adjacent transducers, are capable to  flexibly generate acoustic vortices with different OAM modes. Fig. \ref{Fig9} shows a UCTA composed by $N$ uniformly spaced  elements: by driving each transducer  with a sine wave with frequency $f$ and phase difference $\Delta\varphi=2\pi \ell/N$, the array can generate an acoustic vortex with $|\ell |\leq \lfloor(N- 1)/2\rfloor$, where $\lfloor x \rfloor$ indicates the greatest integer less than or equal to $x$. By precisely controlling the amplitude and the phase of the excitation signals, UCTAs can generate  multiplexed acoustic vortices.
It should be noted that each transducer can be regarded as a point source in low frequency acoustic fields, but not at high frequencies. In high frequency acoustic fields, the directivity of the sound source needs to be considered  because the size of the transducer is larger than the wavelength of the sound wave. Like  UCAAs, UCTAs also require complex  control circuits, and the complexity grows  with the  number of transducers.

\begin{table*}[tb]
\small
\centering
\caption{Comparison of Acoustic OAM Generation Methods.}
  \begin{tabular}{lllll}
  \toprule
  \textbf{Features} &\textbf{SPPs} &\textbf{UCTAs} &\textbf{\tabincell{l}{Spiral diffraction \\ gratings}} &\textbf{Metamaterials} \\
  \midrule
  \specialrule{0em}{2pt}{2pt}
  \textbf{Cost} &Low &High &Low &Low \\
  \specialrule{0em}{2pt}{2pt}
  \textbf{Speed} &Normal &Normal &Normal &Normal \\
  \specialrule{0em}{2pt}{2pt}
  \textbf{Conversion efficiency} &High  &Normal &Low &Relatively high \\
  \specialrule{0em}{2pt}{2pt}
  \textbf{OAM mode} &\tabincell{l}{Single mode; \\Non-pure mode}      &\tabincell{l}{Multiple modes; \\Composite mode}  &Multiple modes  &Single mode\\
  \specialrule{0em}{2pt}{2pt}
  \textbf{Flexibility} &Low &High &Low &Low \\
  \specialrule{0em}{2pt}{2pt}
  \textbf{Working frequency} &One point  &All  &A range  &One point \\
  \specialrule{0em}{2pt}{2pt}
  \textbf{Processing difficulty} &High &Low &High &High \\
  \specialrule{0em}{2pt}{2pt}
  \textbf{System complexity} &Low &High &Low &Low \\
  \specialrule{0em}{2pt}{2pt}
  \textbf{Market readyness} &No &Yes &No &No \\
  \specialrule{0em}{2pt}{2pt}
  \bottomrule
  \label{Table3}
 \end{tabular}
\end{table*}

\begin{figure}[t]
\centering
\includegraphics[scale=0.6]{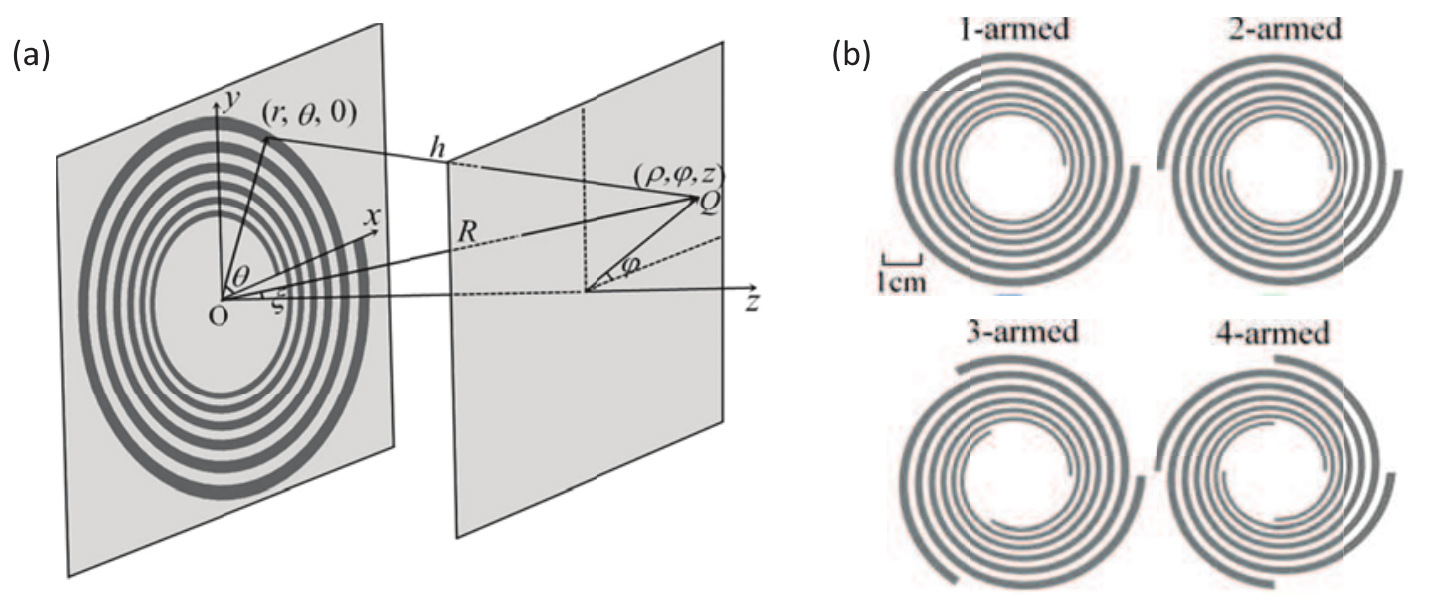}
\caption{(a) The multi-arm coiling slits using logarithmic spiral gratings for generating the stable acoustic vortices. (b) Schematic diagram of the one-armed, two-armed, three-armed, and four-armed coiling slits \cite{Jiang2016Broadband}.}
\label{Fig10}
\end{figure}

Another method for generating acoustic vortices is to employ sub-wavelength \emph{spiral diffraction gratings} depicted on rigid materials, which, due to the diffraction effect,  can generate acoustic vortices in the paraxial region. Higher mode acoustic OAM can be produced by increasing the number of arms of spiral gratings. Using a grating with $m$ arms, an acoustic vortex with the topological charge of $\ell= mn $ is obtained at the $n$th diffraction order. Spiral gratings can be designed as  Archimedes spiral type \cite{Wang2016Particle,Jim2016Formation}, Fresnel spiral type \cite{Jim2018Sharp}, and logarithmic spiral type \cite{Jiang2016Broadband} shown in Fig. \ref{Fig10}. The first two types can only be used for a specific frequency, while the logarithmic spiral type is for a wider frequency range. In polar coordinates, when $m=1$ the two concentric logarithmic spirals are denoted as $r_1=a_1 \exp (b\theta)$ and $r_2=a_2 \exp (b\theta)$, respectively, where $ \theta $ is the angular coordinate, $r_{1}$ and $r_{2}$ are the radial coordinates,  $a_1$ and $a_2$ are the inner radii of the two spirals, $b$ varies with the number of arms, determining the growth rate of the spiral and the width of the spiral grating is $d = r_2-r_1$. As long as the wavelength of the incident acoustic wave  is greater than $2d_{min}$ and less than $2d_{max}$, effective diffraction can occur to obtain the desired acoustic OAM. Since the rigid material blocks a large amount of incident power, such a passive spiral diffraction grating has a low efficiency. Recently, an active spiral grating fabricated using a cellular ferroelectret film  has been proposed to improve conversion efficiency \cite{Muelas2018Generation}.

\begin{figure}[t]
\centering
\includegraphics[scale=0.9]{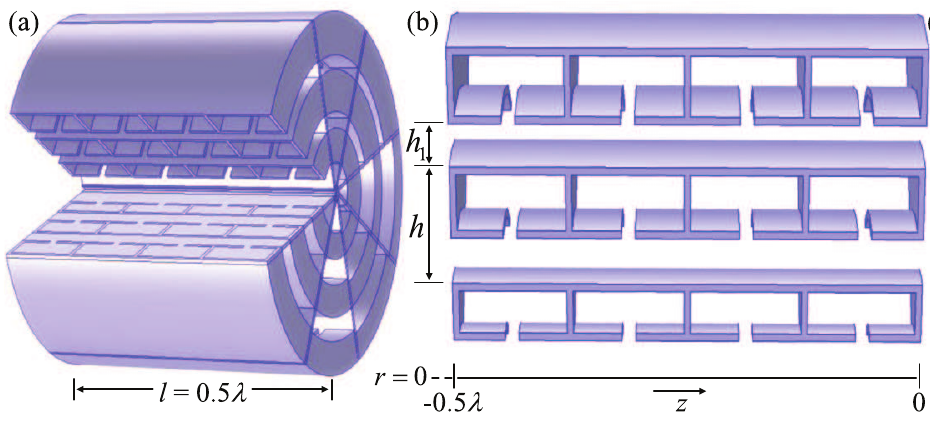}
\caption{(a) Schematic of the assembled planar acoustic resonance layer consisting of eight fanlike sections. The thickness of the layer is $0.5\lambda$. (b) An individual section consisting of three rows of resonators with fixed height $h$ in the radial $r$ direction, sided by pipes of varying height $h_1$ to produce the needed effective wave number \cite{Jiang2016Convert}.}
\label{Fig11}
\end{figure}

In addition to optical OAM and radio OAM, \emph{metamaterial} can also be used to generate acoustic OAM \cite{Ye2016Making,Esfahlani2017Generation,Jiang2016Convert}. Usually, to generate acoustic OAM a metasurface with sub-wavelength thickness is equally divided into $N$ regions. By changing  structural parameters in each region, the phase of the outgoing acoustic wave can be adjusted to be spirally distributed along the azimuthal direction. A planar metamaterial structure for converting planar acoustic waves into vortex acoustic waves using acoustic resonance is shown in Fig. \ref{Fig11}. The metamaterial is divided into eight fanlike sections. Each individual section is configured to be composed of three rows of resonators in the radial direction. Each row consists of four Helmholtz cavities and a straight pipe. The phase of the outgoing sound wave can be expressed as $\phi_{out}=\phi_{in}+k^{(eff)}l$, where $k^{(eff)}$ is  the equivalent wave number and varies with the ratio  of the heights of the pipe and resonator $h_1/h$  and $l$ is the structural thickness. It can be shown that when $k^{(eff)}$ depends on azimuth change, the output phase also changes with azimuth. The cascade of four Helmholtz cavities in each row can  flexibly control  $k^{(eff)}$ from $-k$ to $k$ over the whole azimuth, ie, achieving an exit phase from 0 to $2\pi$, while the combination of the Helmholtz cavity and the straight pipe can get a very high transmittance. The phase difference between adjacent sections of the metamaterial is $\pi/4$,  so that  planar acoustic waves are converted into vortex sound waves with the topological charge of $\ell =1$. This metamaterial structure has the advantages of being highly efficient with a small size and a planar shape. However, the proposed structure can only be used for the generation of acoustic vortices at specific frequencies in the air.

Table \ref{Table3} compares the generation methods of acoustic OAM. Due to the dimension of the structure, SPPs are more suitable for high frequency sound fields and UCTAs are more suitable for low frequency sound fields. Because of their flexibility,  UCTAs are more common in acoustic communications and particle manipulation. Both spiral diffraction gratings and metamaterials are characterized by sub-wavelength dimensions. However, passive diffraction gratings have low efficiency because a large amount of incident sound field energy is blocked and lost, and active diffraction gratings are more expensive. Therefore, metamaterials with small size, low cost, and high efficiency have attracted a lot of attention recently.

\begin{table*}[tb]
\small
\centering
\caption{Summary of OAM Generation Methods.}
  \begin{tabular}{p{3.2cm}ll}
  \toprule
  \textbf{Methods} &\textbf{Introduction and Features} &\textbf{Remarks} \\
  \midrule
  \specialrule{0em}{2pt}{2pt}
  \textbf{Cylindrical lenses} &\tabincell{p{10cm}}{HG modes without OAM can be converted to LG modes with OAM. \\High conversion efficiency; high OAM mode purity; a special incident field is required. }  &\tabincell{l}{Optical OAM} \\
  \specialrule{0em}{2pt}{2pt}
  \cmidrule{1-3}
  \specialrule{0em}{2pt}{2pt}
  \textbf{SPPs}  &\tabincell{p{10cm}}{With spiral thickness $h=\ell\lambda\theta/[2\pi(n-n_0)]$ in optics and electromagnetics and $h=\ell\theta/[2\pi f(c_0^{-1}-c^{-1})]$ in acoustics. \\Simple structure; high precision; high conversion efficiency; specific working frequency.} &\tabincell{l}{Optical OAM \\Radio OAM \\Acoustic OAM} \\
  \specialrule{0em}{2pt}{2pt}
  \cline{1-3}
  \specialrule{0em}{2pt}{2pt}
  \textbf{Holographic gratings}  &\tabincell{p{10cm}}{Interference patterns of plane waves and vortex waves, such as $\ell$-fold forked diffraction gratings, whose transmittance function is related to $\exp(i\ell\theta)$. \\Simple and fast; low efficiency. }  &\tabincell{l}{Optical OAM \\Radio OAM} \\
  \specialrule{0em}{2pt}{2pt}
  \cline{1-3}
  \specialrule{0em}{2pt}{2pt}
  \textbf{SLMs}  &\tabincell{p{10cm}}{A pixelated liquid crystal device loaded with a phase hologram whose transmittance function related to $\exp(i\ell\theta)$. \\High cost; flexible control. }  &\tabincell{l}{Optical OAM} \\
  \specialrule{0em}{2pt}{2pt}
  \cline{1-3}
  \specialrule{0em}{2pt}{2pt}
  \textbf{UCAAs} &\tabincell{p{10cm}}{$N$ elements uniformly distributed on the circumference with the excitation phase difference $2\pi\ell/N$ can generate radio OAM with $|\ell|<N/2$. \\Flexible control; a complicated feed network is required. }  &\tabincell{l}{Radio OAM} \\
  \specialrule{0em}{2pt}{2pt}
  \cline{1-3}
  \specialrule{0em}{2pt}{2pt}
  \textbf{UCTAs} &\tabincell{p{10cm}}{Similar to UCAAs. } &\tabincell{l}{Acoustic OAM} \\
  \specialrule{0em}{2pt}{2pt}
  \cline{1-3}
  \specialrule{0em}{2pt}{2pt}
  \textbf{Spiral reflectors} &\tabincell{p{10cm}}{The reflection surfaces are divided into $N$ discrete regions and can generate OAM with $\ell=2{\mu_0}(N+1)/(\lambda N)$. \\Simple structure; beam convergence effect; it is difficult to determine topological charge values; specific working frequency. } &\tabincell{l}{Radio OAM} \\
  \specialrule{0em}{2pt}{2pt}
  \cline{1-3}
  \specialrule{0em}{2pt}{2pt}
  \textbf{Spiral diffraction gratings} &\tabincell{p{10cm}}{Archimedes spiral type, Fresnel spiral type or logarithmic spiral type gratings depicted on sub-wavelength rigid materials. \\Low efficiency; the topological charge is changed by controlling the number of arms of the grating. } &\tabincell{l}{Acoustic OAM} \\
  \specialrule{0em}{2pt}{2pt}
  \cline{1-3}
  \specialrule{0em}{2pt}{2pt}
  \textbf{Metamaterials} &\tabincell{p{10cm}}{It is composed of sub-wavelength constitutional units and can flexibly manipulate wavefront phase, amplitude and polarization. \\Low cost; small size; easy integration; high conversion efficiency; specific working frequency. } &\tabincell{l}{Optical OAM \\Radio OAM \\Acoustic OAM} \\
  \specialrule{0em}{2pt}{2pt}
  \cline{1-3}
  \specialrule{0em}{2pt}{2pt}
  \textbf{Q-plates} &\tabincell{p{10cm}}{A uniform birefringent liquid crystal or dielectric plate that can generate OAM as a result of optical spin-to-orbital angular momentum conversion. \\High conversion efficiency; fast response; high OAM mode purity; simple chiral control of OAM beams. } &\tabincell{l}{Optical OAM \\ Radio OAM} \\
  \specialrule{0em}{2pt}{2pt}
  \bottomrule
  \label{Table4}
 \end{tabular}
\end{table*}

\begin{figure}[t]
\centering
\includegraphics[scale=0.7]{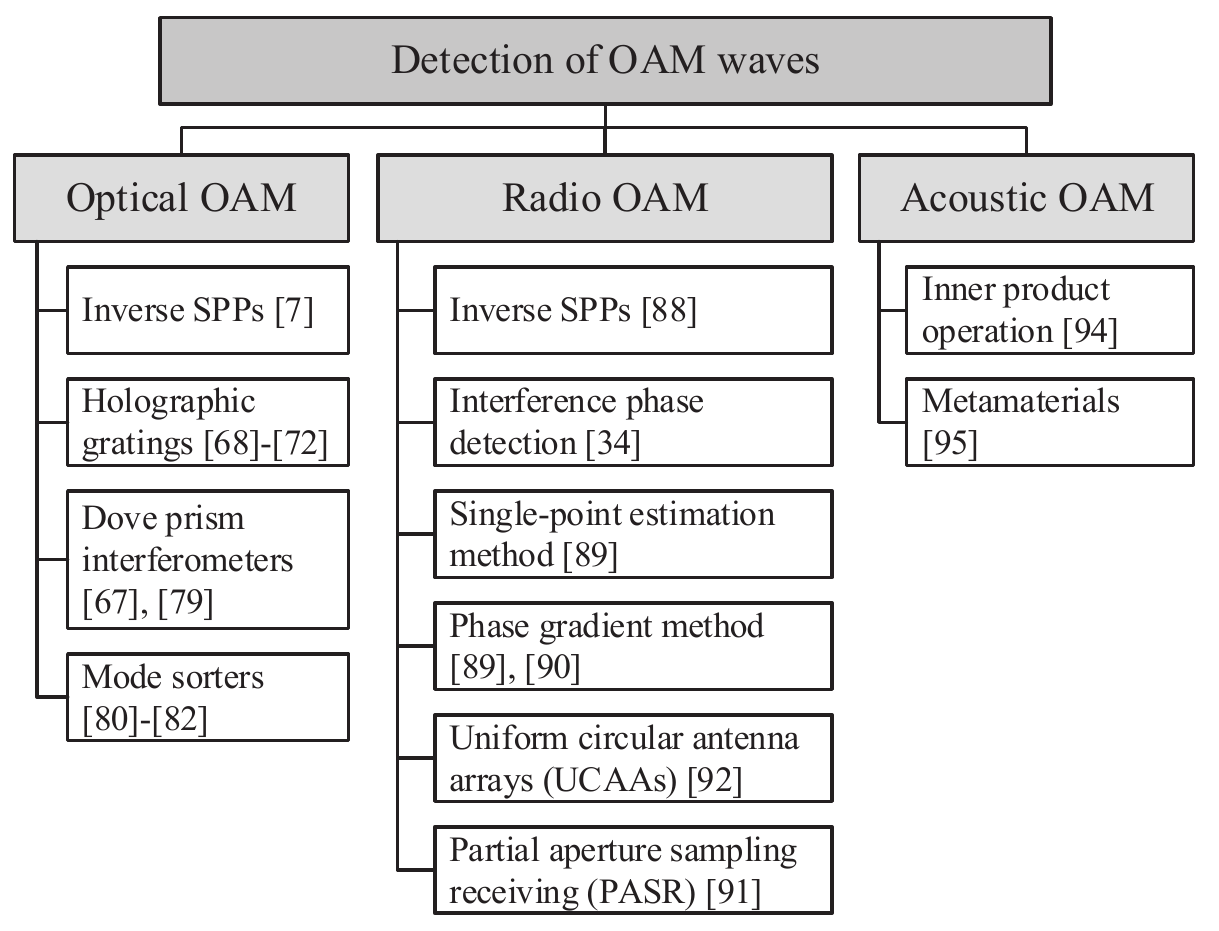}
\caption{Taxonomy diagram of detection of OAM waves.}
\label{Fig32}
\end{figure}

\subsection{Summary and Open Challenges}
Table \ref{Table4} summarizes the most important methods for generating OAM. Optical OAM  generation is a field that has been developed relatively earlier with respect to  radio OAM and acoustic OAM generation. Moreover many methods, such as SPPs and metamaterials, are proposed by analogy with  optical OAM  for radio and acoustic bands. Nevertheless, based on their respective characteristics, specific methods have been proposed to generate radio OAM and acoustic OAM, such as UCAAs and UCTAs.

The methods of generating optical, radio, and acoustic OAM have been compared in their respective subsections. In general, SLMs are relatively optimal in optical OAM generation. SPPs and UCAAs have also proven to be good at generating radio OAM in practice. For acoustic OAM generation, UCTAs are flexible in both air and water. However, the inherent size of the transducer makes UCTAs not suitable for many miniaturized applications. The metamaterials with small size proposed so far can only generate OAM in the air. Nevertheless, underwater acoustic communication is  a research area of great interest and so, finding a new material or structure with low cost, small size and high efficiency to generate a stable acoustic vortex field underwater is an important research direction. In addition, the capacity of focusing the beam for long distance transmissions is essential. How to generate convergent OAM beams is a challenging topic  in optical, radio and acoustic research.

\section{Detection of OAM Waves}
In this section, we  discuss the various methods for detecting OAM waves. At the receiver of a communication system, the different  OAM modes can be  separated easily by exploiting the orthogonality  of  the helical phase fronts. A variety of methods for detecting OAM have been proposed for light and radio waves. In the acoustic frequency range, OAM detection is still in its early days. A taxonomy diagram of the detection methods discussed here is shown in Fig. \ref{Fig32}.

\begin{figure}[t]
\centering
\includegraphics[scale=1]{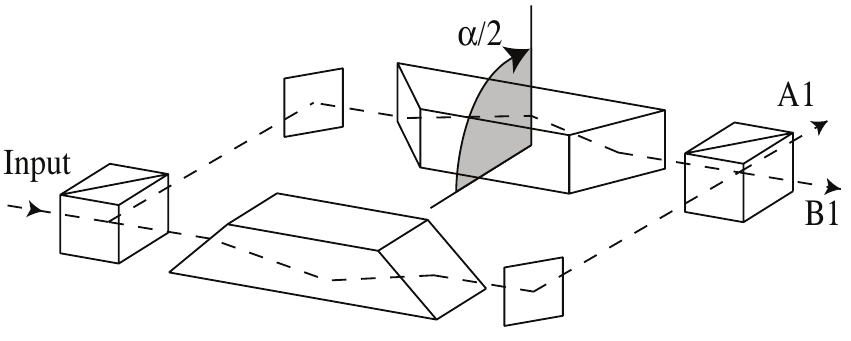}
\caption{A Dove prism interferometer using a Mach-Zehnder interferometer with a Dove prism placed in each arm, where $\alpha/2$ is the relative angle between the Dove prisms. If $\alpha/2=\pi/2$, it sorts photons with even-$\ell$ into port A1 and photons with odd-$\ell$ into port B1 \cite{Leach2002Measuring}.}
\label{Fig12}
\end{figure}

\begin{figure}[t]
\centering
\includegraphics[scale=0.9]{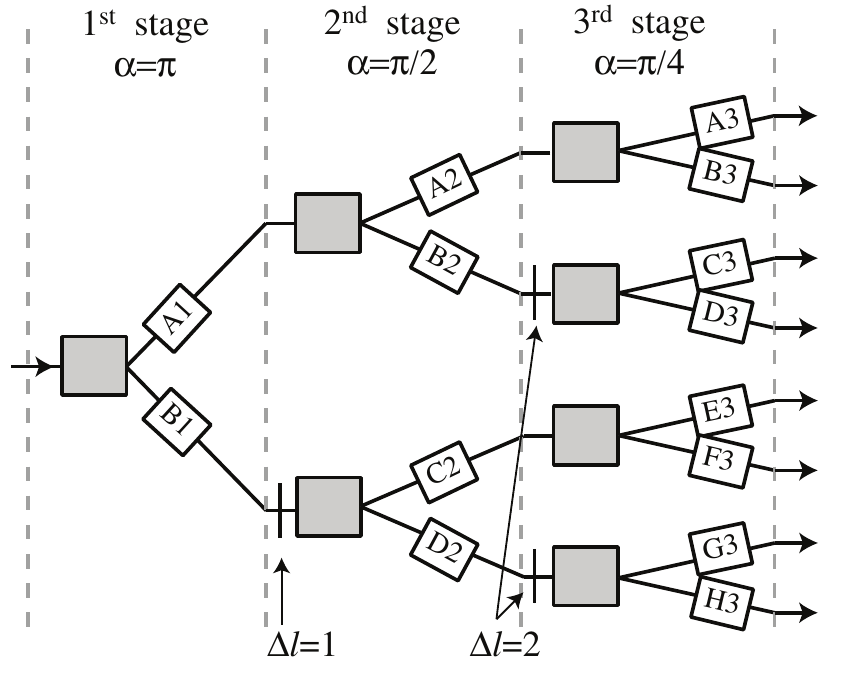}
\caption{The first three stages of a general sorting scheme, where the gray boxes represent Dove prism interferometers. The first stage introduces a phase shift of $\alpha=\pi$ to sort even $\ell$s into port A1 and odd $\ell$s into port B1. The second stage and the third stage introduce phase shifts of $\alpha=\pi/2$ and $\alpha=\pi/4$ to sort $\ell=4n/\ell=4n+2$ and $\ell=8n/\ell=8n+4$, respectively. It should be noted that $\triangle\ell=1$ and $\triangle\ell=2$ holograms are required before sorting in the second and third stages \cite{Leach2002Measuring}.}
\label{Fig13}
\end{figure}

\subsection{Optical OAM}
Being the detection of optical OAM the inverse process of generating OAM,  OAM modes  can be detected  by using an \emph{inverse SPP} designed for the specific mode to be detected \cite{Willner2015Optical}. Moreover, a vortex beam with the topological charge of $\ell$  is converted into planar Gaussian light after being irradiated through a \emph{holographic grating} with a transmittance function related to the anti-helical phase factor $\exp(-i\ell\theta)$ \cite{Mair2001Entanglement}. Using spatial mode filters,  Gaussian light can be separated from other OAM beams, detecting only one specific OAM mode at the time. For multiplexed OAM modes, a series of holographic gratings are required to sequentially detect each single OAM mode, or, alternatively, beam splitter can be combined for parallel detection. In \cite{Gibson2004Free}  and \cite{Gibson2004Increasing} a two-dimensional forked holographic grating, which is the superposition of a vertical grating and a horizontal grating, is designed to simultaneously detect $8$ OAM modes. Two-dimensional Dammann  gratings, combining the characteristics of conventional gratings and Dammann gratings, can achieve equal energy distribution on the designed diffraction order, expanding the detection range of the grating \cite{Zhang2010Extending,Gao2016Integrating}. At present, Dammann holographic gratings are widely used at the receiving end of optical communication systems since they  can perform parallel detection on multiple OAM states. These holographic gratings require high precision and are normally loaded onto SLMs or recorded using advanced grating fabrication techniques. At the same time, their conversion efficiency  is not higher than $1/N$, where $N$ is the total number of detected modes. Recently, \cite{Kai2017Orbital} has proposed a novel phase hologram, designed by modifying the Lin's algorithm \cite{Lin2006Synthesis}, which can be used to achieve almost complete concentration of incident energy to a specified target OAM mode that can be detected more effectively than with conventional forked gratings.

\begin{table*}[tb]
\small
\centering
\caption{Comparison of Optical OAM Detection Methods.}
  \begin{tabular}{lllll}
  \toprule
  \textbf{Features} &\textbf{\tabincell{l}{Inverse SPPs}} &\textbf{Holographic gratings} &\textbf{Dove prism interferometers} &\textbf{Mode sorters} \\
  \midrule
  \specialrule{0em}{2pt}{2pt}
  \textbf{Cost} &Low &Low &Low &Low \\
  \specialrule{0em}{2pt}{2pt}
  \textbf{Conversion efficiency} &High &Low, no higher than $1/N$  &High, near $100\%$  &Relatively high \\
  \specialrule{0em}{2pt}{2pt}
  \textbf{OAM mode} &\tabincell{l}{Single mode\\ detection} &\tabincell{l}{Composite mode detection; \\The number is limited by \\the size of the grating}  &\tabincell{l}{Composite mode sort; \\The more the number, the more\\complicated the device} &\tabincell{l}{Composite mode sort;\\Adjacent OAM mode\\cannot be effectively\\separated} \\
  \specialrule{0em}{2pt}{2pt}
  \textbf{Flexibility} &Low &High &Low &High \\
  \specialrule{0em}{2pt}{2pt}
  \textbf{System complexity} &Low  &Low &High &Low \\
  \specialrule{0em}{2pt}{2pt}
  \textbf{Market readyness} &Yes &Yes &Yes &\tabincell{l}{Yes (If implemented\\on SLMs)} \\
  \specialrule{0em}{2pt}{2pt}
  \bottomrule
  \label{Table5}
 \end{tabular}
\end{table*}

A vortex wave with the topological charge of $\ell$ can interfere with a plane wave to obtain $\ell$ spiral stripes or $\ell$-fold forked stripes. 
As such, there are many methods for detecting optical OAM using interference, like the Young's double-slit interference \cite{Sztul2006Double,Zhou2014Dynamic},  the multipoint interference \cite{Berkhout2008Method}, and the annular aperture interference \cite{Guo2009Characterizing}. However,  with these methods  it is difficult to correctly  detect the intended  OAM mode because of the complexity of the received interference pattern. A simpler method is to use a Mach-Zehnder interferometer with a Dove prism placed in each arm to detect single photon OAM \cite{Leach2002Measuring}, called \emph{Dove prism interferometer}, as shown in Fig. \ref{Fig12}. The angle of rotation between the two Dove prisms is $\alpha/2$, and the rotation angle of the passing beam is $\alpha$. Correspondingly, the phase difference between the two arms is $\ell \alpha$. If $\alpha/2=\pi/2$, by  correctly adjusting the path length of the interferometer, all beams with even $\ell$ have constructive interference  at one output port and  cancellation at the other output port, while for  all beams with odd $\ell$ it is  just the opposite. Therefore, such an interferometer can achieve  OAM mode odd-$\ell$ and even-$\ell$ separation  at the single photon level. As shown in  \cite{Leach2004Interferometric}, the improved interferometer system is capable of detecting a single photon OAM and SAM. By cascading multiple interferometers as shown in Fig. \ref{Fig13}, it is possible to generalize this architecture to detect any number of OAM modes, and the theoretical efficiency is close to $100\%$. In principle, detecting $N$ OAM modes requires $N-1$ interferometers. Therefore, such an interferometer device is too complex for practical applications with a large number of OAM modes.

\begin{figure}[t]
\centering
\includegraphics[scale=0.65]{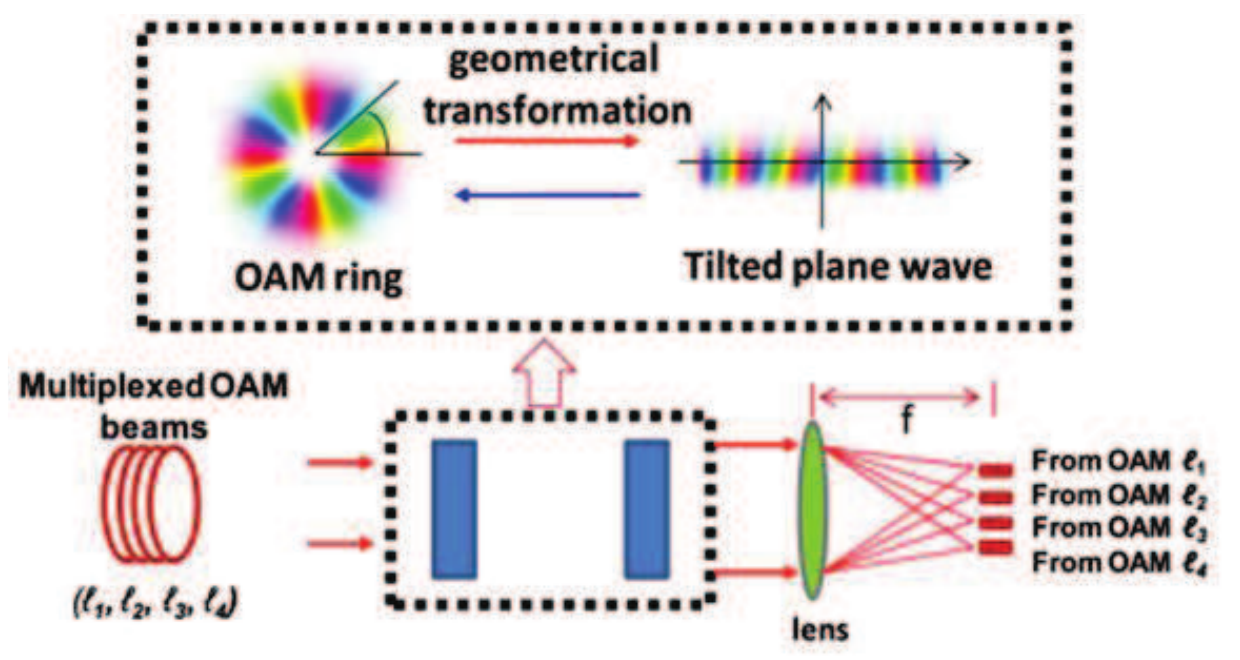}
\caption{Principle of the OAM mode sorter. By using geometric transformations, ring beams with helical wavefronts are mapped to rectangular beams with tilted wavefronts \cite{Willner2015Optical}.}
\label{Fig14}
\end{figure}

After passing through a lens, a plane wave converges into a point, whose lateral position on the focal plane depends on the lateral phase gradient of the plane wave. Note that the two plane waves need at least a phase difference of $2\pi$ to make the distance between the two focal points larger than the Rayleigh resolution limit. Using this idea, \cite{Berkhout2010Efficient} proposes a mode conversion method for sorting OAM modes. The \emph{mode sorter} consists of a converter and a corrector, both implemented on the SLM. The converter implements the conversion from a Cartesian coordinate system to a log-polar coordinate system, so that a point $(x,y)$ on the input plane is converted into a point $(u,v)$ on the output plane, where the conversion law is $u=-a\ln(\sqrt{x^2+y^2}/b)$ and $v=a\arctan(y/x)$, where $a$ and $b$ are two independently chosen parameters. The corrector is used to compensate the phase distortion caused by the optical path length during the coordinate transformation. Thus, a ring beam with a helical phase front on the input plane is mapped to a rectangular beam with a tilted wavefront on the output plane. The rectangular beams with different lateral phase gradients are focused by the lens on different lateral  positions on the focal plane, effectively separating the  different OAM modes. In \cite{Lavery2012Refractive} an OAM mode sorter composed of two custom refractive optical elements is designed to achieve higher conversion efficiency. Nevertheless, there is a limit to the mode sorter due to the overlap between the focal points of two adjacent OAM modes. In \cite{Mirhosseini2013Efficient} it is shown that by using a series of unitary optical transformations, which generate  multiple copies of the rectangular beams, it is possible to achieve a larger separation between two OAM modes, trading a greater efficacy with higher system complexity.

In addition to the above methods, there are other techniques for separating OAM modes. In \cite{Karimi2009Efficient} it is proposed a method based on the use of  $q$-plates. The radius ratio method can be used to detect certain special optical OAM \cite{Jia2013Sidelobe,Long2013Evaluating}. Photonic integrated circuits are well compatible with single-mode fibers while achieving optical OAM multiplexing and demultiplexing \cite{Doerr2012Efficient,Su2012Demonstration}.

Four common methods for detecting optical OAM are compared in Table \ref{Table5}. As mentioned above, also in this case each method has its own advantages and disadvantages. From the perspective of conversion efficiency, the Dove prism interferometer is the best method, but it is bad in terms of flexibility and system complexity, so that it is a not suitable technology for detecting a large number of OAM modes. Inverse SPPs and mode sorters perform better in both respects. A drawback of inverse SPPs is that they  can only detect a specific OAM mode, and it is necessary to combine beam splitters and multiple inverse SPPs to achieve composite mode detection. Mode sorters enable composite mode sort, but they can not effectively separate two adjacent OAM modes. The holographic grating has a low conversion efficiency, but it can be easily loaded onto a SLM. In current optical OAM experimental systems, the holographic grating loaded on a SLM is a relatively common method. In addition, inverse phase holograms are also frequently used in practice. Some novel phase holograms have also been tried to detect optical OAM with high efficiency.

\begin{figure}[t]
\centering
\includegraphics[scale=0.45]{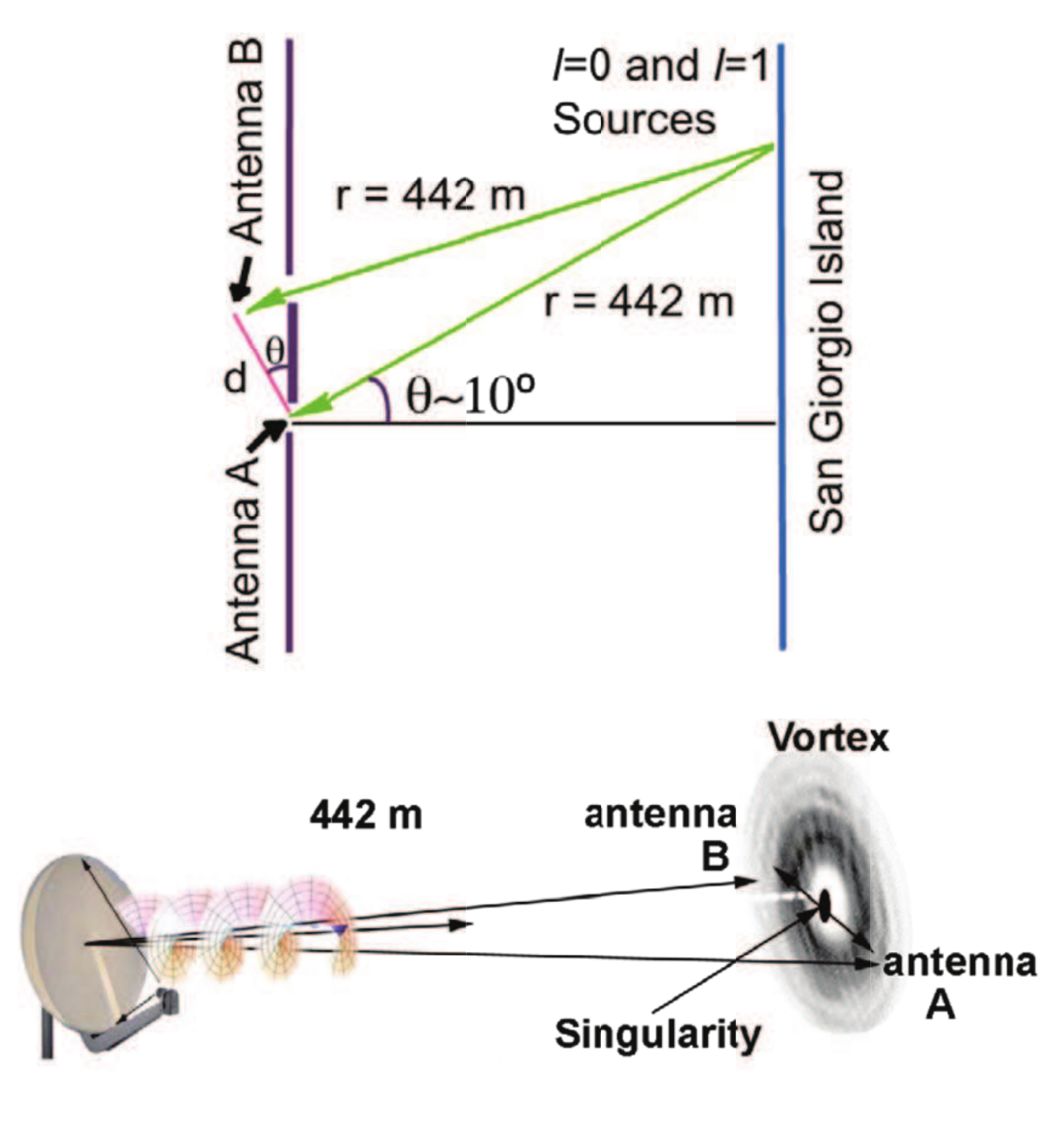}
\caption{Sketch of the famous Venice experiment of radio OAM communications, in which two receive antennas A and B are aligned with the phase singularity \cite{Tamburini2012Encoding}.}
\label{Fig15}
\end{figure}

\subsection{Radio OAM}
As in optics, the detection of OAM at radio frequencies can be achieved by employing carefully designed \emph{inverse SPPs} \cite{Yan2014High}. It is also possible to detect the radio OAM mode by observing the interference pattern of a vortex wave and a plane wave. Nevertheless, many detection methods in optics are no longer applicable at radio frequencies and some new methods have been proposed.

In \cite{Tamburini2012Encoding} the \emph{interference phase detection} is used to separate two electromagnetic beams with topological charges of $\ell=0$ and $\ell=1$. In the experiment set-up, two Yagi antennas A and B are connected by a $180^{\circ}$ phase-shift cable to form a phase interferometer, designed to reduce background interference in a real environment. As shown in Fig. \ref{Fig15}, if the device is fully aligned, the vortex electromagnetic beam with $\ell=1$ will produce a phase difference of $180^{\circ}$ between receive antennas $A$ and $B$. This phase difference is compensated by the cable line phase delay, so that the received interference intensity is maximized. For the beam with $\ell=0$, there is only the phase delay of the cable, and the intensity is minimal. In this way, it is possible to separate beams with different wavefront characteristics. When the interference phase detection method is used for more radio OAM modes, the receiving device becomes very complicated.

\begin{figure}[t]
\centering
\includegraphics[scale=0.35]{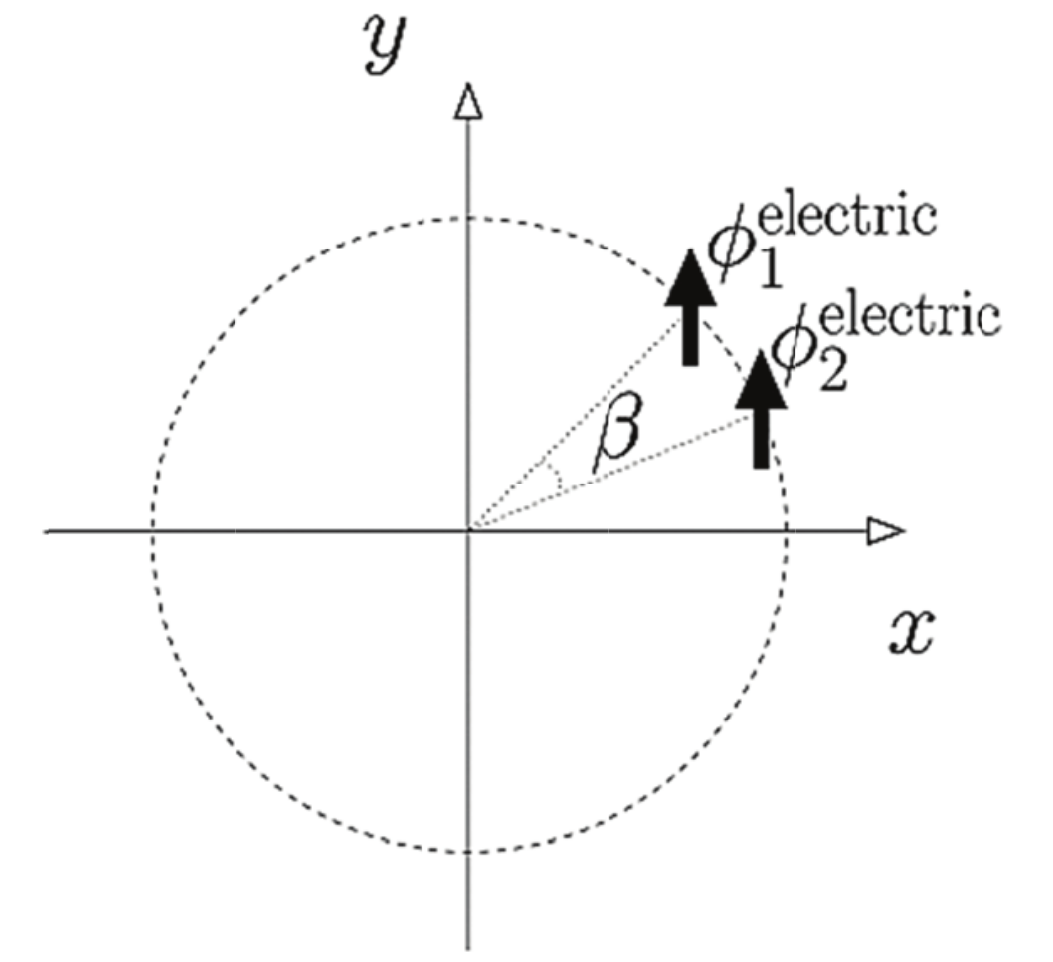}
\caption{Illustration of the phase gradient method. The OAM mode can be estimated by computing the phase difference of the electric field components at two sampling points separated
by an angle $\beta$, i.e.  $\ell=(\phi_1-\phi_2)/\beta$ \cite{Mohammadi2010Orbital1}.}
\label{Fig16}
\end{figure}

In 2010, Mohammadi et al. proposed two methods to estimate far-field OAM modes in radio beams, the \emph{single-point  estimation method} and the \emph{phase gradient method} \cite{Mohammadi2010Orbital1}. The single-point estimation method  uses an approximation for OAM in the  far-field  to estimate the OAM from the  measurements at a single point of the  vertical and the transverse components of the electric field. This method is efficient  for the detection of  OAM with low  mode numbers.
Compared with the single-point estimation method, the phase gradient method is more intuitive and effective. As shown in Fig. \ref{Fig16}, the measurement set-up consists of  two test  points separated by an angle $\beta$ on a  circumference at the receiving end. If the measured phases are $\phi_1$ and $\phi_2$, respectively, the OAM mode will be $\ell=(\phi_1-\phi_2)/\beta$. In order to accurately detect the OAM mode, the angle $\beta$ between the two test points should satisfy the relationship $\beta<\pi/|\ell|$. The phase gradient method requires that the center of the receiving circle is aligned with the center axis of the vortex beam, otherwise there will be large errors. Generally, it can only be used to detect single-mode OAM beams, but with some modifications, it can also achieve accurate measurement of two mixed modes\cite{Xie2017Mode}.

\begin{figure}[t]
\centering
\includegraphics[scale=0.95]{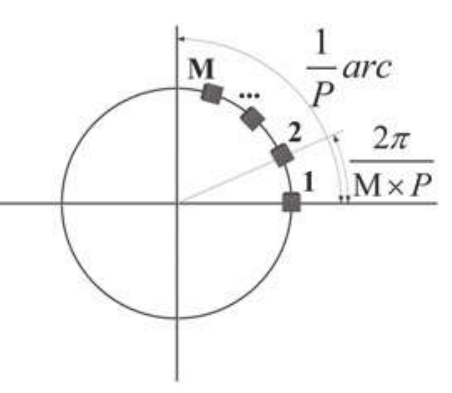}
\caption{The PASR scheme with $M$ antennas uniformly distributed on a $1/P$ arc \cite{Hu2016Simulation}.}
\label{Fig17}
\end{figure}

\begin{table*}[tb]
\small
\centering
\caption{Comparison of Radio OAM Detection Methods.}
  \begin{tabular}{lllllll}
  \toprule
  \textbf{Features} &\textbf{\tabincell{l}{Inverse SPPs}} &\textbf{\tabincell{l}{Interference\\phase detection}} &\textbf{\tabincell{l}{Single-point\\estimation\\ method}} &\textbf{\tabincell{l}{Phase gradient\\method}} &\textbf{UCAAs} &\textbf{PASR} \\
  \midrule
  \specialrule{0em}{2pt}{2pt}
  \textbf{Cost} &Low &Low &Low &Low &High &High \\
  \specialrule{0em}{2pt}{2pt}
  \textbf{Sampling range} &Entire wavefront &Two points  &One point  &Two points &\tabincell{l}{Entire wave-\\front}  &Partial wavefront \\
  \specialrule{0em}{2pt}{2pt}
  \textbf{OAM mode} &\tabincell{l}{Single mode} &\tabincell{l}{Composite mode; \\The higher the n-\\umber, the more\\complicated the\\device}  &\tabincell{l}{Single mode; \\It is more su-\\itable for low\\mode} &\tabincell{l}{Single mode; \\There is a restri-\\ction on the use\\of OAM mode} &\tabincell{l}{Composite\\mode}  &\tabincell{l}{Composite mode;\\There is a restri-\\ctionon the use\\of OAM mode} \\
  \specialrule{0em}{2pt}{2pt}
  \textbf{\tabincell{l}{System comple-\\xity}} &Low &High &Low &Low &High &High \\
  \specialrule{0em}{2pt}{2pt}
  \textbf{\tabincell{l}{Computational\\complexity}} &Low &Low &High &Low &Low &Low \\
  \specialrule{0em}{2pt}{2pt}
  \bottomrule
  \label{Table6}
 \end{tabular}
\end{table*}

UCAAs are used not only for radio OAM generation but also for OAM detection \cite{Wu2014UCA}. The condition for detection is that the two arrays are aligned, i.e, the  cross section of the receiving circular array C should be  perpendicular to the beam axis and its center should coincide with the axis. UCAAs can demultiplex several radio OAM modes. The whole process of OAM demultiplexing can be seen as some sort of  spectral analysis of the received vortex electromagnetic waves. When the mode number $\ell'$ of the circular array C is different from one of  the beam modes, the result of the spectral analysis tends to $0$, conversely, it tends to $1$. By selecting different $\ell'$ values, different information can be extracted from the OAM multiplexed beam. When multiple topological charge values are selected at the same time, composite mode detection can be realized. In general, the detection process needs to receive information of the entire wavefront.
Due to the divergence of OAM beams, a large receiving array is required to capture the effective power of the OAM beam in the far-field, but this might be difficult to achieve.
An answer to the problem of  OAM  detection in the far-field is \emph{partial aperture sampling receiving} (PASR) \cite{Hu2016Simulation}, which  uses partial wavefront information to detect and distinguish different radio OAM modes, as shown in Fig. \ref{Fig17}, and has been verified by computational simulation. The PASR, which is a combination of the partial angular receiving aperture method for OAM demultiplexing in optics \cite{Zheng2015Orbital} and a sampling receiving scheme, has a receiving arc which is $1/P$ of a circumference and employs $M$ antennas evenly distributed on the arc as signal sampling points. The angle between adjacent antennas is $2\pi/MP$. To  ensure strict orthogonality and realize non-crosstalk separation of the two OAM modes $\ell_{n1}$ and $\ell_{n2}$, they  must satisfy simultaneously the two conditions: $\mod(|\ell_{n1}-\ell_{n2}|,P)=0$ and $\mod(|\ell_{n1}-\ell_{n2}|,MP)\neq0$. Although the PASR has certain limitations, it can greatly simplify the scale of the receiving end and is robust to non-ideal OAM beams.

\begin{table*}[tb]
\small
\centering
\caption{Summary of OAM Detection Methods.}
  \begin{tabular}{p{3.2cm}ll}
  \toprule
  \textbf{Methods} &\textbf{Introduction and Features} &\textbf{Remarks} \\
  \midrule
  \specialrule{0em}{2pt}{2pt}
  \textbf{Holographic gratings}  &\tabincell{p{11cm}}{Holographic gratings with $\exp(-i\ell\theta)$ phase factor can convert vortex beams into plane beams, and Daman gratings can realize parallel detection of multiple OAM modes. \\Simple and fast; low conversion efficiency, no higher than $1/N$, where $N$ is the total number of detected modes. }  &\tabincell{l}{Optical OAM} \\
  \specialrule{0em}{2pt}{2pt}
  \cline{1-3}
  \specialrule{0em}{2pt}{2pt}
  \textbf{Dove prism interferometers}  &\tabincell{p{11cm}}{By adjusting the rotation angle of the Dove prism, the odd-even sort of OAM modes can be achieved on a single photon level. \\High conversion efficiency, close to $100\%$; any number of OAM modes can be detected by cascading multiple interferometers, while increasing the complexity of the device.}  &\tabincell{l}{Optical OAM} \\
  \specialrule{0em}{2pt}{2pt}
  \cline{1-3}
  \specialrule{0em}{2pt}{2pt}
  \textbf{Mode sorters} &\tabincell{p{11cm}}{The geometric transformation is used to map the Cartesian coordinate to the log-polar coordinate, and ring beams with helical wavefronts are mapped to rectangular beams with tilted wavefronts, which are focused by lens in different lateral positions. \\High conversion efficiency; adjacent OAM mode cannot be effectively separated. }  &\tabincell{l}{Optical OAM} \\
  \specialrule{0em}{2pt}{2pt}
  \cline{1-3}
  \specialrule{0em}{2pt}{2pt}
  \textbf{Inverse SPPs} &\tabincell{p{11cm}}{SPPs with the topological charge of $-\ell$. \\Simple structure; single mode detection; composite mode detection can be realized by combining multiple inverse SPPs and beam splitters.} &\tabincell{l}{Optical OAM\\Radio OAM} \\
  \specialrule{0em}{2pt}{2pt}
  \cline{1-3}
  \specialrule{0em}{2pt}{2pt}
  \textbf{Interference phase detection} &\tabincell{p{11cm}}{The phase shift cable is used to compensate the phase difference between the receiving antennas, so that the specific OAM mode has high intensity and the other modes have low intensity. \\Intuitive; the receiving device is very complicated when detecting multimodal OAM. } &\tabincell{l}{Radio OAM} \\
  \specialrule{0em}{2pt}{2pt}
  \cline{1-3}
  \specialrule{0em}{2pt}{2pt}
  \textbf{Single-point estimation method} &\tabincell{p{11cm}}{It uses an approximation for OAM in the far-field to estimate the OAM from the measurements at a single point of the vertical and the transverse components of the electric field. \\It is efficient for the detection of OAM with low mode numbers. } &\tabincell{l}{Radio OAM} \\
  \specialrule{0em}{2pt}{2pt}
  \cline{1-3}
  \specialrule{0em}{2pt}{2pt}
  \textbf{Phase gradient method} &\tabincell{p{11cm}}{Estimate the OAM mode by computing the phase difference of the electric field components at two test points separated by an angle $\beta$, i.e. $\ell=(\phi_1-\phi_2)/\beta$. \\Low complexity; small scale; the angle $\beta$ between the test points should satisfy $\beta<\pi/(|\ell|)$. } &\tabincell{l}{Radio OAM} \\
  \specialrule{0em}{2pt}{2pt}
  \cline{1-3}
  \specialrule{0em}{2pt}{2pt}
  \textbf{UCAAs} &\tabincell{p{11cm}}{A receiving circular array C with $\ell=\ell'$ can extract the information carried by the beam with $\ell=\ell'$ from the multiplexed OAM beam. \\The detection process needs to receive information of the entire wavefront. } &\tabincell{l}{Radio OAM} \\
  \specialrule{0em}{2pt}{2pt}
  \cline{1-3}
  \specialrule{0em}{2pt}{2pt}
  \textbf{PASR} &\tabincell{p{11cm}}{Demultiplexing is implemented by utilizing partial wavefront information and the orthogonality between OAM modes. \\Small scale; any two OAM modes $\ell_{n1}$ and $\ell_{n2}$ must satisfy $\mod(|\ell_{n1}-\ell_{n2}|,P)=0$ and $\mod(|\ell_{n1}-\ell_{n2}|,MP)\neq0$. } &\tabincell{l}{Radio OAM} \\
  \specialrule{0em}{2pt}{2pt}
  \cline{1-3}
  \specialrule{0em}{2pt}{2pt}
  \textbf{Metamaterials} &\tabincell{p{11cm}}{$M$ metamaterials with $\ell=-1$ can  convert the component with the mode number $\ell=m$ into a plane wave. \\Realtime; high conversion efficiency; low flexibility; specific working frequency; it is not suitable for detection of high-mode OAM. } &\tabincell{l}{Acoustic OAM} \\
  \specialrule{0em}{2pt}{2pt}
  \bottomrule
  \label{Table7}
   \end{tabular}
\end{table*}

Table \ref{Table6} compares the common detection methods for radio OAM. Unlike  the case of optical OAM detection, these methods often require  a heavy computational load so that also  computational complexity is one of the parameters evaluated  in Table \ref{Table6}. For multiple OAM modes detection, the system complexity of interference phase detection is very high. The single-point estimation method extrapolates  the OAM mode in the far field from the measurements of the two components of the electric field, so its computational complexity is rather high. Therefore, these two methods are not recommended for practical implementations. In term of complexity, the phase gradient method is probably the best method, but it can only detect single mode OAM. Although UCAAs need to receive the entire wavefront information, it can be used for composite OAM mode detection. The PASR  can be seen as an improvement of UCAAs. It has proved its advantages through computational simulation, but further experimental verification is required. In radio OAM communications, inverse SPPs are widely used to detect OAM modes. In the particular case of  line-of-sight (LOS) wireless communications, UCAAs have proved to be more suitable for demultiplexing composite OAM modes.

\subsection{Acoustic OAM}
As mentioned above, research on OAM detection  at  acoustic frequencies is still in an early stage. Due to the orthogonality between the OAM eigenstates, Shi et al. use the inner product operation to separate the multiplexed OAM modes  received by a transducer array \cite{Shi2017High}. Recently, Jiang et al. used passive metamaterials to detect acoustic OAM \cite{Jiang2018Twisted}. The structure of such passive acoustic metamaterials has been described in Section III. After that an acoustic beam is passed through a metamaterial with the topological charge of $\ell=-1$, the mode number of all multiplexed OAM modes will be reduced by $1$. Therefore, $m$ metamaterials with $\ell=-1$ can convert the component with the mode number $\ell = m$ into a plane wave and a non-zero value can be detected at the ``dark'' core point of the OAM beam. This non-zero value contains the information carried by the OAM with $\ell=m$. This method can realize real-time demodulation of  OAM beams and has high conversion efficiency but low flexibility and is not suitable for detection of high-mode OAM.

\subsection{Summary and Open Challenges}
Table \ref{Table7} summarizes the most common  methods employed  for the detection of OAM modes.
As seen from the table, for optical OAM and radio OAM, many detection methods have been proposed, while further research, investigation and testing  is needed for developing effective acoustic OAM detection.

In general, using the inverse SPP is the easiest way to detect optical OAM and radio OAM. In optical OAM communications, holographic gratings or phase holograms loaded on SLMs, which can detect multiple OAM modes simultaneously, are used in most cases. In radio OAM communications, in addition to SPPs, UCAAs have the potential to demultiplex composite OAM modes. However, OAM beam divergence increases the aperture of the receiving UCAA. Reducing the aperture and scale of the far field receiving array is extremely challenging. The PASR and the phase gradient method are promising solutions to reduce the scale of the receiver. But how to apply them to OAM mode demultiplexing requires further research in the coming years.

\begin{figure*}[t]
\centering
\includegraphics[scale=0.8]{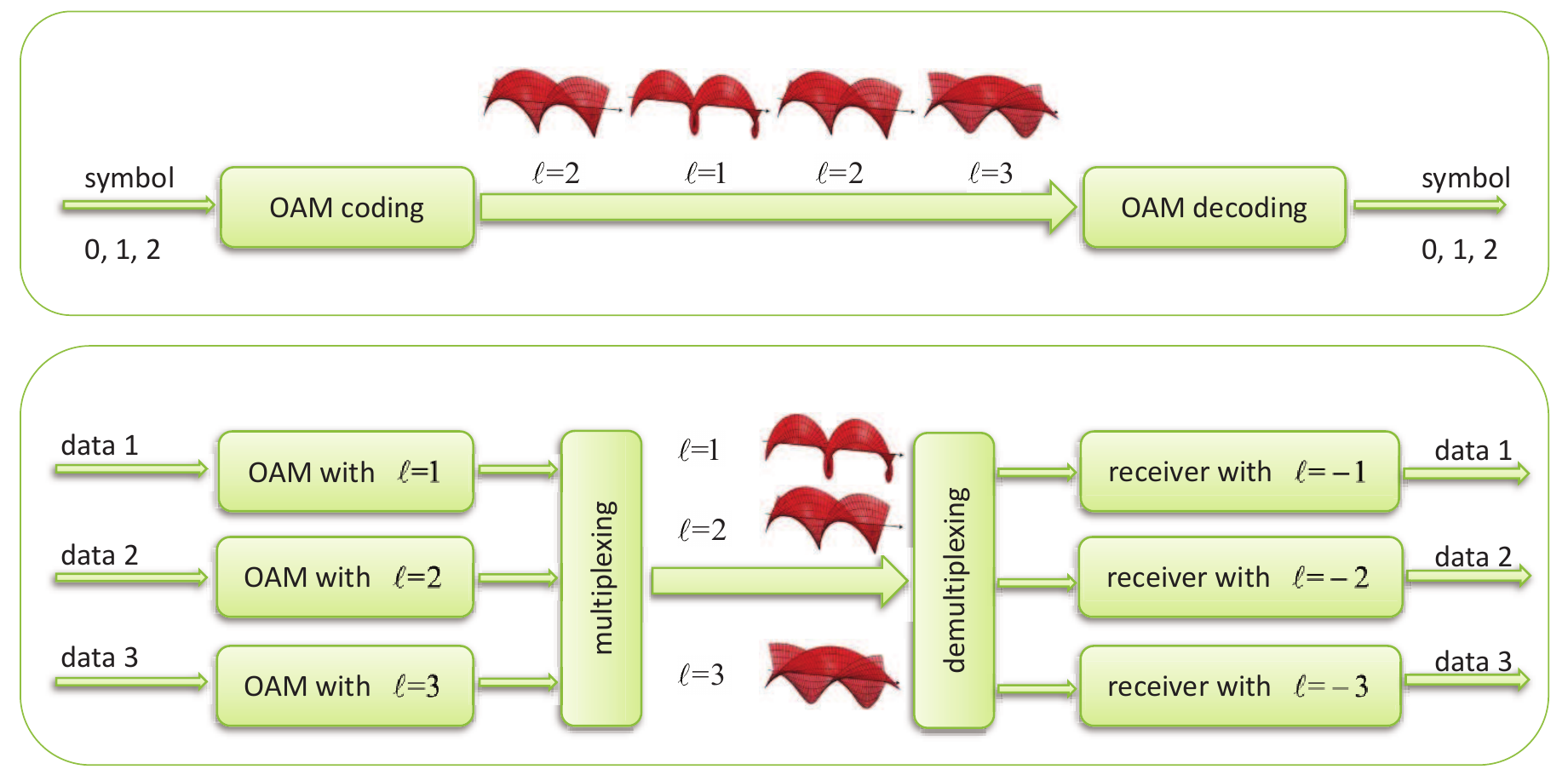}
\caption{Schematic diagram of OAM-SK and OAM-DM communications: in OAM-SK communication (top) the information symbols are encoded in the topological charges of OAM beams, while in OAM-DM communication (bottom) the information is multiplexed in multimodal OAM beams.}
\label{Fig18}
\end{figure*}

\section{Applications of OAM in Communications}
This section focuses on the applications of OAM in communications. Since OAM modes with different values of $\ell$ are mutually orthogonal, vortex beams carrying different OAM modes  can provide independent communication channels for efficient information transmission. As shown in Fig. \ref{Fig18}, there are two main strategies to transfer information with OAM: \emph{OAM shift keying} (OAM-SK), where the information is encoded in the value $\ell$ of the OAM beam, and \emph{OAM division multiplexing} (OAM-DM), where the information is multiplexed in multimodal OAM beams. OAM-DM uses the vortex beams as the carrier of information  so that data is loaded onto different OAM modes and then multiplexed and transmitted coaxially through a single aperture. At the receiver, the beams are collected by another aperture and then demultiplexed and detected for data recovery. Compared with OAM-SK, OAM-DM system has higher spectral efficiency and exhibits lower bit error rates.

OAM-DM as a mode-division multiplexing (MDM) is a subset of space-division multiplexing (SDM) and can  be compatible with different modulation formats, such as $M$-ary amplitude-shift keying ($M$-ASK), $M$-ary phase-shift keying ($M$-PSK) and $M$-ary quadrature amplitude modulation ($M$-QAM), as well as other multiplexing techniques such as frequency-division/wavelength-division multiplexing (FDM/WDM) and polarization-division multiplexing (PDM), thereby further improving the communication system capacity from another dimension.
In the OAM-DM system, $N$ OAM waves carrying information are multiplexed, and the obtained field can be expressed as
\begin{equation} \label{satisfy2}
U_{MUX}(r, \theta, t) = \sum_{p=1}^N S_{p}(t)A_{p}(r)e^{i\ell_{p}\theta},
\end{equation}
where $S_p(t)$ is the modulated data signal on the $p$th OAM mode and $A_p(r)$ is the complex electric field amplitude of the $p$th OAM mode. Then,
at the receiving end, the multiplexed OAM wave is multiplied by an anti-helical phase factor $\exp(-i\ell_{q}\theta)$, which is achieved by using an inverse SPP or an antenna array. The OAM wave with $\ell=\ell_{q}$ is converted into a plane wave and can be easily separated from other vortex waves with $\ell=\ell_{p}-\ell_{q}$, thus achieving demultiplexing. Eventually, the data information carried by the OAM wave with $\ell=\ell_{q}$ can be obtained.

\begin{figure*}[t]
\centering
\includegraphics[scale=0.65]{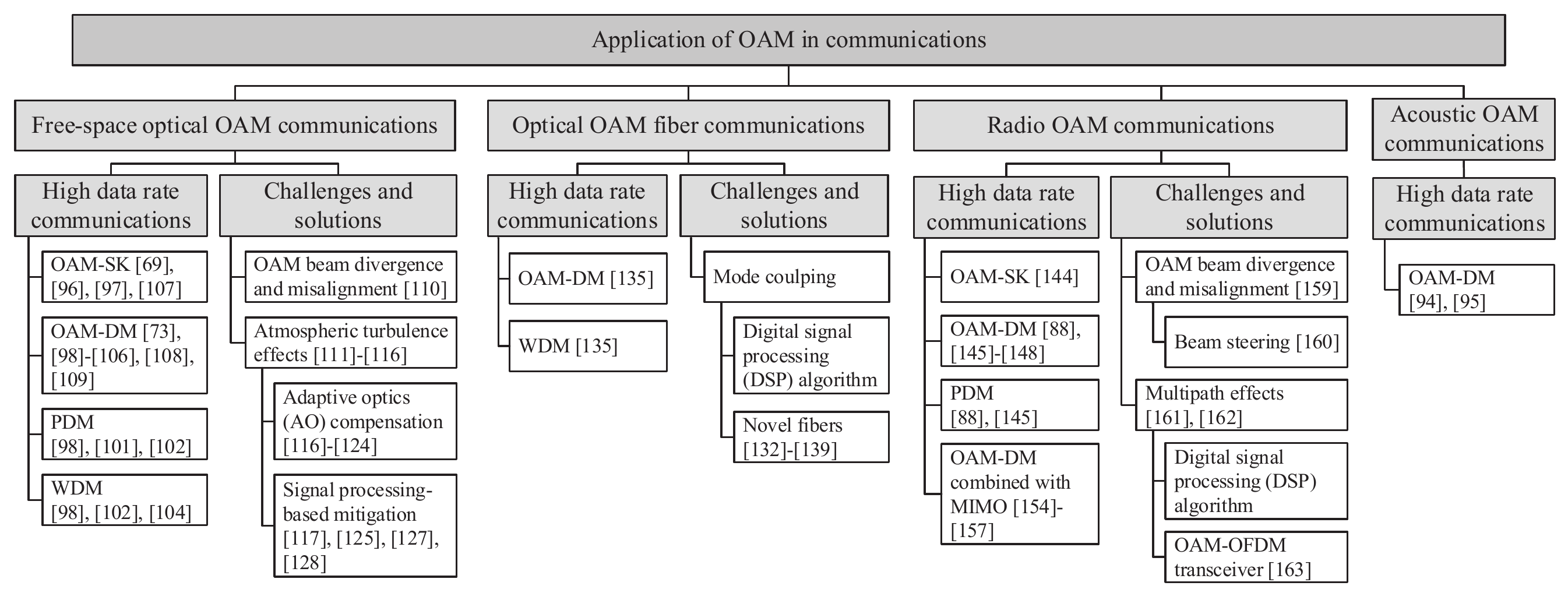}
\caption{Taxonomy diagram of applications of OAM in communications.}
\label{Fig33}
\end{figure*}

OAM communications still face many challenges and technical issues, such as misalignment and OAM beam divergence. OAM waves are required to be transmitted and received coaxially in OAM-DM system, so that precise alignment is required between the transmitter and receiver, i.e. the center of the receiver coincides with the center of the transmitted beam and the receiver is perpendicular to the line connecting the centers. Moreover, since optical components usually have a limited aperture size, OAM beam divergence during propagation can result in received power loss in the far field, limiting the link achievable distance. The atmospheric turbulence, mode coupling and multipath effects  should to be considered in free-space optical links, fiber links and radio communication links, respectively. For acoustic OAM communications, it is clear that underwater creatures and turbulence have a great impact on links, although there is currently no research to discuss them.
A taxonomy diagram of the applications discussed in this  section, together with main challenges and solutions for each application,  is shown in Fig. \ref{Fig33}.

\subsection{Free-Space Optical OAM Communications}
In 2004, Gibson et al. \cite{Gibson2004Free} demonstrated for the first time that OAM-SK modulation  can be used for free-space optical communications with good results. The transmitter  maps the information data on  the OAM beams' topological charge, which belongs to the set  $\mathcal{L}=\{-16, -12, -8, -4, +4, +8, +12, +16\}$. At the receiving end, two vertically stacked forked gratings are used to detect the transmitted OAM modes. To compensate for small perturbations in alignment, a Gaussian beam with $\ell=0$ is used as a reference signal. In \cite{Krenn2014Communication} $16$ superimposed OAM modes are employed to transmit information over a $3$ km intra-city link. At the receiver the beam topological charge is recovered by a non-coherent detection scheme aided by an artificial neural network. This experimental scheme has recently been extended to a $143$ km free-space link between the two Canary Islands of La Palma and Tenerife \cite{Krenn2016Twisted}.

\begin{figure}[t]
\centering
\includegraphics[scale=0.95]{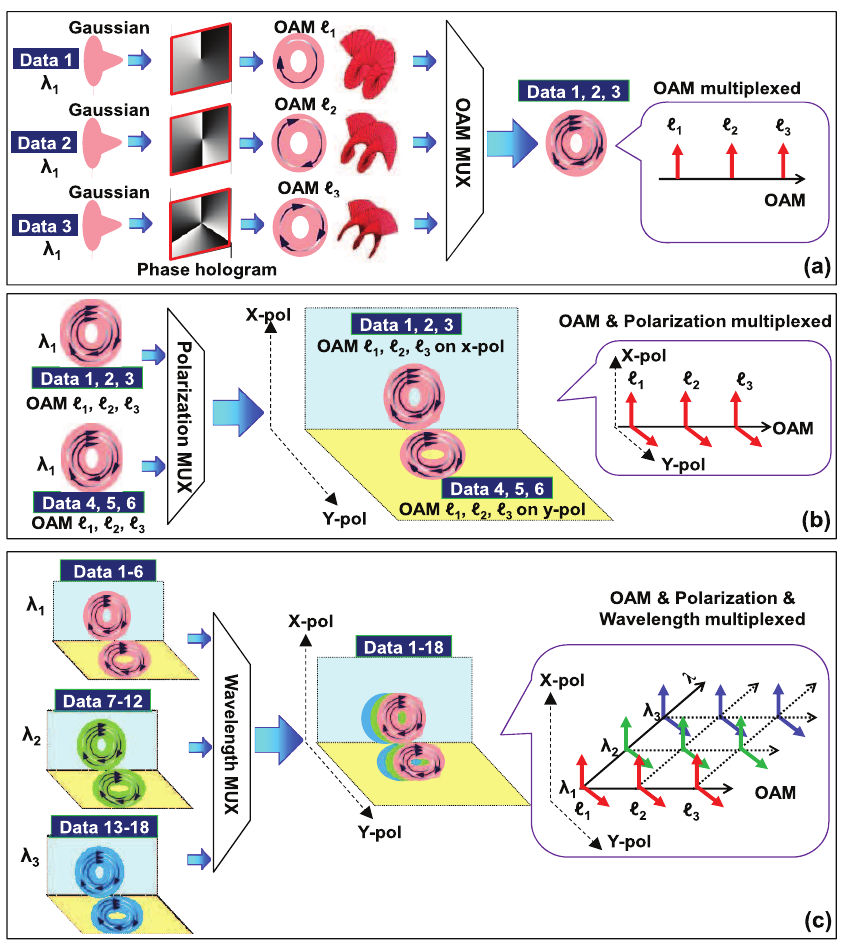}
\caption{Concept of using three-dimensional multiplexing to increase the multiplexed data channels. (a), (b), and (c) are performed successively to achieve OAM-DM, PDM, and WDM, respectively \cite{Huang2014100}.}
\label{Fig19}
\end{figure}

If an OAM-SK system with $L$ different topological charges can transmit up to $\log_{2}(L)$ bits per beam, an OAM-DM system that multiplexes $L$ modes can transmit up to $L$ bits per beam. For example, consider the case with $\mathcal{L}=\{\ell_{0},\ell_{1}\}$, i.e., $L=2$. With OAM-SK it is possible to transmit one bit per beam (0 if $\ell=\ell_{0}$ and 1 if $\ell=\ell_{1}$). With OAM-DM,  the on and off states of each OAM mode can represent either a  $``1"$ or a  $``0"$, so that the two-bit set of $\{00, 01, 10, 11\}$ can be mapped on the multiplexed modes $\{0 0, 0\ell_{1}, \ell_{0} 0, \ell_{0} \ell_{1}\}$.  Using this idea, \cite{Lin2007Multiplexing} transmits the information in free space by multiplexing four OAM modes employing a phase hologram loaded on a SLM. Subsequently, this method is used to implement the encoding of two-dimensional images \cite{Kai2017Orbital, Zou2018High}.

In order to meet the ever-increasing demands for higher data rates, OAM multiplexing can be combined with different modulation formats and different multiplexing techniques to achieve high-speed communication in multiple dimension. A data link multiplexing the signal in three dimensions is shown in Fig. \ref{Fig19}. Three OAM beams with $\mathcal{L}=\{\ell_1,\ell_2, \ell_3\}$, carrying the  modulation information data 1, data 2 and data 3, are multiplexed into one multimodal beam. A second OAM beam with the same set of modes $\mathcal{L}$ carrying the data streams data 4, data 5 and data 6 is multiplexed with the first one employing orthogonal polarizations. Finally, three polarization-and-OAM-multiplexed beams carrying three different sets of information streams, data 1-6, data 7-12 and  data 13-18, are transmitted at three different wavelengths $\lambda_{1},\lambda_{2}$ and $\lambda_{3}$.  In a recent laboratory experiment \cite{Wang2012Terabit}, a transmission rate of 2.56 Tbps and a spectral efficiency of 95.7 bps/Hz have been achieved by using $20\times4$ Gbps 16-QAM signals on 8 OAM modes, 2 polarization states, and two sets of concentric rings. In  \cite{Huang2014100} a transmission rate of 100.8 Tbps is achieved by transmitting 100 Gbps quadrature phase-shift keying (QPSK) signals on 12 OAM modes, 2 polarization states, and 42 wavelengths. In a similar fashion, in \cite{Wang2014N} a transmission rate of 1.036 Pbps with a spectral efficiency of 112.6 bps/Hz has been achieved by multiplexing 54.139 Gbps OFDM-8QAM signals over 368 wavelengths, 2 polarization states and  26 OAM modes. In addition to these lab-scale high-rate data  experiments, a 400 Gbps transmission rate has been achieved  over an outdoor link of about 120m  by using 100 Gbps QPSK signals on 4 OAM modes \cite{Ren2016Experimental}.

Multimodal LG beams with different topological charges are often employed to provide  a set of orthogonal OAM modes for information transmission.
Most of the literature on the LG beam consider azimuthal index $|\ell|>0$ and radial index $p=0$. Because the radial index $p$ can be used as a radial degree of freedom just as $\ell$ can provide an azimuthal degree of freedom, LG beams with $p>0$ have received some attention \cite{Abderrahmen2016Optical,Xie2016Experimental} recently. Indeed, LG beams with different $p$ and $\ell$  form a complete set of orthogonal mode bases. By multiplexing LG beams with different $p$, the system capacity and transmission rate can be improved. In addition to LG beams, Bessel beams with non-diffractive properties, because of their self-healing properties after encountering an obstruction,  also have a great application potential in free-space optical communications \cite{Gatto2011Free,Du2015High,Ahmed2016Mode}. Perfect vortex beams,  obtained by the Fourier transformation of Bessel beams, have the attractive feature that the beam radius is independent of the OAM mode and have recently   used with success in a free-space optical communication link \cite{Zhu2017Free}.

\begin{figure}[t]
\centering
\includegraphics[scale=1.2]{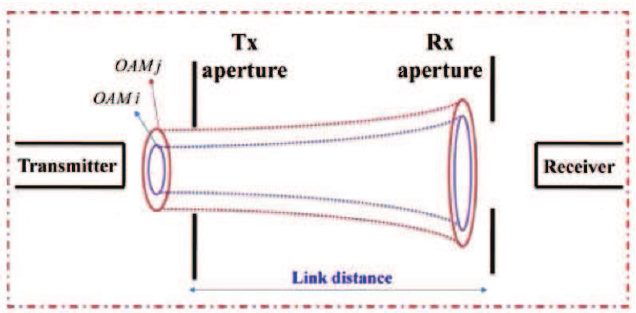}
\caption{Concept of OAM beam divergence in free-space optical communications. Tx, transmitted; Rx, receiver \cite{Willner2015Performance}.}
\label{Fig20}
\end{figure}

\begin{figure}[t]
\centering
\includegraphics[scale=1.2]{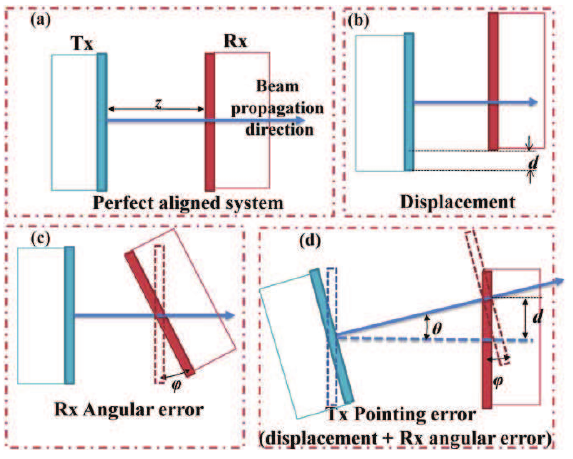}
\caption{Alignment between the transmitter and the receiver for (a) a perfectly aligned system, (b) a system with lateral displacement $d$, (c) a system with a receiver angular error $\varphi$, and (d) a system with a transmitter pointing error $\theta$. Tx, transmitted; Rx, receiver \cite{Willner2015Performance}.}
\label{Fig21}
\end{figure}

Due to the unique wavefront of  OAM beams, there are some challenges in designing an OAM-based communication link in free space, such as \emph{beam divergence} and \emph{misalignement}. Based on diffraction theory, it is known that divergence occurs when a collimated OAM beam propagates in free space. Divergence increases as OAM mode number increases. Fig. \ref{Fig20} shows the concept of OAM beam divergence in a free space communication link. It can be seen that divergence leads to system power loss, especially when the size of the receiving aperture is limited.
A suitable transmitted beam size can be designed for a specific OAM mode number and transmitted distance to achieve the minimum received beam diameter \cite{Willner2015Performance}. However, in long-distance propagation, a larger receiving aperture is still expected to capture more received power.
Besides beam divergence, the misalignment between transmitter and receiver,  which can result in both power loss and mode crosstalk, also needs to be considered. Misalignment errors generally include lateral displacement and receiver angle errors \cite{Willner2015Performance}, as shown in Fig. \ref{Fig21}. It can be found through simulation that  large lateral displacements and large receiver angle errors cause a high power leakage into other modes, resulting in severe crosstalk. Misalignment effects can be mitigated by increasing  OAM mode spacing, but this will result in higher received power loss due to beam divergence. The trade-off between system power loss and crosstalk  needs to be considered in link design. When the system power loss dominates (i.e. small lateral displacement and receiver angular error), small mode spacing is used, and when crosstalk dominates (i.e. large lateral displacement and receiver angular error), large mode spacing is used.

\begin{figure}[t]
\centering
\includegraphics[scale=1.2]{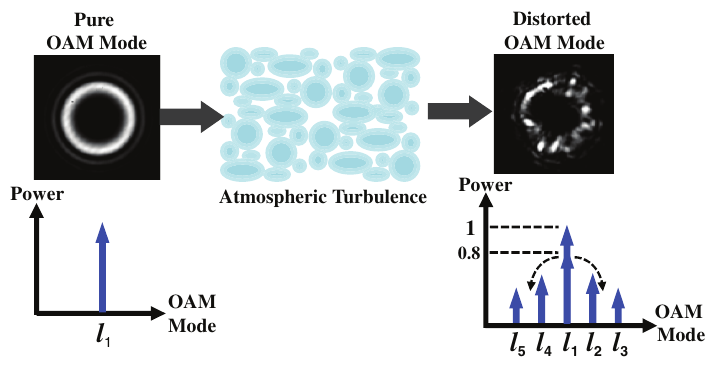}
\caption{Concept diagram of the effects of atmospheric turbulence on an OAM beam. A distorted OAM mode can be decomposed into multiple OAM modes \cite{Ren2013Atmospheric}.}
\label{Fig22}
\end{figure}

For a free-space optical OAM link another critical challenge is \emph{atmospheric turbulence}, caused by changes of the atmosphere refractive index due to temperature and pressure non-uniformities. Atmospheric turbulence  can destroy the helical phase front of vortex beams. Fig. \ref{Fig22} shows its effects  on an OAM beam, including received energy fluctuations and crosstalk between OAM channels. The effects have been quantitatively analyzed using the Kolmogorov spectral statistical model \cite{Paterson2005Atmospheric,Tyler2009Influence} and experimentally verified by simulating turbulence in  laboratory \cite{Ren2013Atmospheric,Rodenburg2012Influence}. In \cite{Ren2013Atmospheric}, a thin phase screen plate mounted on a rotating stage and placed in the middle of the optical path is used as a turbulence emulator. The pseudorandom phase distribution due to the thin plate obeys the Kolmogorov spectral statistics and is characterized by a parameter $r_0$. The strength of the simulated turbulence can be varied by using a plate with a different $r_0$ or by adjusting the size of the beam incident on the plate or the number of passes through the plate.
It is found that, as the turbulence strength increases, the power of the transmitted OAM mode leaks into neighboring modes and for strong turbulence tends to be equally distributed among modes, which will result in severe crosstalk at the receiver. The effects of atmospheric turbulence on different vortex beams, including LG and Bessel beams, have also been numerically calculated by analyzing their OAM spectra \cite{Fu2016Influences}
and it has been shown that Bessel beams suffer more than LG beams in passing through atmosphere turbulence. The non-diffractive property of Bessel beams is affected by  a  strong turbulence \cite{Li2017Adaptive}.

\begin{figure*}[t]
\centering
\includegraphics[scale=0.5]{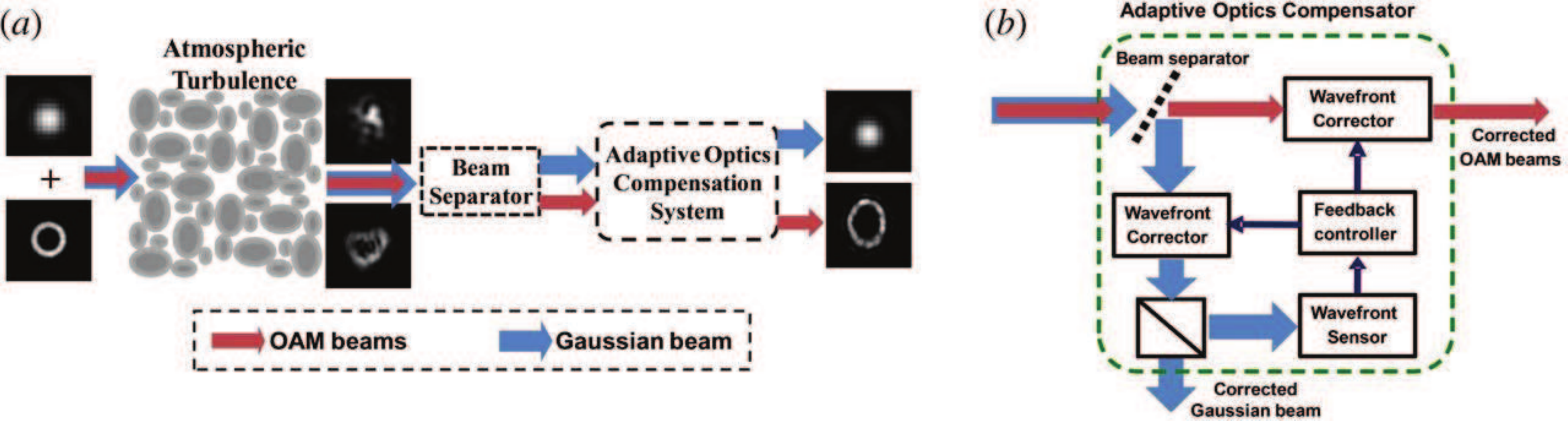}
\caption{(a) Schematic of an AO compensation system for OAM beams using a Gaussian probe beam for wavefront sensing, and (b) detailed implementation of the AO system \cite{Li2018Atmospheric}.}
\label{Fig29}
\end{figure*}

Many approaches for mitigating the turbulence effects have been proposed and are divided into two main categories: \emph{adaptive optics (AO) compensation} and \emph{signal processing-based mitigation} \cite{Li2018Atmospheric}.
AO compensation corrects the distorted OAM wavefronts in the optical domain, and  signal processing-based mitigation utilizes digital signal processing (DSP) algorithms, such as MIMO equalization, to reduce the signal degradation in the electrical domain.

AO compensation systems usually work in a closed-loop configuration. A typical working iteration is to first sense the wavefront of the distorted OAM beam and  then, on the base of the received feedback,  generate an error correction pattern and apply it to the beam to undo the distortion. Based on whether there is a wavefront sensor (WFS) to measure the distorted OAM wavefront, AO compensation can be further divided into WFS-based and non-WFS types.

The main challenge for the WFS-based AO compensation is the difficulty to correctly measure the wavefront of the helical phase front of OAM beams. Recently, a modified AO system that employs a Gaussian probe beam  has been proposed to overcome this problem \cite{Ren2014Adaptive}. As shown in Fig. \ref{Fig29}, the Gaussian probe and the multiplexed OAM beams are coaxially propagated through the atmospheric turbulence so that they all suffer the same distortion.  At the receiver, the distorted Gaussian probe is separated and sent to the WFS for the estimation of  wavefront distortion  and the retrieval of the required correction pattern.  At the transmitter the two wavefront controllers are updated with the fedback correction pattern. It should be noted that, for efficient separation, the Gaussian probe should occupy a separate orthogonal channel, which can be an orthogonal polarization \cite{Ren2014Adaptive} or a separate wavelength \cite{Ren2015Turbulence}. However, the use of an orthogonal polarization channel sacrifices the polarization degree of freedom for multiplexing. It has been shown that using a separate wavelength channel, the compensation performance degrades slowly with the increase in the probe's wavelength offset from the OAM beams \cite{Ren2015Turbulence}. Moreover, the AO system shown in Fig. \ref{Fig29} can be used not only to post-compensate but also to pre-compensate the distorted multiplexed OAM beams in the bidirectional free-space optical communication link \cite{Ren2014Adaptive1}. The AO compensation has been experimentally demonstrated for multiple Bessel beams through atmospheric turbulence and obstructions. Fortunately, the self-healing property of Bessel beams can  be recuperated after turbulence compensation \cite{Li2017Adaptive}.

In addition to WFS sensing, the distortion wavefronts can also be retrieved from  measured intensity profiles by using phase retrieval algorithms, such as the Gerchberg-Saxton (GS) algorithm \cite{Ren2012Correction,Fu2016Pre,Fu2017Non} and stochastic-parallel-gradient-descent (SPGD) algorithm \cite{Xie2015Phase}.
The GS algorithm, a well-known iterative algorithm, can be used to analyze the original and distorted probe Gaussian intensity patterns to obtain a correction phase pattern, which can be used for pre-compensation of the distorted OAM beam through the same turbulence \cite{Fu2016Pre}. If the complex amplitude of the transmitted vortex beam, including the ``doughnut'' amplitude and the helical phase (or mixed helical phase), is known, this method can be generalized to pre-turbulence compensation without the addition of a Gaussian probe \cite{Fu2017Non}. Moreover, the phase correction pattern can also be derived by using the distorted OAM beam intensity pattern  with the SPGD algorithm \cite{Xie2015Phase}.

Signal processing-based algorithms at the receiver can also help mitigate atmospheric turbulence effects and partially shift the complexity of the optical subsystem to the electrical domain,  providing a complementary approach to AO compensation systems.  A $4\times4$ adaptive MIMO equalization system has been implemented in a $4$-channel OAM multiplexed free-space optical link to reduce crosstalk caused by weak turbulence \cite{Huang2014Crosstalk}. The used MIMO DSP is similar to the one that has been used in  few-mode and multi-mode fiber multiplexing systems for mitigating the mode coupling effects \cite{Richardson2013Space}. Experimental results show that all data channels can be recovered and the system power penalty is reduced after MIMO equalization. However, MIMO equalization is not universally useful. Under strong turbulence distortion, outage may occur because crosstalk between channels exceeds a certain threshold or severe power attenuation causes some channels to be barely detectable, in which case MIMO equalization will be ineffective \cite{Winzer2011MIMO}. A modified method is proposed in \cite{Ren2016Atmospheric} to improve system performance under strong turbulence. The proof-of-concept experiment shows that, in an OAM-based spatial diversity free-space optical link, the OAM channel can be recovered under strong turbulence distortion by using a diversity reception strategy assisted with MIMO equalization \cite{Ren2016Atmospheric}.

\subsection{Optical OAM Fiber Communications}
As mentioned above, OAM free-space optical communication links are greatly affected by beam divergence and atmospheric turbulence. In recent years,  research has focused on OAM-based optical fiber communications, which do not need to take these effects into account. Conventional optical fiber communications utilize SDM by using multicore fibers (MCFs) \cite{Zhu2011112, Sakaguchi2012Space} and  few-mode fibers (FMFs) \cite{Randel20116} to increase system capacity and spectral efficiency. Generally, multicore fibers require complex manufacturing, while multi-mode fibers face the problem of mode coupling caused by random perturbations or other nonidealities.  OAM modes, just like linear polarization (LP) modes in fibers, can be used as an orthogonal basis for data transmission in FMFs. Like  LP modes, also OAM modes  face the challenge of \emph{mode coupling}, which leads to channel crosstalk. One possible solution is to use DSP algorithms based on complex MIMO equalization. The other is to use a specially designed FMF, which is called a vortex fiber. The vortex fiber lifts the near-degeneracy between  OAM modes and the parasitic TE$_{01}$ and TM$_{01}$ by modifying the fiber refractive index profile, minimizing mode crosstalk \cite{Ramachandran2009Generation}. Therefore, by using vortex fibers, OAM multiplexing technology has the potential to increase the throughput of optical fiber communication systems with low complexity DSP algorithms or even without any DSP algorithm.

Using a vortex fiber with a characteristic high-index ring around the fiber core, Bozinovic et al. have performed experiments of OAM mode transmission in fibers $20$ m and $900$ m long  \cite{Bozinovic2012Control, Bozinovic2011Long}. In the experiments, a microbend grating is used before the vortex fiber to achieve the conversion from the fundamental modes to the desired HE$_{21}$ modes. The linear combination of the odd and even modes of HE$_{21}$ with a $\pm\pi/2$ phase shift between them results in  OAM modes with $\ell=\pm1$.
In a later experiment, two OAM modes with $\ell=\pm1$ and two LP modes, each mode carrying a $50$ Gbaud QPSK signal, are simultaneously propagated to achieve a transmission rate of $400$ Gbps in a $1.1$ km long vortex fiber \cite{Bozinovic2013Terabit}. At the output of the fiber, the measured mode crosstalk between two OAM modes is approximately $-20$dB. In addition, WDM has also been added to further increase the capacity of the system. By using two OAM modes over $10$ wavelengths in a vortex fiber, $20$ channels, each transmitting a $20$ Gbaud $16$-QAM signal, are established and a transmission rate of $1.6$ Tbps is achieved. These experiments show that OAM can provide an additional degree of freedom for data multiplexing in fiber networks.

\begin{figure}[t]
\centering
\includegraphics[scale=0.85]{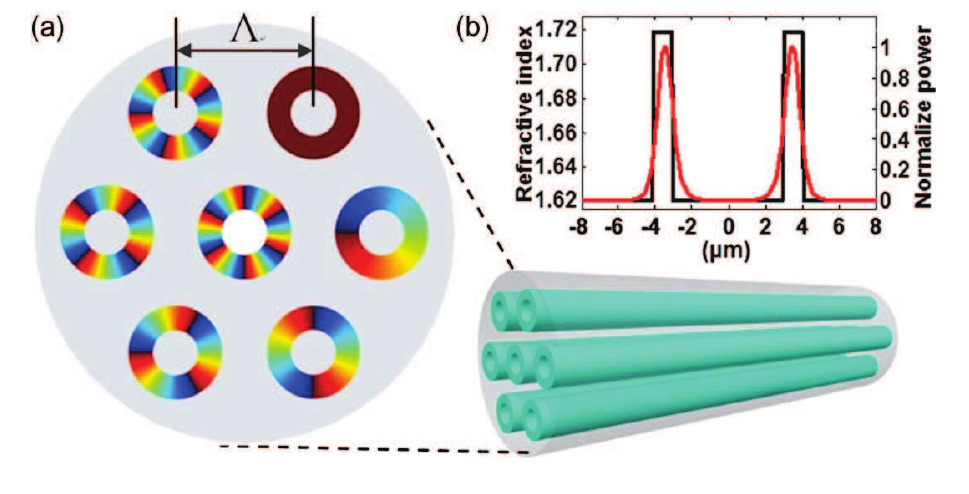}
\caption{(a) 3D structure and cross-section of multi-OAM-mode multi-ring fiber (MOMRF), in which the ring-to-ring distance is $\Lambda$. (b) Index profile of single ring (black) and mode profile of TE$_{01}$ mode (red) in the ring \cite{Li2013Multi}.}
\label{Fig27}
\end{figure}

At present, research on OAM-based optical fiber communications mainly focuses on the design of  fibers that supports stable transmission of multiple OAM modes. In 2012, Birnbaum et al. designed a ring fiber with $0.05$ up-doping to support up to $10$ OAM modes, while maintaining radial single-mode conditions \cite{Yue2012Mode}. In 2013, Li et al. designed a multi-OAM-mode multi-ring fiber (MOMRF) that supports multi-mode OAM transmissions \cite{Li2013Multi}. As shown in Fig. \ref{Fig27}, the fiber consists of seven rings, each supporting $18$ OAM modes. Mode crosstalk and inter-ring crosstalk are reduced by increasing the effective refractive index and the distance between rings. In addition, it is compatible with  WDM  and advanced multilevel amplitude/phase modulation formats, which makes it possible to realize a total transmission capacity in the range of the petabits-per-second and hundredbits-per-second-per-hertz aggregate spectral efficiency. In addition to these ring fibers, some new types of microstructured fibers have also been proposed for multi-mode OAM transmission \cite{Zhang2016A,Zhou2016Design}.

\begin{figure*}[t]
\centering
\includegraphics[scale=1]{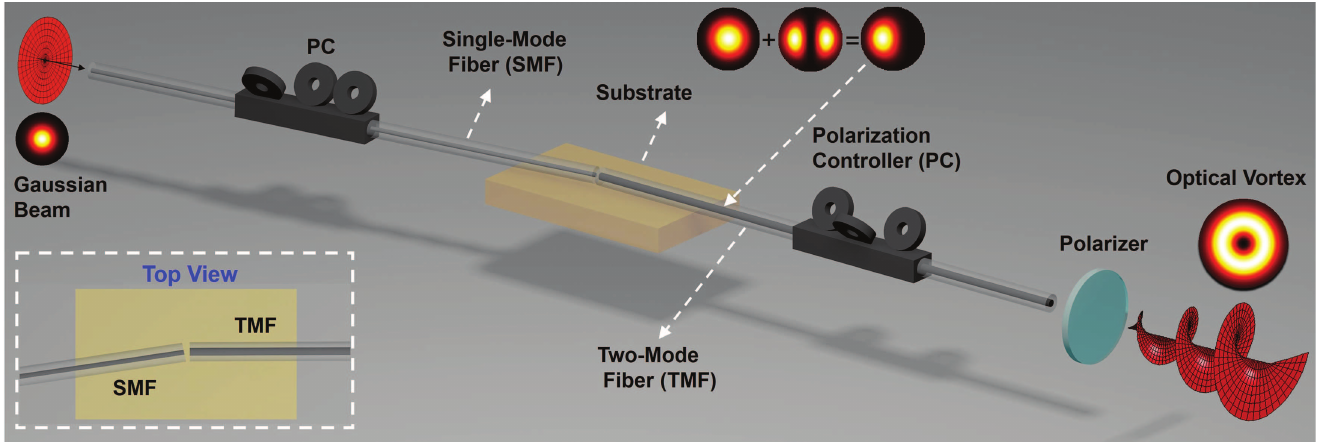}
\caption{Controllable broadband fiber-based OAM converter. The converter consists of a SMF, a TMF wih specific offsets and tilt angles, two PCs and a polarizer, and the input end of TMF is stuck to a standard SMF \cite{Li2018Generation}.}
\label{Fig28}
\end{figure*}

Recently, some studies have explored the potential of the conversion between  LP modes and  OAM modes in fibers. This can be used as a complement to the OAM generation method, and it is easier to couple than the methods mentioned above. In \cite{Zeng2016Experimental}, Zeng et al. designed a novel all-fiber OAM generator, which is cascaded by a mode-selective coupler and a few mode-polarization maintaining fiber, to convert LP$_{01}$ mode to OAM mode.  In \cite{Li2017Superposing}  a series of LP$_{11}$ modes with microphase difference distribution are generated by twisting a few-mode fiber long period grating (FMF-LPG), these LP$_{11}$ modes are  then superimposed in a fiber to generate  OAM modes. In  \cite{Li2018Generation} Li et al. proposed and demonstrated a controllable broadband fiber-based OAM converter. As shown in Fig. \ref{Fig28}, the converter consists of a single-mode fiber (SMF), a two-mode fiber (TMF) with specific offsets and tilt angles, two polarization controllers (PCs) and a polarizer. The input end of TMF is stuck to a standard SMF. By adjusting the two PCs, an input fundamental mode LP$_{01}$ can be selectively converted to  high order LP$_{11}$ modes or OAM modes with $\ell=\pm1$. The purity of the OAM mode is ensured by adjusting the state of the two PCs and the polarizer. The experimental results show that the extinction ratio (used to evaluate the mode purity) of the generated OAM mode and other modes is greater than $20$ dB in a wide wavelength range ($1480$ nm to $1640$ nm).

\subsection{Radio OAM Communications}
The application of OAM to radio communications is expected to be a possible solution to the problem of  spectrum scarcity and it has, accordingly,  received widespread attention.
The first OAM-based wireless communication experiment that successfully separated two radio signals at the same frequency was presented in  \cite{Tamburini2012Encoding}.  A Yagi antenna and a spiral parabolic antenna were used to transmit a plane electromagnetic wave and a vortex electromagnetic wave, respectively,  at a distance of 442 meters. Subsequently, a 4 Gbps uncompressed video transmission link over a 60 GHz OAM radio channel has been implemented \cite{Mahmouli20134}.
In the RF or millimeter wave bands, provided that there is a LOS link so to guarantee the correct transmitter-receiver alignment, OAM can be used to encode information \cite{Allen2014Wireless}, and can also be multiplexed and combined with other technologies  \cite{Yan2014High, Yan201632, Zhao2016A}. In \cite{Yan2014High} are reported the results of an  experiment in the 28 GHz band that has achieved a 32 Gbit/s capacity and about 16 bit/s/Hz spectral efficiency  by combining  4 OAM modes each carrying  4 Gbit/s 16-QAM modulated signals with 2 polarization states.
In recent experiments, multiple radio OAM beams  carrying multiple data streams have been generated by employing some specially designed structures, such as thin metamaterial plates based on rectangular apertures \cite{ Zhao2015Experimental } and dual OAM mode antennas based on ring resonators \cite{Hui2015Multiplexed}.

\begin{figure}[t]
\centering
\includegraphics[scale=1.1]{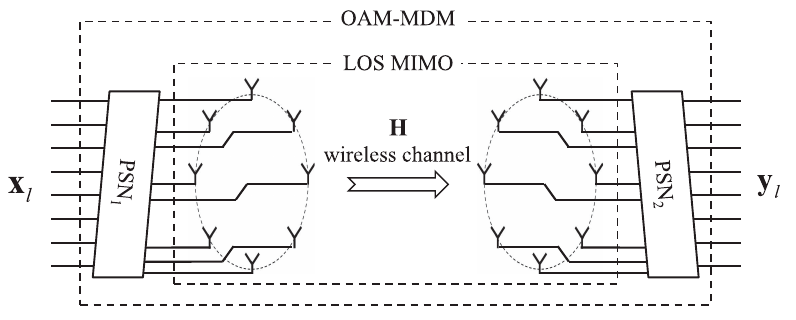}
\caption{System model for a UCAA-based OAM-MDM in a LOS MIMO channel. The OAM-MDM system uses PSN$_1$ and PSN$_2$ for OAM modes multiplexing and demultiplexing, respectively. The vectors $\textbf{x}_l$ and $\textbf{y}_l$ are the complex transmitted and received data vectors, respectively, and the matrix $\textbf{\textrm{H}}$ is the LOS MIMO channel matrix \cite{Zhang2017Mode}.}
\label{Fig23}
\end{figure}

It is well known that MIMO is a key technology that can greatly increase the capacity of wireless communication systems by using multiple transmitting and receiving antennas. The relationship between MIMO SDM and OAM multiplexing has caused a heated debate in the academia. It has been pointed out that radio OAM systems based on UCAA are a subset of MIMO systems and do not really provide additional capacity gain \cite{Edfors2012Is, Tamagnone2012Comment}. But in practice there are  differences between the two approaches. For example,  in a keyhole channel  traditional MIMO systems have very poor performance, while  UCAA-based OAM benefits of the fact that the keyhole does not change the helical phase structure of OAM \cite{Chen2018On}. In terms of capacity over a LOS link, a UCAA-based OAM system is equivalent to a traditional MIMO system from the perspective of channel spatial multiplexing \cite{Oldoni2015Space}. However, OAM receivers have a lower complexity because, unlike traditional MIMO systems, the inherent orthogonality between OAM modes can mitigate inter-channel interference \cite{Zhang2017Mode}.  A sketch of  a system employing OAM-MDM in a LOS MIMO channel is shown in Fig. \ref{Fig23}, where phase shifters networks (PSNs) are used to  multiplex OAM modes at the transmitter and demultiplex OAM modes with a reduced computational load at the receiver.
Moreover, as described in Section III, radio OAM can be generated not only by employing antenna arrays, but also by using SPPs or spiral parabolic antennas or other single transmit antennas. Combining OAM with traditional MIMO technology can result in higher capacity gains \cite{Zhang2016The, Ge2017Millimeter} or provide a more flexible system design \cite{Ren2014Experimental, Ren2017Line}. A $2\times2$ antenna aperture architecture, where each aperture multiplexes two OAM modes, is implemented in the  28 GHz band to achieve a 16 Gbit/s transmission rate. At the receiving end, MIMO signal processing is employed to mitigate inter-channel interference \cite{Ren2017Line}.
Such a system architecture incorporating OAM multiplexing and MIMO technology can also be used for free-space optical communication links \cite{Ren2015Free}.

As it happens for optical OAM, radio OAM communication systems are affected by   \emph{beam divergence} and \emph{misalignment} between the transmitter and the receiver. Since wavelengths at radio frequencies are much larger than at light frequencies, the  effects of atmospheric turbulence on radio OAM decrease and can be usually neglected. Unfortunately,   for the same reason  the problem of beam divergence becomes more important  and represents a bigger challenge in  wireless communication systems than in free-space optical communication systems,  limiting the achievable distance of  OAM  radio links. Like in free-space optical  communications, radio OAM communication systems also require perfect alignment between the transmitter and the receiver \cite{Zhang2013On}. To prevent large performance degradation due  to misalignment, one possible solution in  UCAA-based OAM links is to use a beam steering approach \cite{Chen2018Beam}. By adding additional phases, the UCAA can transmit/receive OAM beams in the desired steering direction.
In the non-parallel case receive beam steering can compensate the phase change caused by oblique angle at the receive UCAA. The off-axis case, where the transmitter and receiver are parallel but not around the same axis, can be decomposed into two non-parallel cases by introducing a virtual UCAA that is perpendicular to and in the middle of the connection between the transmit and receive UCAA centers. Therefore, in the off-axis or the more general misalignment cases,  transmit and receive beam steering need to be used simultaneously. The beam steering approach has been demonstrated through simulation \cite{Chen2018Beam} and the results show that after applying  beam steering in the non-parallel and off-axis case, the misaligned OAM channel capacity shows almost no loss with respect to the case of perfect alignment.

It is well known that \emph{multipath effects} caused by beam spreading and reflections of surrounding objects must be considered in wireless communication systems. In OAM-based wireless communications, the multipath effects caused by specular reflection from a plane parallel to the propagation path have been discussed in \cite{Yan2015Experimental} and \cite{Yan2016Multipath}. Both simulation and experimental results show that OAM channels with larger $\ell$ have stronger intra- and inter-channel interference due to OAM beam divergence. DSP equalization algorithms have the potential to mitigate multipath effects. Another potential solution is to use orthogonal frequency-division multiplexing (OFDM) technology in radio OAM systems. Recently, a transceiver architecture for broadband OAM-OFDM wireless communication systems has been proposed in \cite{Chen2018A}. The transmitter uses a baseband digital two-dimensional Fast Fourier Transform ($2$-D FFT) algorithm and a UCAA to generate OAM-OFDM signals. After multipath channel propagation, the signal is collected by another UCAA and processed using an inverse FFT (IFFT) algorithm. By using the  $2$-D FFT/IFFT algorithms, system implementation complexity can be reduced and  multipath effects are mitigated.

\begin{figure}[t]
\centering
\includegraphics[scale=0.4]{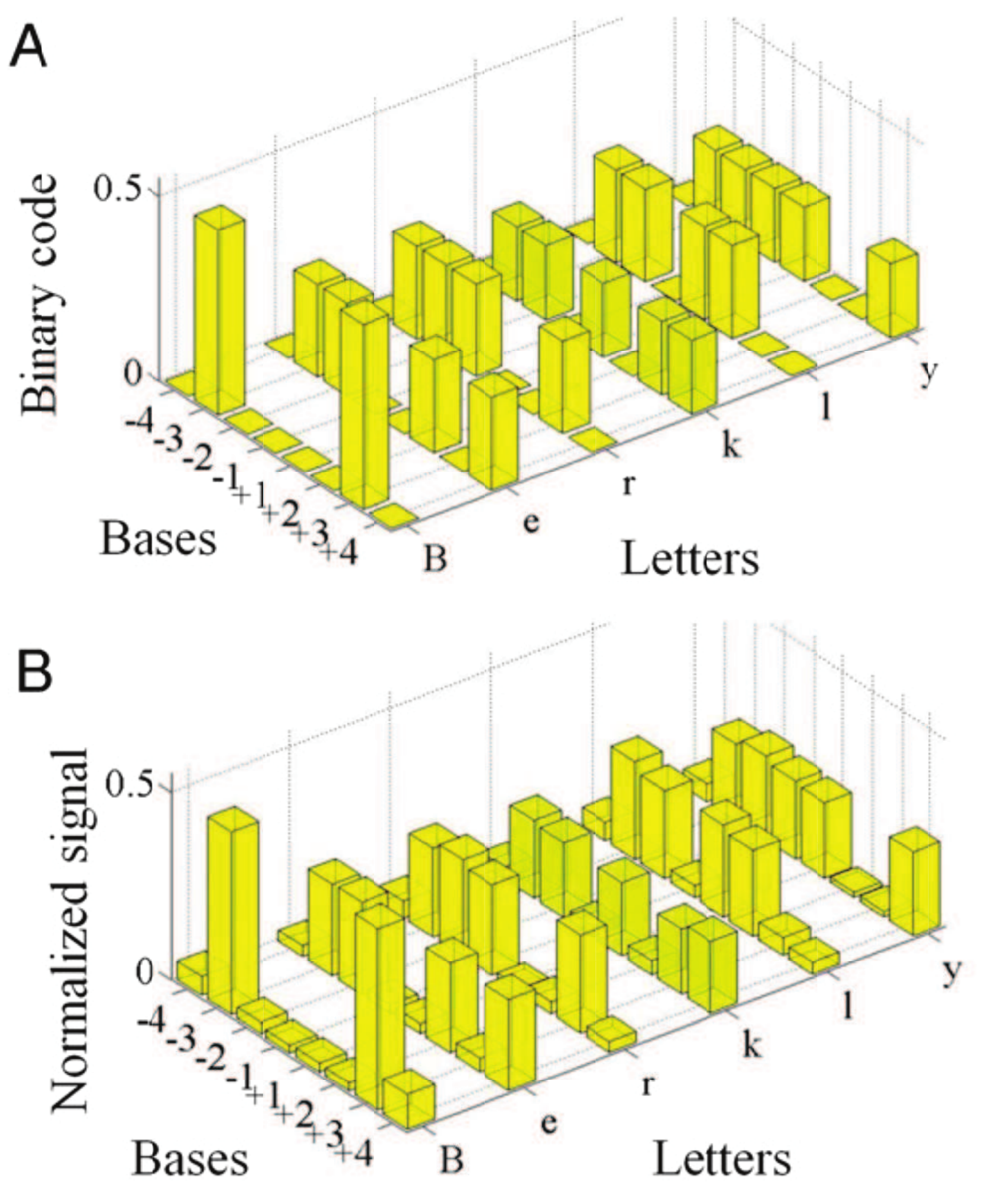}
\caption{(A) Encoding of the letters of the word `Berkly'. Each letter is mapped on $1$ byte ($8$ bits) of information, and each byte contains the same amount of total amplitude and this amplitude amount is equally distributed into the vortex beams. (B) Decoding after transmission with 8 OAM modes \cite{Shi2017High}.}
\label{Fig24}
\end{figure}

\subsection{Acoustic OAM Communications}
In the field of acoustics, OAM-based communication technology is still in its infancy. In 2017, Shi et al. \cite{Shi2017High} used $8$ acoustic OAM modes to transmit the letters of the word `Berkly'  in ASCII binary protocol, achieving a high spectral efficiency of $8.0\pm0.4$ bit/s/Hz. In details, each OAM mode carries $1$-bit information of the eight bits that map each letter, as shown in Fig. \ref{Fig24} (A). These $8$ OAM orthogonal bases are multiplexed  and transmitted by a single transducer array. At the receiver, another transducer array  receives the signal and then demultiplexes it by exploiting the orthogonality between OAM modes. The recovered signal is shown in Fig. \ref{Fig24} (B). Instead of using OAM modes to encode data, Jiang et al. directly loaded the data onto the acoustic OAM channels and established a real-time information transmission system based on passive metamaterials \cite{Jiang2018Twisted}. These studies have shown that in acoustic communications, OAM can be a viable alternative to more traditional technologies.

Since microwaves and mid- and far-infrared radiations are strongly  absorbed underwater, and light is easily obstructed and  scattered by small particles in the oceans, sound waves are the only information carrier for underwater long-distance (more than 200 meters) communications. Accordingly,  OAM  can be applied to underwater acoustic communications to further increase system capacity and spectrum efficiency. In addressing link design for underwater acoustic communications, underwater creatures and turbulence effects should also be considered in addition to the inherent OAM beam divergence and misalignment problems. However, very little research and experimental data  are available  in current scientific literature. Nevertheless, considered the potential capacity gains of OAM, theoretical analysis and experimental verification are expected to be developed in the near future.

\subsection{Discussion and Lessons Learned}
OAM has shown great potential in the communications community due to the inherent orthogonality between modes. Integrating OAM into existing communication systems is expected to further improve spectral efficiency and meet ever-increasing data rate demands. Moreover, the typical circular symmetry of OAM waves makes them very easy to be included in many communication system components. However, OAM integration presents some challenges and technical issues, such as beam divergence, misalignment, and  atmospheric turbulence effects in free-space optical links, mode coupling in fiber links, and multipath effects in radio communication links. These technical issues and their solutions must be considered when building an OAM-based communication system.

Analyzing OAM implementation on the basis of the frequency bands used,  we can summarize the most important characteristics of OAM systems:
\begin{itemize}
\item For free-space optical communication systems, SLMs loaded with phase holograms are recommended in transmitters and receivers to flexibly multiplex and demultiplex OAM beams. The effects of atmospheric turbulence on free-space optical OAM links must be considered in practical environments. Therefore, AO compensation and DSP-based mitigation techniques play a significant role in improving system performance. Optical OAM fiber communication systems can use common optical OAM generation and detection methods, such as SPPs and SLMs. Mode conversion in fibers can also be used to generate OAM. It should be noted that common OAM generators and detectors are required to be well compatible with fibers. Mode coupling is a major challenge in optical OAM fiber links. Novel fibers that support stable transmission of multiple OAM modes and reduce mode crosstalk should be developed in the future.
\item In radio and acoustic communication systems, the flexibility of UCAAs and UCTAs makes them perform well in generating and detecting OAM waves.  However, for generating high frequency OAM waves, SPPs are more suitable than UCAAs and UCTAs. In radio OAM communication systems, OAM beam divergence and misalignment between the transmitter and receiver will greatly limit link achievable distance and degrade system performance.  OAM beam convergence and   system misalignment compensation schemes are required for long-distance transmission. Acoustic OAM communication system is still in an initial stage and the feasibility of acoustic OAM communications, especially underwater, need further experimental verification. The impact of the underwater environment on OAM links also needs further research.
\end{itemize}

After considering all possible challenges and technical issues, we are  at a stage where  experiments and research  on OAM systems should be more oriented to practice testing than to develop proof of concept. At present, most OAM experiments are performed in the controlled environments of laboratories. Testing beyond laboratory distances and practical deployment of OAM systems should be the focus of  future research activities. Another important issue is that, since all  communication systems keep  moving towards low costs, small sizes and high data rates, all  OAM system components  must achieve a reduction in cost and size in the near future.

\section{Application of OAM in Particle Manipulation and Imaging}

\subsection{Optical and Acoustical Particle Manipulation}
OAM has found applications also outside the field of optical, radio and acoustic communications. Thanks to the work of Nobel laureate Arthur Ashkin, light has since long been used to trap small particles in what we call \emph{optical tweezers}.  Optical tweezers use a single tightly-focused beam with a large enough gradient force around the focus to overcome the linear momentum of the light, attracting the particles toward the center \cite{Ashkin1986Observation}. In 1995, \cite{He1995Direct} has shown that,  by transferring the OAM carried by photons to small particles, lasers with OAM can be employed to achieve the rotation of the trapped particles, transforming the tweezers into \emph{optical spanners} that cause the objects to spin \cite{Simpson1997Mechanical}. When a circularly polarized LG beam with $\ell=1$ is used to interact with the particles, the total angular momentum is either $\sim 0.06\hbar$  or  $\sim 2.06\hbar$ per photon depending on the spin direction of the beam, resulting in the stopping or the rotation of the particles. Similarly, using the same method but with a different beam with $\ell=3$, two different rotational velocities of the particles can be observed \cite{Friese1996Optical}. In the above work, the size of the particles is larger than the beam, and the particles are considered to be trapped on or near the axis. When the size is smaller than the ring beam, the particles are generally trapped in the off-axis area. In this case, the spin of the trapped particles depends on the SAM of the ring beam, and the ring beam with the helical phase structure associated with OAM imposes an azimuthal scattering force on the particle to rotate the particle around the beam axis \cite{O2002Intrinsic,Garc2003Observation}.

When interacting with the absorbing particles, OAM carried by acoustic vortices can also be transferred to the particles, exerting a torque on them. This mechanical effect can be applied in the form of \emph{acoustic tweezers} and \emph{acoustic spanners}. Acoustic vortices can manipulate larger size particles than optical vortices due to larger wavelength, and have great application prospects in the field of ultrasonic medicine. Courtney et al. implemented acoustic tweezers using first-order and superimposed high-order Bessel acoustic vortices \cite{Courtney2013Dexterous,Courtney2014Independent}. Their experiment uses a transducer array to generate the required Bessel acoustic vortex field and can control the movement of the vortex center by adjusting the drive signal of the array element, achieving particle capture at different locations. Unlike optical vortices, acoustic vortices cannot carry SAM and,  when vortex beams act on particles, only OAM transfer occurs. In \cite{Volke2008Transfer} it is experimentally demonstrated that acoustic vortices in free space can transmit acoustic OAM to an object and make it rotate. The experiment uses a torsion pendulum to measure the angular momentum of  acoustic vortices and compares the effective acoustic torque obtained with different topological charges. Moreover, in \cite{Anh2012Acoustic} the amount of OAM transmitted by an acoustic vortex to an absorbing disk in a viscous liquid has been quantitatively measured.

From the perspective of simplicity and flexibility, optical OAM for particle manipulation  is usually generated using SPPs and SLMs, while acoustic OAM for particle manipulation  are generated using UCTAs.

\subsection{Optical and Radar Imaging}
OAM applied to an optical or radar imaging system can break the limits of  resolution or sensitivity. The stimulated emission depletion (STED) microscope uses a ring beam with a helical phase distribution to suppress the fluorescence of particles around the scanning point, achieving a resolution beyond the diffraction limit \cite{Hell1994Breaking}. Applying OAM to diffraction tomography can also  make a significant breakthrough with respect to the diffraction limits associated with traditional techniques \cite{Li2013Beating}.

Inserting a spiral phase mask into the Fourier plane of an optical imaging system can effectively convert the point spread function of the system into an annular intensity cross section with an $\exp(i\ell\theta)$ phase distribution  \cite{Maurer2011What}. This can be used to observe the light around bright objects \cite{Swartzlander2001Peering} or to achieve spiral phase contrast imaging that is often used to observe bright edges of phase objects \cite{F2005Spiral,Wang2015Gradual}. In \cite{Swartzlander2008Astronomical}  this method is used to suppress the brightness of stars and make orbital planets more visible.
In addition, when a helical phase beam is used as reference wave of an interference system, the interference pattern is a chiral spiral stripe, so that the protrusions and depressions of the sample surface can be distinguished by observing the direction of rotation of the spiral stripe \cite{F2005Spiralinterferometry}.

\begin{figure}[t]
\centering
\includegraphics[scale=1]{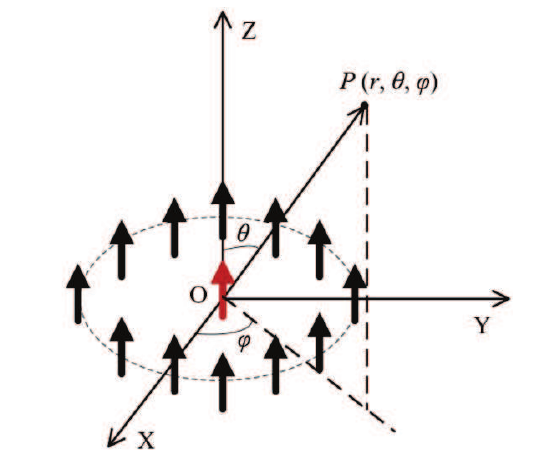}
\caption{Schematic diagram of radar target detection based on vortex electromagnetic waves. The target position $P$ is denoted by $(r, \theta, \varphi)$. In the multiple-input multiple-output mode, all antenna in the UCAA (black arrows) are used to receive the echo signal, while only a single antenna (red arrow) is used in the multiple-input single-output mode \cite{Liu2015Orbital}.}
\label{Fig25}
\end{figure}

In 2013, \cite{Guo2013Electromagnetic} applied OAM to the field of radar imaging, and proposed the idea that vortex electromagnetic waves have the potential for radar target imaging. The paper developed an echo model of the ideal point scattering target under vortex electromagnetic wave illumination and implemented radar imaging using the back projection algorithm and the filtered FFT based algorithm. Afterwards, \cite{Liu2015Orbital} established echo signal models for multiple-input multiple-output and multiple-input single-output systems.
The OAM-based radar imaging system model is shown in Fig. \ref{Fig25}. Assume that the target is made up of $M$ ideal scattering points, denoted by $P_m(r_m,\theta_m,\varphi_m)$ and corresponding radar cross-section $\sigma_m$, $m=1,\cdots,M$. In the multiple-input single-output mode, a single antenna is set at the original point to receive the echo signal. The normalized echo signal is expressed as
\begin{equation} \label{satisfy10}
s(k,\ell)=\sum_{m=1}^{M}\frac{\sigma_{m}}{r_{m}^{2}}e^{i2kr_{m}}e^{i\ell\varphi_{m}}J_{\ell}(ka\sin\theta_{m}),
\end{equation}
where $k=2{\pi}f/c$ is the wave number, $J_{\ell}(ka\sin\theta)$ is the Bessel function of the first kind, and $a$ is the radius of the UCAA. As can be seen from \eqref{satisfy10}, the OAM mode number $\ell$ and the azimuth $\varphi$ satisfy the dual relationship, and so do the frequency $k$ and target range $r$. Thus, the 2-D FFT can be used in estimating the range and azimuth of targets. This shows that the OAM-based radar imaging system has the prospect of acquiring the azimuth information of target.

\begin{figure}[t]
\centering
\includegraphics[scale=1]{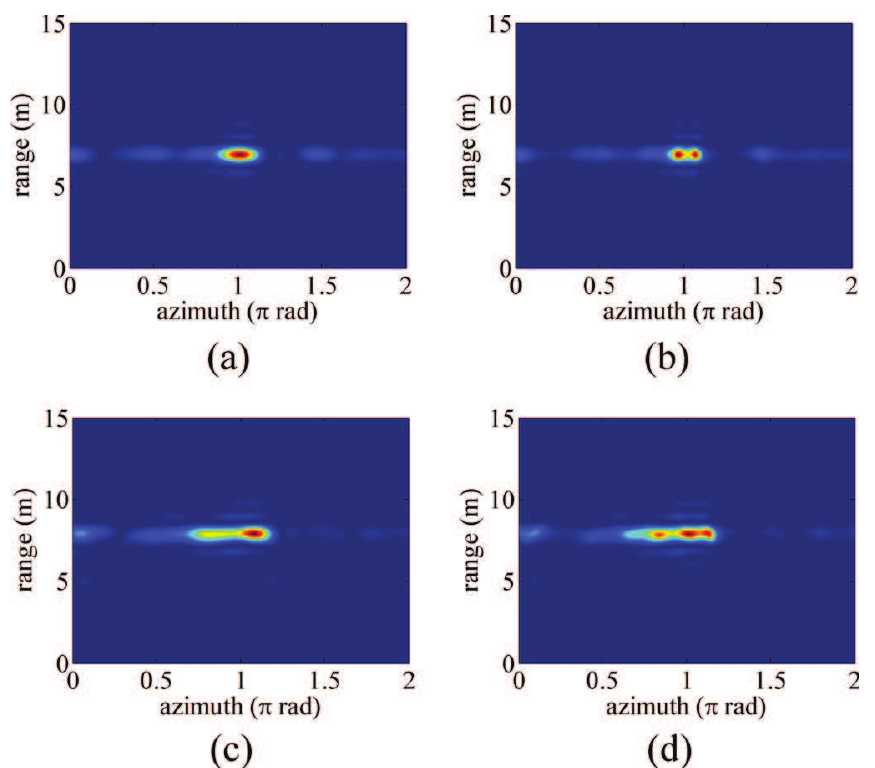}
\caption{Comparisons of imaging results between the traditional array imaging and the electromagnetic vortex imaging: (a) traditional array imaging of two reflectors, (b) electromagnetic vortex imaging of two reflectors, (c) traditional array imaging of three reflectors, and (d) electromagnetic vortex imaging of three reflectors \cite{Liu2017Super}.}
\label{Fig26}
\end{figure}

OAM-based radar detection provides a new set of ideas and solutions for the development of accurate target imaging.
Due to its helical phase structure, a vortex electromagnetic wave carrying OAM can be regarded as a traditional plane wave illuminating a target from multiple consecutive angles, which is equivalent to realizing continuous sampling in a two-dimensional space in a short time, obtaining a good degree of spatial diversity. Therefore, applying multimodal OAM to radar imaging enables azimuthal imaging without relative motion \cite{Yuan2016Electromagnetic}. Moreover, when vortex electromagnetic waves with different helical phase structures hit the target in space, the phase differences associated to the various OAM modes make the backscattering characteristics different and  improve  the radar cross-section  diversity gain \cite{Zhang2017Large}. In conventional radar imaging, the azimuth resolution is generally enhanced by increasing the aperture size, while  vortex electromagnetic waves can provide a high resolution that is not limited by the array aperture, as it has been verified by a proof of concept experiment in \cite{Liu2017Super}. The comparison of the results of traditional array imaging and electromagnetic vortex imaging is shown in Fig. \ref{Fig26},  where it is shown that the  imaging resolution is clearly  improved by using vortex electromagnetic waves. Similarly, it has been demonstrated that applying vortex electromagnetic waves to synthetic aperture radar (SAR) imaging can also achieve higher azimuthal resolution than planar electromagnetic waves \cite{Bu2018Implementation}.

Recently, research on OAM-based radar imaging has mainly focused on azimuth imaging algorithms. Due to the existence of the Rayleigh limit, traditional radar imaging algorithms  can only provide limited azimuth resolution. Some efficient algorithms have been proposed for OAM-based super-resolution radar imaging,
such as UCAA-based autoregressive models and power spectral density (PSD) estimation algorithm \cite{Liu2016Generation}, UCAA-based echo models and multiple signal classification (MUSIC) algorithm \cite{Lin2016Super,Lin2016Improved}, estimating signal parameter via rotational invariance techniques (ESPRIT) algorithm \cite{Chen2018Orbital} and least squares algorithms \cite{Yuan2017Beam,Yuan2016Orbital}.

There are also many challenges in OAM-based radar imaging.
Because the center of vortex electromagnetic beams is null and the angular direction of the main lobe varies with the OAM modes, it is difficult to  simultaneously illuminate the target with vortex electromagnetic beams of different modes, which results in  limited echo energy. The difference of echo intensity will lead to amplitude modulation of the echo signal, resulting in a deterioration of the imaging capability. These effects can be mitigated by carefully designing the radius and excitation of the UCAA to adjust the main lobes of the vortex electromagnetic waves with different modes \cite{Yuan2016Electromagnetic,Liu2016Generation,Yuan2017Beam}.
In addition, detecting dynamic targets is also challenging, which requires more in-depth research.

Optical OAM beams for imaging is generated using SPPs and SLMs, while radio OAM beams  for radar imaging are generated using UCAAs. OAM beam divergence and beam steering must be considered in order to make OAM waves with multiple modes illuminate the same target. More in-depth research on topics  such as high-precision azimuth imaging and dynamic target detection are the object of  future research.

\section{Conclusions and Perspectives}
In recent years, OAM has been extensively studied in many fields of application, especially in the field of communications. A large number of studies have shown that it is possible to multiplex a set of orthogonal OAM modes on the same frequency channel so to achieve a high spectral efficiency. In this paper, we provide a comprehensive overview of the generation and detection methods of optical, radio, and acoustic OAM, together with their applications in communications, particle manipulation and imaging.
Particular attention has been devoted to the issues that still obstruct or limit  the application of OAM in practice.

OAM technology is on the brink of been deployed in many communication systems but there are still some problems that need to be solved. For example, many of the experiments related to OAM multiplexing still use bulky and expensive components, which are not suitable for practical implementation and large-scale use. How to generate and detect high-order and multi-mode OAM beams with a finite-size aperture is challenging. Future success of  OAM  heavily relies on the development of system components such as transmitters, multiplexers, demultiplexers and receivers. These components are required to provide reductions in cost and size and to be compatible with existing technologies. Moreover, the state of the propagation channel has a large impact on OAM-based links and the problems arising in some complex and severe channel conditions need to be addressed and solved.
In spite of the many challenges in applying OAM to communication systems, on the  base of the research progress surveyed in this paper, we remain optimistic that these challenges will be gradually resolved in the future and that OAM will help to make breakthroughs in future communications and radar target detection systems.

\bibliographystyle{IEEEtran}
\bibliography{IEEEabrv,OAM}

\begin{thebibliography}{100}
\providecommand{\url}[1]{#1}
\csname url@samestyle\endcsname
\providecommand{\newblock}{\relax}
\providecommand{\bibinfo}[2]{#2}
\providecommand{\BIBentrySTDinterwordspacing}{\spaceskip=0pt\relax}
\providecommand{\BIBentryALTinterwordstretchfactor}{4}
\providecommand{\BIBentryALTinterwordspacing}{\spaceskip=\fontdimen2\font plus
\BIBentryALTinterwordstretchfactor\fontdimen3\font minus
  \fontdimen4\font\relax}
\providecommand{\BIBforeignlanguage}[2]{{%
\expandafter\ifx\csname l@#1\endcsname\relax
\typeout{** WARNING: IEEEtran.bst: No hyphenation pattern has been}%
\typeout{** loaded for the language `#1'. Using the pattern for}%
\typeout{** the default language instead.}%
\else
\language=\csname l@#1\endcsname
\fi
#2}}
\providecommand{\BIBdecl}{\relax}
\BIBdecl

\bibitem{Allen1992Orbital}
L.~Allen, M.~W. Beijersbergen, R.~J. Spreeuw, and J.~P. Woerdman, ``Orbital
  angular momentum of light and the transformation of {Laguerre-Gaussian} laser
  modes,'' \emph{Phys. Rev. A: At. Mol. Opt. Phys.}, vol.~45, no.~11, pp.
  8185--8189, 1992.

\bibitem{Thid2007Utilization}
B.~Thid\'e, H.~Then, J.~Sj\"oholm, K.~Palmer, J.~Bergman, T.~D. Carozzi, Y.~N.
  Istomin, N.~H. Ibragimov, and R.~Khamitova, ``Utilization of photon orbital
  angular momentum in the low-frequency radio domain,'' \emph{Phys. Rev.
  Lett.}, vol.~99, no.~8, p. 087701, 2007.

\bibitem{Broadbent1979Acoustic}
E.~G. Broadbent and D.~W. Moore, ``Acoustic destabilization of vortices,''
  \emph{Philos. Trans. R. Soc. London}, vol. 290, no. 1372, pp. 353--371, 1979.

\bibitem{Hefner1999An}
B.~T. Hefner and P.~L. Marston, ``An acoustical helicoidal wave transducer with
  applications for the alignment of ultrasonic and underwater systems,''
  \emph{J. Acoust. Soc. Am.}, vol. 106, no.~6, pp. 3313--3316, 1999.

\bibitem{Yao2011Orbital}
A.~M. Yao and M.~J. Padgett, ``Orbital angular momentum: origins, behavior and
  applications,'' \emph{Adv. Opt. Photonics}, vol.~3, no.~2, pp. 161--204,
  2011.

\bibitem{Willner2017Recent}
A.~E. Willner, Y.~Ren, G.~Xie, Y.~Yan, L.~Li, Z.~Zhao, J.~Wang, M.~Tur, A.~F.
  Molisch, and S.~Ashrafi, ``Recent advances in high-capacity free-space
  optical and radio-frequency communications using orbital angular momentum
  multiplexing,'' \emph{Phil. Trans. R. Soc. A}, vol. 375, no. 2087, p.
  20150439, 2017.

\bibitem{Willner2015Optical}
A.~E. Willner, H.~Huang, Y.~Yan, Y.~Ren, N.~Ahmed, G.~Xie, C.~Bao, L.~Li,
  Y.~Cao, Z.~Zhao, J.~Wang, M.~P.~J. Lavery, M.~Tur, S.~Ramachandran, A.~F.
  Molisch, N.~Ashrafi, and S.~Ashrafi, ``Optical communications using orbital
  angular momentum beams,'' \emph{Adv. Opt. Photonics}, vol.~7, no.~1, pp.
  66--106, 2015.

\bibitem{Mohammadi2010Orbital}
S.~M. Mohammadi, L.~K.~S. Daldorff, J.~E.~S. Bergman, R.~L. Karlsson,
  B.~Thid\'e, K.~Forozesh, T.~D. Carozzi, and B.~Isham, ``Orbital angular
  momentum in radio--a system study,'' \emph{{IEEE} Trans. Antennas Propag.},
  vol.~58, no.~2, pp. 565--572, 2010.

\bibitem{Vaughan1983Temporal}
J.~M. Vaughan and D.~V. Willetts, ``Temporal and interference fringe analysis
  of {TEM$_{01}$*} laser modes,'' \emph{J. Opt. Soc. Am.}, vol.~73, no.~8, pp.
  1018--1021, 1983.

\bibitem{Kano2011Generation}
K.~Kano, Y.~Kozawa, and S.~Sato, ``Generation of a purely single transverse
  mode vortex beam from a {He-Ne} laser cavity with a spot-defect mirror,''
  \emph{Int. J. Opt.}, vol. 2012, pp. 1--6, 2012.

\bibitem{Beijersbergen1993Astigmatic}
M.~W. Beijersbergen, L.~Allen, H.~van~der Veen, and J.~P. Woerdman,
  ``Astigmatic laser mode converters and transfer of orbital angular
  momentum,'' \emph{Opt. Commun.}, vol.~96, no.~1, pp. 123--132, 1993.

\bibitem{Beijersbergen1994Helical}
M.~W. Beijersbergen, R.~P.~C. Coerwinkel, M.~Kristensen, and J.~P. Woerdman,
  ``Helical-wavefront laser beams produced with a spiral phaseplate,''
  \emph{Opt. Commun.}, vol. 112, no.~5, pp. 321--327, 1994.

\bibitem{Yu1990Laser}
V.~Y. Bazhenov, M.~V. Vasnetsov, and M.~S. Soskin, ``Laser beams with screw
  dislocations in their wavefronts,'' \emph{JETP Lett.}, vol.~52, no.~8, pp.
  429--431, 1990.

\bibitem{Bazhenov1992Screw}
V.~Y. Bazhenov, M.~S. Soskin, and M.~V. Vasnetsov, ``Screw dislocations in
  light wavefronts,'' \emph{J. Mod. Opt.}, vol.~39, no.~5, pp. 985--990, 1992.

\bibitem{Heckenberg1992Generation}
N.~R. Heckenberg, R.~Mcduff, C.~P. Smith, and A.~G. White, ``Generation of
  optical phase singularities by computer-generated holograms,'' \emph{Opt.
  Lett.}, vol.~17, no.~3, pp. 221--223, 1992.

\bibitem{Curtis2002Dynamic}
J.~E. Curtis, B.~A. Koss, and D.~G. Grier, ``Dynamic holographic optical
  tweezers,'' \emph{Opt. Commun.}, vol. 207, no.~1, pp. 169--175, 2002.

\bibitem{Maurer2007Tailoring}
C.~Maurer, A.~Jesacher, S.~F\"urhapter, S.~Bernet, and M.~Ritsch-Marte,
  ``Tailoring of arbitrary optical vector beams,'' \emph{New J. Phys.}, vol.~9,
  no.~3, p.~78, 2007.

\bibitem{Mirhosseini2013Rapid}
M.~Mirhosseini, O.~S. Maga\~{n}a Loaiza, C.~Chen, B.~Rodenburg, M.~Malik, and
  R.~W. Boyd, ``Rapid generation of light beams carrying orbital angular
  momentum,'' \emph{Opt. Express}, vol.~21, no.~25, pp. 30\,196--30\,203, 2013.

\bibitem{Mitchell2016High}
K.~J. Mitchell, S.~Turtaev, M.~J. Padgett, T.~\v{C}i\v{z}m\'{a}r, and D.~B.
  Phillips, ``High-speed spatial control of the intensity, phase and
  polarisation of vector beams using a digital micro-mirror device,''
  \emph{Opt. Express}, vol.~24, no.~25, pp. 29\,269--29\,282, 2016.

\bibitem{Yu2011Light}
N.~Yu, P.~Genevet, M.~A. Kats, F.~Aieta, J.~P. Tetienne, F.~Capasso, and
  Z.~Gaburro, ``Light propagation with phase discontinuities: generalized laws
  of reflection and refraction,'' \emph{Science}, vol. 334, no. 6054, pp.
  333--337, 2011.

\bibitem{Genevet2012Ultra}
P.~Genevet, N.~Yu, F.~Aieta, J.~Lin, M.~A. Kats, R.~Blanchard, M.~O. Scully,
  Z.~Gaburro, and F.~Capasso, ``Ultra-thin plasmonic optical vortex plate based
  on phase discontinuities,'' \emph{Appl. Phys. Lett.}, vol. 100, no.~1, p.
  013101, 2012.

\bibitem{Karimi2014Generating}
E.~Karimi, S.~A. Schulz, I.~D. Leon, H.~Qassim, J.~Upham, and R.~W. Boyd,
  ``Generating optical orbital angular momentum at visible wavelengths using a
  plasmonic metasurface,'' \emph{Light: Sci. Appl.}, vol.~3, p. e167, 2014.

\bibitem{Zhao2013Metamaterials}
Z.~Zhao, J.~Wang, S.~Li, and A.~E. Willner, ``Metamaterials-based broadband
  generation of orbital angular momentum carrying vector beams,'' \emph{Opt.
  Lett.}, vol.~38, no.~6, pp. 932--934, 2013.

\bibitem{Wang2015Ultra}
W.~Wang, Y.~Li, Z.~Guo, R.~Li, J.~Zhang, A.~Zhang, and S.~Qu, ``Ultra-thin
  optical vortex phase plate based on the metasurface and the angular momentum
  transformation,'' \emph{J. Opt.}, vol.~17, no.~4, p. 045102, 2015.

\bibitem{Marrucci2011Spin}
L.~Marrucci, E.~Karimi, S.~Slussarenko, B.~Piccirillo, E.~Santamato, E.~Nagali,
  and F.~Sciarrino, ``Spin-to-orbital conversion of the angular momentum of
  light and its classical and quantum applications,'' \emph{J. Opt.}, vol.~13,
  no.~6, p. 064001, 2011.

\bibitem{Marrucci2006Optical}
L.~Marrucci, C.~Manzo, and D.~Paparo, ``Optical spin-to-orbital angular
  momentum conversion in inhomogeneous anisotropic media,'' \emph{Phys. Rev.
  Lett.}, vol.~96, no.~16, p. 163905, 2006.

\bibitem{Hasman2002Space}
Z.~Bomzon, G.~Biener, V.~Kleiner, and E.~Hasman, ``Space-variant
  {Pancharatnam-Berry} phase optical elements with computer-generated
  subwavelength gratings,'' \emph{Opt. Lett.}, vol.~27, no.~13, pp. 1141--1143,
  2002.

\bibitem{Biener2002Formation}
G.~Biener, A.~Niv, V.~Kleiner, and E.~Hasman, ``Formation of helical beams by
  use of {Pancharatnam-Berry} phase optical elements,'' \emph{Opt. Lett.},
  vol.~27, no.~21, pp. 1875--1877, 2002.

\bibitem{Zhu2014Experimental}
L.~Zhu, X.~Wei, J.~Wang, Z.~Zhang, Z.~Li, H.~Zhang, S.~Li, K.~Wang, and J.~Liu,
  ``Experimental demonstration of basic functionalities for {0.1-THz} orbital
  angular momentum ({OAM}) communications,'' in \emph{Proc. Opt. Fiber Commun.
  Conf. Expo.}, 2014, p. M3K.7.

\bibitem{Bennis2013Flat}
A.~Bennis, R.~Niemiec, C.~Brousseau, K.~Mahdjoubi, and O.~Emile, ``Flat plate
  for {OAM} generation in the millimeter band,'' in \emph{Proc. 7th Eur. Conf.
  Antennas Propag.}, 2013, pp. 3203--3207.

\bibitem{Cheng2014Generation}
L.~Cheng, W.~Hong, and Z.-C. Hao, ``Generation of electromagnetic waves with
  arbitrary orbital angular momentum modes,'' \emph{Sci. Rep.}, vol.~4, p.
  4814, 2014.

\bibitem{Mahmouli2012Orbital}
F.~E. Mahmouli and S.~Walker, ``Orbital angular momentum generation in a
  {60GHz} wireless radio channel,'' in \emph{Proc. 20th Telecommun. Forum},
  2012, pp. 315--318.

\bibitem{Tamburini2011Experimental}
F.~Tamburini, E.~Mari, B.~Thid\'e, C.~Barbieri, and F.~Romanato, ``Experimental
  verification of photon angular momentum and vorticity with radio
  techniques,'' \emph{Appl. Phys. Lett.}, vol.~99, no.~20, p. 204102, 2011.

\bibitem{Tamburini2012Encoding}
F.~Tamburini, E.~Mari, A.~Sponselli, B.~Thid\'e, A.~Bianchini, and F.~Romanato,
  ``Encoding many channels in the same frequency through radio vorticity: first
  experimental test,'' \emph{New J. Phys.}, vol.~14, no.~3, p. 033001, 2012.

\bibitem{Turnbull1996The}
G.~A. Turnbull, D.~A. Robertson, G.~M. Smith, L.~Allen, and M.~J. Padgett,
  ``The generation of free-space {Laguerre-Gaussian} modes at millimetre-wave
  frequencies by use of a spiral phaseplate,'' \emph{Opt. Commun.}, vol. 127,
  no.~4, pp. 183--188, 1996.

\bibitem{Yan2014Demonstration}
Y.~Yan, G.~Xie, H.~Huang, M.~J. Lavery, N.~Ahemd, C.~Bao, Y.~Ren, A.~F.
  Molisch, M.~Tur, M.~Padgett, and A.~E. Willner, ``Demonstration of 8-mode
  {32-Gbit/s} millimeter-wave free-space communication link using 4
  orbital-angular-momentum modes on 2 polarizations,'' in \emph{Proc. IEEE Int.
  Conf. Commun.}, 2014, pp. 4850--4855.

\bibitem{Deng2013Generation}
C.~Deng, W.~Chen, Z.~Zhang, Y.~Li, and Z.~Feng, ``Generation of {OAM} radio
  waves using circular {Vivaldi} antenna array,'' \emph{Int. J. Antennas
  Propag.}, vol. 2013, no.~2, pp. 607--610, 2013.

\bibitem{Bai2015Generation}
X.~Bai, X.~Liang, R.~Jin, and J.~Geng, ``Generation of {OAM} radio waves with
  three polarizations using circular horn antenna array,'' in \emph{Proc. 9th
  Eur. Conf. Antennas Propag.}, 2015.

\bibitem{Gaffoglio2016OAM}
R.~Gaffoglio, A.~Cagliero, A.~D. Vita, and B.~Sacco, ``{OAM} multiple
  transmission using uniform circular arrays: Numerical modeling and
  experimental verification with two digital television signals,'' \emph{Radio
  Sci.}, vol.~51, no.~6, pp. 645--658, 2016.

\bibitem{Bai2013Generation}
Q.~Bai, A.~Tennant, B.~Allen, and M.~U. Rehman, ``Generation of orbital angular
  momentum ({OAM}) radio beams with phased patch array,'' in \emph{Proc.
  Loughborough Antennas Propag. Conf.}, 2013, pp. 410--413.

\bibitem{Sun2016The}
X.~Sun, Y.~Du, Y.~Fan, and M.~Sun, ``The design of array antenna based on
  multi-modal {OAM} vortex electromagnetic wave,'' in \emph{Proc. Prog.
  Electromagn. Res. Symp.}, 2016, pp. 2786--2791.

\bibitem{Lin2017Theoretical}
M.~Lin, Y.~Gao, P.~Liu, and J.~Liu, ``Theoretical analyses and design of
  circular array to generate orbital angular momentum,'' \emph{{IEEE} Trans.
  Antennas Propag.}, vol.~65, no.~7, pp. 3510--3519, 2017.

\bibitem{Tennant2013Generation}
A.~Tennant and B.~Allen, ``Generation of radio frequency {OAM} radiation modes
  using circular time-switched and phased array antennas,'' in \emph{Proc.
  Loughborough Antennas Propag. Conf.}, 2012.

\bibitem{Yu2016Design}
S.~Yu, L.~Li, G.~Shi, C.~Zhu, X.~Zhou, and Y.~Shi, ``Design, fabrication, and
  measurement of reflective metasurface for orbital angular momentum vortex
  wave in radio frequency domain,'' \emph{Appl. Phys. Lett.}, vol. 108, no.~12,
  p. 121903, 2016.

\bibitem{Chen2016Artificial}
M.~L.~N. Chen, L.~J. Jiang, and W.~E.~I. Sha, ``Artificial perfect electric
  conductor-perfect magnetic conductor anisotropic metasurface for generating
  orbital angular momentum of microwave with nearly perfect conversion
  efficiency,'' \emph{J. Appl. Phys.}, vol. 119, no.~6, p. 064506, 2016.

\bibitem{Yu2016Generating}
S.~Yu, L.~Li, G.~Shi, C.~Zhu, and Y.~Shi, ``Generating multiple orbital angular
  momentum vortex beams using a metasurface in radio frequency domain,''
  \emph{Appl. Phys. Lett.}, vol. 108, no.~24, p. 241901, 2016.

\bibitem{Yu2016Dual}
S.~Yu, L.~Li, and G.~Shi, ``Dual-polarization and dual-mode orbital angular
  momentum radio vortex beam generated by using reflective metasurface,''
  \emph{Appl. Phys. Express}, vol.~9, no.~8, p. 082202, 2016.

\bibitem{Zhang2018Transforming}
Y.~Zhang, L.~Yang, H.~Wang, X.~Zhang, and X.~Jin, ``Transforming surface wave
  to propagating {OAM} vortex wave via flat dispersive metasurface in radio
  frequency,'' \emph{{IEEE} Antennas Wireless Propag. Lett.}, vol.~17, no.~1,
  pp. 172--175, 2018.

\bibitem{Kou2016Generation}
N.~Kou, S.~Yu, and L.~Li, ``Generation of high-order {Bessel} vortex beam
  carrying orbital angular momentum using multilayer amplitude-phase-modulated
  surfaces in radiofrequency domain,'' \emph{Appl. Phys. Express}, vol.~10,
  no.~1, p. 016701, 2016.

\bibitem{Maccalli2013q}
S.~Maccalli, G.~Pisano, S.~Colafrancesco, B.~Maffei, M.~W. Ng, and M.~Gray,
  ``Q-plate for millimeter-wave orbital angular momentum manipulation,''
  \emph{Appl. Opt.}, vol.~52, no.~4, pp. 635--639, 2013.

\bibitem{Wunenburger2015Acoustic}
R.~Wunenburger, J.~I.~V. Lozano, and E.~Brasselet, ``Acoustic orbital angular
  momentum transfer to matter by chiral scattering,'' \emph{New J. Phys.},
  vol.~17, no.~10, p. 103022, 2015.

\bibitem{Gspan2004Optoacoustic}
S.~Gspan, A.~Meyer, S.~Bernet, and M.~Ritsch-Marte, ``Optoacoustic generation
  of a helicoidal ultrasonic beam,'' \emph{J. Acoust. Soc. Am.}, vol. 115,
  no.~3, pp. 1142--1146, 2004.

\bibitem{Hefner1998Acoustical}
B.~T. Hefner and P.~L. Marston, ``Acoustical helicoidal waves and
  {Laguerre-Gaussian} beams: Applications to scattering and to angular momentum
  transport,'' \emph{J. Acoust. Soc. Am.}, vol. 103, no.~5, pp. 2971--2971,
  1998.

\bibitem{Ealo2011Airborne}
J.~L. Ealo, J.~C. Prieto, and F.~Seco, ``Airborne ultrasonic vortex generation
  using flexible ferroelectrets,'' \emph{IEEE T. Ultrason. Ferr.}, vol.~58,
  no.~8, pp. 1651--1657, 2011.

\bibitem{Marchiano2005Synthesis}
R.~Marchiano and J.~L. Thomas, ``Synthesis and analysis of linear and nonlinear
  acoustical vortices,'' \emph{Phys. Rev. E}, vol.~71, no.~6, p. 066616, 2005.

\bibitem{Volke2008Transfer}
K.~Volke-Sep\'ulveda, A.~O. Santill\'an, and R.~R. Boullosa, ``Transfer of
  angular momentum to matter from acoustical vortices in free space,''
  \emph{Phys. Rev. Lett.}, vol. 100, no.~2, p. 024302, 2008.

\bibitem{Yang2013Phase}
L.~Yang, Q.~Ma, J.~Tu, and D.~Zhang, ``Phase-coded approach for controllable
  generation of acoustical vortices,'' \emph{J. Appl. Phys.}, vol. 113, no.~15,
  p. 154904, 2013.

\bibitem{Li2017Deep}
Y.~Li, G.~Guo, Q.~Ma, J.~Tu, and D.~Zhang, ``Deep-level stereoscopic multiple
  traps of acoustic vortices,'' \emph{J. Appl. Phys.}, vol. 121, no.~16, p.
  164901, 2017.

\bibitem{Jiang2016Broadband}
X.~Jiang, J.~Zhao, S.~Liu, B.~Liang, X.~Zou, J.~Yang, C.~Qiu, and J.~Cheng,
  ``Broadband and stable acoustic vortex emitter with multi-arm coiling
  slits,'' \emph{Appl. Phys. Lett.}, vol. 108, no.~20, p. 203501, 2016.

\bibitem{Wang2016Particle}
T.~Wang, M.~Ke, W.~Li, Q.~Yang, C.~Qiu, and Z.~Liu, ``Particle manipulation
  with acoustic vortex beam induced by a brass plate with spiral shape
  structure,'' \emph{Appl. Phys. Lett.}, vol. 109, no.~12, p. 123506, 2016.

\bibitem{Jim2016Formation}
N.~Jim\'enez, R.~Pic\'o, V.~S\'anchez-Morcillo, V.~Romero-Garc\'{\i}a, L.~M.
  Garc\'{\i}a-Raffi, and K.~Staliunas, ``Formation of high-order acoustic
  {Bessel} beams by spiral diffraction gratings,'' \emph{Phys. Rev. E},
  vol.~94, no.~5, p. 053004, 2016.

\bibitem{Jim2018Sharp}
N.~Jim\'enez, V.~Romero-Garc\'{\i}a, L.~M. Garc\'{\i}a-Raffi, F.~Camarena, and
  K.~Staliunas, ``Sharp acoustic vortex focusing by {Fresnel}-spiral zone
  plates,'' \emph{Appl. Phys. Lett.}, vol. 112, no.~20, p. 204101, 2018.

\bibitem{Muelas2018Generation}
R.~D. Muelas-Hurtado, J.~L. Ealo, J.~F. Pazos-Ospina, and K.~Volke-Sep\'ulveda,
  ``Generation of multiple vortex beam by means of active diffraction
  gratings,'' \emph{Appl. Phys. Lett.}, vol. 112, no.~8, p. 084101, 2018.

\bibitem{Jiang2016Convert}
J.~Xue, L.~Yong, L.~Bin, C.~Jianchun, and Z.~Likun, ``Convert acoustic
  resonances to orbital angular momentum,'' \emph{Phys. Rev. Lett.}, vol. 117,
  no.~3, p. 034301, 2016.

\bibitem{Ye2016Making}
L.~Ye, C.~Qiu, J.~Lu, K.~Tang, H.~Jia, M.~Ke, S.~Peng, and Z.~Liu, ``Making
  sound vortices by metasurfaces,'' \emph{AIP Advances}, vol.~6, no.~8, p.
  085007, 2016.

\bibitem{Esfahlani2017Generation}
H.~Esfahlani, H.~Lissek, and J.~R. Mosig, ``Generation of acoustic helical
  wavefronts using metasurfaces,'' \emph{Phys. Rev. B}, vol.~95, no.~2, p.
  024312, 2017.

\bibitem{Leach2002Measuring}
J.~Leach, M.~J. Padgett, S.~M. Barnett, S.~Franke-Arnold, and J.~Courtial,
  ``Measuring the orbital angular momentum of a single photon,'' \emph{Phys.
  Rev. Lett.}, vol.~88, no.~25, p. 257901, 2002.

\bibitem{Mair2001Entanglement}
A.~Mair, A.~Vaziri, G.~Weihs, and A.~Zeilinger, ``Entanglement of the orbital
  angular momentum states of photons,'' \emph{Nature}, vol. 412, no. 6844, pp.
  313--316, 2001.

\bibitem{Gibson2004Free}
G.~Gibson, J.~Courtial, M.~Vasnetsov, M.~J. Padgett, S.~Franke-Arnold, S.~M.
  Barnett, and V.~Pas¡¯Ko, ``Free-space information transfer using light beams
  carrying orbital angular momentum,'' \emph{Opt. Express}, vol.~12, no.~22,
  pp. 5448--5456, 2004.

\bibitem{Gibson2004Increasing}
G.~Gibson, J.~Courtial, M.~Vasnetsov, S.~Barnett, S.~Franke-Arnold, and
  M.~Padgett, ``Increasing the data density of free-space optical
  communications using orbital angular momentum,'' in \emph{Proc. SPIE}, 2004,
  pp. 367--373.

\bibitem{Zhang2010Extending}
N.~Zhang, X.~C. Yuan, and R.~E. Burge, ``Extending the detection range of
  optical vortices by dammann vortex gratings,'' \emph{Opt. Lett.}, vol.~35,
  no.~20, pp. 3495--3497, 2010.

\bibitem{Gao2016Integrating}
C.~Gao, S.~Zhang, S.~Fu, and T.~Wang, ``Integrating 5{\texttimes}5 dammann
  gratings to detect orbital angular momentum states of beams with the range of
  $-$24 to $+$24,'' \emph{Appl. Opt.}, vol.~55, no.~7, pp. 1514--1517, 2016.

\bibitem{Kai2017Orbital}
C.~Kai, P.~Huang, F.~Shen, H.~Zhou, and Z.~Guo, ``Orbital angular momentum
  shift keying based optical communication system,'' \emph{IEEE Photonics J.},
  vol.~9, no.~2, pp. 1--10, 2017.

\bibitem{Lin2006Synthesis}
J.~Lin, X.~Yuan, S.~H. Tao, and R.~E. Burge, ``Synthesis of multiple collinear
  helical modes generated by a phase-only element,'' \emph{J. Opt. Soc. Am. A},
  vol.~23, no.~5, pp. 1214--1218, 2006.

\bibitem{Sztul2006Double}
H.~I. Sztul and R.~R. Alfano, ``Double-slit interference with
  {Laguerre-Gaussian} beams,'' \emph{Opt. Lett.}, vol.~31, no.~7, pp.
  999--1001, 2006.

\bibitem{Zhou2014Dynamic}
H.~Zhou, L.~Shi, X.~Zhang, and J.~Dong, ``Dynamic interferometry measurement of
  orbital angular momentum of light,'' \emph{Opt. Lett.}, vol.~39, no.~20, pp.
  6058--6061, 2014.

\bibitem{Berkhout2008Method}
G.~C. Berkhout and M.~W. Beijersbergen, ``Method for probing the orbital
  angular momentum of optical vortices in electromagnetic waves from
  astronomical objects,'' \emph{Phys. Rev. Lett.}, vol. 101, no.~10, p. 100801,
  2008.

\bibitem{Guo2009Characterizing}
G.~Chengshan, L.~Leilei, and W.~Huitian, ``Characterizing topological charge of
  optical vortices by using an annular aperture,'' \emph{Opt. Lett.}, vol.~34,
  no.~23, pp. 3686--3688, 2009.

\bibitem{Leach2004Interferometric}
J.~Leach, J.~Courtial, K.~Skeldon, S.~M. Barnett, S.~Franke-Arnold, and M.~J.
  Padgett, ``Interferometric methods to measure orbital and spin, or the total
  angular momentum of a single photon,'' \emph{Phys. Rev. Lett.}, vol.~92,
  no.~1, p. 013601, 2004.

\bibitem{Berkhout2010Efficient}
G.~C. Berkhout, M.~P. Lavery, J.~Courtial, M.~W. Beijersbergen, and M.~J.
  Padgett, ``Efficient sorting of orbital angular momentum states of light,''
  \emph{Phys. Rev. Lett.}, vol. 105, no.~15, p. 153601, 2010.

\bibitem{Lavery2012Refractive}
M.~P.~J. Lavery, D.~J. Robertson, G.~C.~G. Berkhout, G.~D. Love, M.~J. Padgett,
  and J.~Courtial, ``Refractive elements for the measurement of the orbital
  angular momentum of a single photon,'' \emph{Opt. Express}, vol.~20, no.~3,
  pp. 2110--2115, 2012.

\bibitem{Mirhosseini2013Efficient}
M.~Mirhosseini, M.~Malik, Z.~Shi, and R.~W. Boyd, ``Efficient separation of the
  orbital angular momentum eigenstates of light,'' \emph{Nature Commun.},
  vol.~4, p. 2781, 2013.

\bibitem{Karimi2009Efficient}
E.~Karimi, B.~Piccirillo, E.~Nagali, L.~Marrucci, and E.~Santamato, ``Efficient
  generation and sorting of orbital angular momentum eigenmodes of light by
  thermally tuned q-plates,'' \emph{Appl. Phys. Lett.}, vol.~94, no.~23, p.
  231124, 2009.

\bibitem{Jia2013Sidelobe}
P.~Jia, Y.~Yang, C.~J. Min, H.~Fang, and X.-C. Yuan, ``Sidelobe-modulated
  optical vortices for free-space communication,'' \emph{Opt. Lett.}, vol.~38,
  no.~4, pp. 588--590, 2013.

\bibitem{Long2013Evaluating}
J.~Long, R.~Liu, F.~Wang, Y.~Wang, P.~Zhang, H.~Gao, and F.~Li, ``Evaluating
  {Laguerre-Gaussian} beams with an invariant parameter,'' \emph{Opt. Lett.},
  vol.~38, no.~16, pp. 3047--3049, 2013.

\bibitem{Doerr2012Efficient}
N.~K. Fontaine, C.~R. Doerr, and L.~L. Buhl, ``Efficient multiplexing and
  demultiplexing of free-space orbital angular momentum using photonic
  integrated circuits,'' in \emph{Proc. Opt. Fiber Commun. Conf. Expo.}, 2012,
  p. OTu1I.2.

\bibitem{Su2012Demonstration}
T.~Su, R.~P. Scott, S.~S. Djordjevic, N.~K. Fontaine, D.~J. Geisler, X.~Cai,
  and S.~J.~B. Yoo, ``Demonstration of free space coherent optical
  communication using integrated silicon photonic orbital angular momentum
  devices,'' \emph{Opt. Express}, vol.~20, no.~9, pp. 9396--9402, 2012.

\bibitem{Yan2014High}
Y.~Yan, G.~Xie, M.~P.~J. Lavery, H.~Huang, N.~Ahmed, C.~Bao, Y.~Ren, Y.~Cao,
  L.~Li, Z.~Zhao, A.~F. Molisch, M.~Tur, M.~J. Padgett, and A.~E. Willner,
  ``High-capacity millimetre-wave communications with orbital angular momentum
  multiplexing,'' \emph{Nature Commun.}, vol.~5, p. 4876, 2014.

\bibitem{Mohammadi2010Orbital1}
S.~M. Mohammadi, L.~K.~S. Daldorff, K.~Forozesh, B.~Thid\'e, J.~E.~S. Bergman,
  B.~Isham, R.~Karlsson, and T.~D. Carozzi, ``Orbital angular momentum in
  radio: Measurement methods,'' \emph{Radio Sci.}, vol.~45, no.~4, pp. 1--14,
  2010.

\bibitem{Xie2017Mode}
M.~Xie, X.~Gao, M.~Zhao, W.~Zhai, W.~Xu, J.~Qian, M.~Lei, and S.~Huang, ``Mode
  measurement of a dual-mode radio frequency orbital angular momentum beam by
  circular phase gradient method,'' \emph{{IEEE} Antennas Wireless Propag.
  Lett.}, vol.~16, pp. 1143--1146, 2017.

\bibitem{Hu2016Simulation}
Y.~Hu, S.~Zheng, Z.~Zhang, H.~Chi, X.~Jin, and X.~Zhang, ``Simulation of
  orbital angular momentum radio communication systems based on partial
  aperture sampling receiving scheme,'' \emph{IET Microwaves, Antennas
  Propag.}, vol.~10, no.~10, pp. 1043--1047, 2016.

\bibitem{Wu2014UCA}
H.~Wu, Y.~Yuan, Z.~Zhang, and J.~Cang, ``{UCA}-based orbital angular momentum
  radio beam generation and reception under different array configurations,''
  in \emph{Proc. 6th Int. Conf. Wireless Commun. Signal Process.}, 2014.

\bibitem{Zheng2015Orbital}
S.~Zheng, X.~Hui, J.~Zhu, H.~Chi, X.~Jin, S.~Yu, and X.~Zhang, ``Orbital
  angular momentum mode-demultiplexing scheme with partial angular receiving
  aperture,'' \emph{Opt. Express}, vol.~23, no.~9, pp. 12\,251--12\,257, 2015.

\bibitem{Shi2017High}
C.~Shi, M.~Dubois, Y.~Wang, and X.~Zhang, ``High-speed acoustic communication
  by multiplexing orbital angular momentum,'' \emph{Proc. Natl. Acad. Sci.},
  vol. 114, no.~28, pp. 7250--7253, 2017.

\bibitem{Jiang2018Twisted}
X.~Jiang, B.~Liang, J.~Cheng, and C.~Qiu, ``Twisted acoustics:
  Metasurface-enabled multiplexing and demultiplexing,'' \emph{Adv. Mater.},
  vol.~30, no.~18, p. 1800257, 2018.

\bibitem{Krenn2014Communication}
M.~Krenn, R.~Fickler, M.~Fink, J.~Handsteiner, M.~Malik, T.~Scheidl, R.~Ursin,
  and A.~Zeilinger, ``Communication with spatially modulated light through
  turbulent air across vienna,'' \emph{New J. Phys.}, vol.~16, no.~11, p.
  113028, 2014.

\bibitem{Krenn2016Twisted}
M.~Krenn, J.~Handsteiner, M.~Fink, R.~Fickler, R.~Ursin, M.~Malik, and
  A.~Zeilinger, ``Twisted light transmission over 143 km,'' \emph{Proc. Natl.
  Acad. Sci.}, vol. 113, no.~48, pp. 13\,648--13\,653, 2016.

\bibitem{Huang2014100}
H.~Huang, G.~Xie, Y.~Yan, N.~Ahmed, Y.~Ren, Y.~Yue, D.~Rogawski, M.~J. Willner,
  B.~I. Erkmen, and K.~M. Birnbaum, ``100 {Tbit/s} free-space data link enabled
  by three-dimensional multiplexing of orbital angular momentum, polarization,
  and wavelength,'' \emph{Opt. Lett.}, vol.~39, no.~2, pp. 197--200, 2014.

\bibitem{Lin2007Multiplexing}
J.~Lin, X.~Yuan, S.~H. Tao, and R.~E. Burge, ``Multiplexing free-space optical
  signals using superimposed collinear orbital angular momentum states,''
  \emph{Appl. Opt.}, vol.~46, no.~21, pp. 4680--4685, 2007.

\bibitem{Zou2018High}
L.~Zou, X.~Gu, and L.~Wang, ``High-dimensional free-space optical
  communications based on orbital angular momentum coding,'' \emph{Opt.
  Commun.}, vol. 410, pp. 333--337, 2018.

\bibitem{Wang2012Terabit}
J.~Wang, J.-Y. Yang, I.~M. Fazal, N.~Ahmed, Y.~Yan, H.~Huang, Y.~Ren, Y.~Yue,
  S.~Dolinar, M.~Tur, and A.~E. Willner, ``Terabit free-space data transmission
  employing orbital angular momentum multiplexing,'' \emph{Nat. Photonics},
  vol.~6, no.~7, pp. 488--496, 2012.

\bibitem{Wang2014N}
J.~Wang, S.~Li, M.~Luo, J.~Liu, L.~Zhu, C.~Li, D.~Xie, Q.~Yang, S.~Yu, J.~Sun,
  X.~Zhang, W.~Shieh, and A.~E. Willner, ``N-dimentional multiplexing link with
  {1.036-Pbit/s} transmission capacity and 112.6-bit/s/{Hz} spectral efficiency
  using {OFDM-8QAM} signals over 368 {WDM} pol-muxed 26 {OAM} modes,'' in
  \emph{Proc. Eur. Conf. Opt. Commun.}, 2014, pp. 1--3.

\bibitem{Ren2016Experimental}
Y.~Ren, Z.~Wang, P.~Liao, L.~Li, G.~Xie, H.~Huang, Z.~Zhao, and Y.~Yan,
  ``Experimental characterization of a 400 {Gbit/s} orbital angular momentum
  multiplexed free-space optical link over 120 m,'' \emph{Opt. Lett.}, vol.~41,
  no.~3, pp. 622--625, 2016.

\bibitem{Abderrahmen2016Optical}
A.~Trichili, C.~Rosales-Guzm\'an, A.~Dudley, B.~Ndagano, A.~B. Salem, M.~Zghal,
  and A.~Forbes, ``Optical communication beyond orbital angular momentum,''
  \emph{Sci. Rep.}, vol.~6, p. 27674, 2016.

\bibitem{Xie2016Experimental}
G.~Xie, Y.~Ren, Y.~Yan, H.~Huang, N.~Ahmed, L.~Li, Z.~Zhao, C.~Bao, M.~Tur,
  S.~Ashrafi, and A.~E. Willner, ``Experimental demonstration of a {200-Gbit/s}
  free-space optical link by multiplexing {Laguerre-Gaussian} beams with
  different radial indices,'' \emph{Opt. Lett.}, vol.~41, no.~15, pp.
  3447--3450, 2016.

\bibitem{Gatto2011Free}
A.~Gatto, M.~Tacca, P.~Martelli, P.~Boffi, and M.~Martinelli, ``Free-space
  orbital angular momentum division multiplexing with {Bessel} beams,''
  \emph{J. Opt.}, vol.~13, no.~6, p. 064018, 2011.

\bibitem{Du2015High}
J.~Du and J.~Wang, ``High-dimensional structured light coding/decoding for
  free-space optical communications free of obstructions,'' \emph{Opt. Lett.},
  vol.~40, no.~21, pp. 4827--4830, 2015.

\bibitem{Ahmed2016Mode}
N.~Ahmed, Z.~Zhao, L.~Li, H.~Huang, M.~P.~J. Lavery, P.~Liao, Y.~Yan, Z.~Wang,
  G.~Xie, Y.~Ren, A.~Almaiman, A.~J. Willner, S.~Ashrafi, A.~F. Molisch,
  M.~Tur, and A.~E. Willner, ``Mode-division-multiplexing of multiple
  {Bessel-Gaussian} beams carrying orbital-angular-momentum for
  obstruction-tolerant free-space optical and millimetre-wave communication
  links,'' \emph{Sci. Rep.}, vol.~6, p. 22082, 2016.

\bibitem{Zhu2017Free}
F.~Zhu, S.~Huang, W.~Shao, J.~Zhang, M.~Chen, W.~Zhang, and J.~Zeng,
  ``Free-space optical communication link using perfect vortex beams carrying
  orbital angular momentum ({OAM}),'' \emph{Opt. Commun.}, vol. 396, pp.
  50--57, 2017.

\bibitem{Willner2015Performance}
G.~Xie, L.~Li, Y.~Ren, H.~Huang, Y.~Yan, N.~Ahmed, Z.~Zhao, M.~P.~J. Lavery,
  N.~Ashrafi, S.~Ashrafi, R.~Bock, M.~Tur, A.~F. Molisch, and A.~E. Willner,
  ``Performance metrics and design considerations for a free-space optical
  orbital-angular-momentum-multiplexed communication link,'' \emph{Optica},
  vol.~2, no.~4, pp. 357--365, 2015.

\bibitem{Paterson2005Atmospheric}
C.~Paterson, ``Atmospheric turbulence and orbital angular momentum of single
  photons for optical communication,'' \emph{Phys. Rev. Lett.}, vol.~94,
  no.~15, p. 153901, 2005.

\bibitem{Tyler2009Influence}
G.~A. Tyler and R.~W. Boyd, ``Influence of atmospheric turbulence on the
  propagation of quantum states of light carrying orbital angular momentum,''
  \emph{Opt. Lett.}, vol.~34, no.~2, pp. 142--144, 2009.

\bibitem{Ren2013Atmospheric}
Y.~Ren, H.~Huang, G.~Xie, N.~Ahmed, Y.~Yan, B.~I. Erkmen, N.~Chandrasekaran,
  M.~P.~J. Lavery, N.~K. Steinhoff, M.~Tur, S.~Dolinar, M.~Neifeld, M.~J.
  Padgett, R.~W. Boyd, J.~H. Shapiro, and A.~E. Willner, ``Atmospheric
  turbulence effects on the performance of a free space optical link employing
  orbital angular momentum multiplexing,'' \emph{Opt. Lett.}, vol.~38, no.~20,
  pp. 4062--4065, 2013.

\bibitem{Rodenburg2012Influence}
B.~Rodenburg, M.~P.~J. Lavery, M.~Malik, M.~N. O'Sullivan, M.~Mirhosseini,
  D.~J. Robertson, M.~Padgett, and R.~W. Boyd, ``Influence of atmospheric
  turbulence on states of light carrying orbital angular momentum,'' \emph{Opt.
  Lett.}, vol.~37, no.~17, pp. 3735--3737, 2012.

\bibitem{Fu2016Influences}
S.~Fu and C.~Gao, ``Influences of atmospheric turbulence effects on the orbital
  angular momentum spectra of vortex beams,'' \emph{Photonics Res.}, vol.~4,
  no.~5, pp. B1--B4, 2016.

\bibitem{Li2017Adaptive}
S.~Li and J.~Wang, ``Adaptive free-space optical communications through
  turbulence using self-healing {Bessel} beams,'' \emph{Sci. Rep.}, vol.~7, p.
  43233, 2017.

\bibitem{Li2018Atmospheric}
S.~Li, S.~Chen, C.~Gao, A.~E. Willner, and J.~Wang, ``Atmospheric turbulence
  compensation in orbital angular momentum communications: Advances and
  perspectives,'' \emph{Opt. Commun.}, vol. 408, pp. 68--81, 2018.

\bibitem{Ren2014Adaptive}
Y.~Ren, G.~Xie, H.~Huang, C.~Bao, Y.~Yan, N.~Ahmed, M.~P.~J. Lavery, B.~I.
  Erkmen, S.~Dolinar, M.~Tur, M.~A. Neifeld, M.~J. Padgett, R.~W. Boyd, J.~H.
  Shapiro, and A.~E. Willner, ``Adaptive optics compensation of multiple
  orbital angular momentum beams propagating through emulated atmospheric
  turbulence,'' \emph{Opt. Lett.}, vol.~39, no.~10, pp. 2845--2848, 2014.

\bibitem{Ren2015Turbulence}
Y.~Ren, G.~Xie, H.~Huang, L.~Li, N.~Ahmed, Y.~Yan, M.~P.~J. Lavery, R.~Bock,
  M.~Tur, M.~A. Neifeld, R.~W. Boyd, J.~H. Shapiro, and A.~E. Willner,
  ``Turbulence compensation of an orbital angular momentum and
  polarization-multiplexed link using a data-carrying beacon on a separate
  wavelength,'' \emph{Opt. Lett.}, vol.~40, no.~10, pp. 2249--2252, 2015.

\bibitem{Ren2014Adaptive1}
Y.~Ren, G.~Xie, H.~Huang, N.~Ahmed, Y.~Yan, L.~Li, C.~Bao, M.~P.~J. Lavery,
  M.~Tur, M.~A. Neifeld, R.~W. Boyd, J.~H. Shapiro, and A.~E. Willner,
  ``Adaptive-optics-based simultaneous pre- and post-turbulence compensation of
  multiple orbital-angular-momentum beams in a bidirectional free-space optical
  link,'' \emph{Optica}, vol.~1, no.~6, pp. 376--382, 2014.

\bibitem{Ren2012Correction}
Y.~Ren, H.~Huang, J.-Y. Yang, Y.~Yan, N.~Ahmed, Y.~Yue, A.~E. Willner,
  K.~Birnbaum, J.~Choi, B.~Erkmen, and S.~Dolinar, ``Correction of phase
  distortion of an {OAM} mode using {GS} algorithm based phase retrieval,'' in
  \emph{Proc. Conf. Lasers Electro-Opt.}, 2012, p. CF3I.4.

\bibitem{Fu2016Pre}
S.~Fu, S.~Zhang, T.~Wang, and C.~Gao, ``Pre-turbulence compensation of orbital
  angular momentum beams based on a probe and the {Gerchberg-Saxton}
  algorithm,'' \emph{Opt. Lett.}, vol.~41, no.~14, pp. 3185--3188, 2016.

\bibitem{Fu2017Non}
S.~Fu, T.~Wang, S.~Zhang, Z.~Zhang, Y.~Zhai, and C.~Gao, ``Non-probe
  compensation of optical vortices carrying orbital angular momentum,''
  \emph{Photonics Res.}, vol.~5, no.~3, pp. 251--255, 2017.

\bibitem{Xie2015Phase}
G.~Xie, Y.~Ren, H.~Huang, M.~P.~J. Lavery, N.~Ahmed, Y.~Yan, C.~Bao, L.~Li,
  Z.~Zhao, Y.~Cao, M.~Willner, M.~Tur, S.~J. Dolinar, R.~W. Boyd, J.~H.
  Shapiro, and A.~E. Willner, ``Phase correction for a distorted orbital
  angular momentum beam using a {Zernike} polynomials-based
  stochastic-parallel-gradient-descent algorithm,'' \emph{Opt. Lett.}, vol.~40,
  no.~7, pp. 1197--1200, 2015.

\bibitem{Huang2014Crosstalk}
H.~Huang, Y.~Cao, G.~Xie, Y.~Ren, Y.~Yan, C.~Bao, N.~Ahmed, M.~A. Neifeld,
  S.~J. Dolinar, and A.~E. Willner, ``Crosstalk mitigation in a free-space
  orbital angular momentum multiplexed communication link using $4\times4$
  {MIMO} equalization,'' \emph{Opt. Lett.}, vol.~39, no.~15, pp. 4360--4363,
  2014.

\bibitem{Richardson2013Space}
D.~J. Richardson, J.~M. Fini, and L.~E. Nelson, ``Space-division multiplexing
  in optical fibres,'' \emph{Nat. Photonics}, vol.~7, no.~5, pp. 354--362,
  2013.

\bibitem{Winzer2011MIMO}
P.~J. Winzer and G.~J. Foschini, ``{MIMO} capacities and outage probabilities
  in spatially multiplexed optical transport systems,'' \emph{Opt. Express},
  vol.~19, no.~17, pp. 16\,680--16\,696, 2011.

\bibitem{Ren2016Atmospheric}
Y.~Ren, Z.~Wang, G.~Xie, L.~Li, A.~J. Willner, Y.~Cao, Z.~Zhao, Y.~Yan,
  N.~Ahmed, N.~Ashrafi, S.~Ashrafi, R.~Bock, M.~Tur, and A.~E. Willner,
  ``Atmospheric turbulence mitigation in an {OAM}-based {MIMO} free-space
  optical link using spatial diversity combined with {MIMO} equalization,''
  \emph{Opt. Lett.}, vol.~41, no.~11, pp. 2406--2409, 2016.

\bibitem{Zhu2011112}
B.~Zhu, T.~F. Taunay, M.~Fishteyn, X.~Liu, S.~Chandrasekhar, M.~F. Yan, J.~M.
  Fini, E.~M. Monberg, and F.~V. Dimarcello, ``{112-Tb/s} space-division
  multiplexed {DWDM} transmission with {14-b/s/Hz} aggregate spectral
  efficiency over a 76.8-km seven-core fiber,'' \emph{Opt. Express}, vol.~19,
  no.~17, pp. 16\,665--16\,671, 2011.

\bibitem{Sakaguchi2012Space}
J.~Sakaguchi, Y.~Awaji, N.~Wada, A.~Kanno, T.~Kawanishi, T.~Hayashi, T.~Taru,
  T.~Kobayashi, and M.~Watanabe, ``Space division multiplexed transmission of
  109-{Tb/s} data signals using homogeneous seven-core fiber,'' \emph{J.
  Lightw. Technol.}, vol.~30, no.~4, pp. 658--665, 2012.

\bibitem{Randel20116}
S.~Randel, R.~Ryf, A.~Sierra, P.~J. Winzer, A.~H. Gnauck, C.~A. Bolle, R.~J.
  Essiambre, D.~W. Peckham, A.~Mccurdy, and R.~Lingle, ``6{\texttimes}56-{Gb/s}
  mode-division multiplexed transmission over 33-km few-mode fiber enabled by
  6{\texttimes}6 {MIMO} equalization,'' \emph{Opt. Express}, vol.~19, no.~17,
  pp. 16\,697--16\,707, 2011.

\bibitem{Ramachandran2009Generation}
S.~Ramachandran, P.~Kristensen, and M.~F. Yan, ``Generation and propagation of
  radially polarized beams in optical fibers,'' \emph{Opt. Lett.}, vol.~34,
  no.~16, pp. 2525--2527, 2009.

\bibitem{Bozinovic2012Control}
N.~Bozinovic, P.~Kristensen, S.~Ramachandran, and S.~Golowich, ``Control of
  orbital angular momentum of light with optical fibers,'' \emph{Opt. Lett.},
  vol.~37, no.~13, pp. 2451--2453, 2012.

\bibitem{Bozinovic2011Long}
N.~Bozinovic, P.~Kristensen, and S.~Ramachandran, ``Long-range
  fiber-transmission of photons with orbital angular momentum,'' in \emph{Proc.
  Conf. Lasers Electro-Opt.}, 2011, p. CTuB1.

\bibitem{Bozinovic2013Terabit}
N.~Bozinovic, Y.~Yue, Y.~Ren, M.~Tur, P.~Kristensen, H.~Huang, A.~E. Willner,
  and S.~Ramachandran, ``Terabit-scale orbital angular momentum mode division
  multiplexing in fibers,'' \emph{Science}, vol. 340, no. 6140, pp. 1545--1548,
  2013.

\bibitem{Yue2012Mode}
Y.~Yue, Y.~Yan, N.~Ahmed, J.-Y. Yang, L.~Zhang, Y.~Ren, H.~Huang, K.~M.
  Birnbaum, B.~I. Erkmen, S.~Dolinar, M.~Tur, and A.~E. Willner, ``Mode
  properties and propagation effects of optical orbital angular momentum
  ({OAM}) modes in a ring fiber,'' \emph{IEEE Photonics J.}, vol.~4, no.~2, pp.
  535--543, 2012.

\bibitem{Li2013Multi}
S.~Li and J.~Wang, ``Multi-orbital-angular-momentum multi-ring fiber for
  high-density space-division multiplexing,'' \emph{IEEE Photonics J.}, vol.~5,
  no.~5, p. 7101007, 2013.

\bibitem{Zhang2016A}
H.~Zhang, W.~Zhang, L.~Xi, X.~Tang, X.~Zhang, and X.~Zhang, ``A new type
  circular photonic crystal fiber for orbital angular momentum mode
  transmission,'' \emph{{IEEE} Photon. Technol. Lett.}, vol.~28, no.~13, pp.
  1426--1429, 2016.

\bibitem{Zhou2016Design}
G.~Zhou, G.~Zhou, C.~Chen, M.~Xu, C.~Xia, and Z.~Hou, ``Design and analysis of
  a microstructure ring fiber for orbital angular momentum transmission,''
  \emph{IEEE Photonics J.}, vol.~8, no.~2, pp. 1--12, 2016.

\bibitem{Zeng2016Experimental}
X.~Zeng, Y.~Li, Q.~Mo, W.~Li, Y.~Tian, Z.~Liu, and J.~Wu, ``Experimental
  investigation of {LP$_{11}$} mode to {OAM} conversion in few
  mode-polarization maintaining fiber and the usage for all fiber {OAM}
  generator,'' \emph{IEEE Photonics J.}, vol.~8, no.~4, pp. 1--7, 2016.

\bibitem{Li2017Superposing}
Y.~Li, L.~Jin, H.~Wu, S.~Gao, Y.-H. Feng, and Z.~Li, ``Superposing multiple
  {LP} modes with micro phase difference distributed along fiber to generate
  {OAM} mode,'' \emph{IEEE Photonics J.}, vol.~9, no.~2, pp. 1--9, 2017.

\bibitem{Li2018Generation}
S.~Li, Z.~Xu, R.~Zhao, L.~Shen, C.~Du, and J.~Wang, ``Generation of orbital
  angular momentum beam using fiber-to-fiber butt coupling,'' \emph{IEEE
  Photonics J.}, vol.~10, no.~4, pp. 1--7, 2018.

\bibitem{Mahmouli20134}
F.~E. Mahmouli and S.~D. Walker, ``{4-Gbps} uncompressed video transmission
  over a {60-GHz} orbital angular momentum wireless channel,'' \emph{IEEE
  Wireless Commun. Lett.}, vol.~2, no.~2, pp. 223--226, 2013.

\bibitem{Allen2014Wireless}
B.~Allen, A.~Tennant, Q.~Bai, and E.~Chatziantoniou, ``Wireless data encoding
  and decoding using {OAM} modes,'' \emph{Electron. Lett.}, vol.~50, no.~3, pp.
  232--233, 2014.

\bibitem{Yan201632}
Y.~Yan, L.~Li, Z.~Zhao, G.~Xie, Z.~Wang, Y.~Ren, N.~Ahmed, S.~Sajuyigbe,
  S.~Talwar, M.~Tur, N.~Ashrafi, S.~Ashrafi, A.~F. Molisch, and A.~E. Willner,
  ``{32-Gbit/s 60-GHz} millimeter-wave wireless communication using orbital
  angular momentum and polarization multiplexing,'' in \emph{Proc. IEEE Int.
  Conf. Commun.}, 2016.

\bibitem{Zhao2016A}
Z.~Zhao, Y.~Yan, L.~Li, G.~Xie, Y.~Ren, N.~Ahmed, Z.~Wang, C.~Liu, A.~J.
  Willner, P.~Song, H.~Hashemi, H.~Yao, D.~Macfarlane, R.~Henderson,
  N.~Ashrafi, S.~Ashrafi, S.~Talwar, S.~Sajuyigbe, M.~Tur, A.~F. Molisch, and
  A.~E. Willner, ``A dual-channel 60 {GHz} communications link using patch
  antenna arrays to generate data-carrying orbital-angular-momentum beams,'' in
  \emph{Proc. IEEE Int. Conf. Commun.}, 2016.

\bibitem{Zhao2015Experimental}
Z.~Zhao, Y.~Ren, G.~Xie, Y.~Yan, L.~Li, H.~Huang, C.~Bao, N.~Ahmed, M.~P.
  Lavery, C.~Zhang, N.~Ashrafi, S.~Ashrafi, S.~Talwar, S.~Sajuyigbe, M.~Tur,
  A.~F. Molisch, and A.~E. Willner, ``Experimental demonstration of {16-Gbit/s}
  millimeter-wave communications link using thin metamaterial plates to
  generate data-carrying orbital-angular-momentum beams,'' in \emph{Proc. IEEE
  Int. Conf. Commun.}, 2015, pp. 1392--1397.

\bibitem{Hui2015Multiplexed}
X.~Hui, S.~Zheng, Y.~Chen, Y.~Hu, X.~Jin, H.~Chi, and X.~Zhang, ``Multiplexed
  millimeter wave communication with dual orbital angular momentum ({OAM}) mode
  antennas,'' \emph{Sci. Rep.}, vol.~5, p. 10148, 2015.

\bibitem{Zhang2017Mode}
W.~Zhang, S.~Zheng, X.~Hui, R.~Dong, X.~Jin, H.~Chi, and X.~Zhang, ``Mode
  division multiplexing communication using microwave orbital angular momentum:
  An experimental study,'' \emph{{IEEE} Trans. Wireless Commun.}, vol.~16,
  no.~2, pp. 1308--1318, 2017.

\bibitem{Edfors2012Is}
O.~Edfors and A.~J. Johansson, ``Is orbital angular momentum {(OAM)} based
  radio communication an unexploited area?'' \emph{{IEEE} Trans. Antennas
  Propag.}, vol.~60, no.~2, pp. 1126--1131, 2012.

\bibitem{Tamagnone2012Comment}
M.~Tamagnone, C.~Craeye, and J.~Perruisseau-Carrier, ``Comment on `{Encoding}
  many channels on the same frequency through radio vorticity: first
  experimental test','' \emph{New J. Phys.}, vol.~14, no.~11, p. 118001, 2012.

\bibitem{Chen2018On}
R.~Chen, H.~Xu, W.~Yang, X.~Wang, and J.~Li, ``On the performance of {OAM} in
  keyhole channels,'' \emph{IEEE Wireless Commun. Lett.}, pp. 1--4, 2018.

\bibitem{Oldoni2015Space}
M.~Oldoni, F.~Spinello, E.~Mari, G.~Parisi, C.~G. Someda, F.~Tamburini,
  F.~Romanato, R.~A. Ravanelli, P.~Coassini, and B.~Thid\'e, ``Space-division
  demultiplexing in orbital-angular-momentum-based {MIMO} radio systems,''
  \emph{{IEEE} Trans. Antennas Propag.}, vol.~63, no.~10, pp. 4582--4587, 2015.

\bibitem{Zhang2016The}
Z.~Zhang, S.~Zheng, Y.~Chen, X.~Jin, H.~Chi, and X.~Zhang, ``The capacity gain
  of orbital angular momentum based multiple-input-multiple-output system,''
  \emph{Sci. Rep.}, no.~6, p. 25418, 2016.

\bibitem{Ge2017Millimeter}
X.~Ge, R.~Zi, X.~Xiong, Q.~Li, and L.~Wang, ``Millimeter wave communications
  with {OAM-SM} scheme for future mobile networks,'' \emph{{IEEE} J. Sel. Areas
  Commun.}, vol.~35, no.~9, pp. 2163--2177, 2017.

\bibitem{Ren2014Experimental}
Y.~Ren, L.~Li, G.~Xie, Y.~Yan, Y.~Cao, H.~Huang, N.~Ahemd, M.~J. Lavery,
  Z.~Zhao, C.~Zhang, M.~Tur, M.~Padgett, G.~Caire, A.~F. Molisch, and A.~E.
  Willner, ``Experimental demonstration of 16 {Gbit/s} millimeter-wave
  communications using {MIMO} processing of 2 {OAM} modes on each of two
  transmitter/receiver antenna apertures,'' in \emph{Proc. IEEE Global Commun.
  Conf.}, 2014, pp. 3821--3826.

\bibitem{Ren2017Line}
Y.~Ren, L.~Li, G.~Xie, Y.~Yan, Y.~Cao, H.~Huang, N.~Ahmed, Z.~Zhao, P.~Liao,
  C.~Zhang, G.~Caire, A.~F. Molisch, M.~Tur, and A.~E. Willner, ``Line-of-sight
  millimeter-wave communications using orbital angular momentum multiplexing
  combined with conventional spatial multiplexing,'' \emph{{IEEE} Trans.
  Wireless Commun.}, vol.~16, no.~5, pp. 3151--3161, 2017.

\bibitem{Ren2015Free}
Y.~Ren, Z.~Wang, G.~Xie, L.~Li, Y.~Cao, C.~Liu, P.~Liao, Y.~Yan, N.~Ahmed,
  Z.~Zhao, A.~Willner, N.~Ashrafi, S.~Ashrafi, R.~D. Linquist, R.~Bock, M.~Tur,
  A.~F. Molisch, and A.~E. Willner, ``Free-space optical communications using
  orbital-angular-momentum multiplexing combined with {MIMO}-based spatial
  multiplexing,'' \emph{Opt. Lett.}, vol.~40, no.~18, pp. 4210--4213, 2015.

\bibitem{Zhang2013On}
Y.~Zhang, W.~{Feng}, and N.~{Ge}, ``On the restriction of utilizing orbital
  angular momentum in radio communications,'' in \emph{8th International
  Conference on Communications and Networking in China}, 2013, pp. 271--275.

\bibitem{Chen2018Beam}
R.~Chen, H.~Xu, M.~Moretti, and J.~Li, ``Beam steering for the misalignment in
  {UCA}-based {OAM} communication systems,'' \emph{IEEE Wireless Commun.
  Lett.}, vol.~7, no.~4, pp. 582--585, 2018.

\bibitem{Yan2015Experimental}
Y.~Yan, L.~Li, G.~Xie, C.~Bao, P.~Liao, H.~Huang, Y.~Ren, N.~Ahmed, Z.~Zhao,
  M.~P. Lavery, N.~Ashrafi, S.~Ashrafi, S.~Talwar, S.~Sajuyigbe, M.~Tur, A.~F.
  Molisch, and A.~E. Willner, ``Experimental measurements of multipath-induced
  intra- and inter-channel crosstalk effects in a millimeter-wave
  communications link using orbital-angular-momentum multiplexing,'' in
  \emph{Proc. IEEE Int. Conf. Commun.}, 2015, pp. 1370--1375.

\bibitem{Yan2016Multipath}
Y.~Yan, L.~Li, G.~Xie, C.~Bao, P.~Liao, H.~Huang, Y.~Ren, N.~Ahmed, Z.~Zhao,
  Z.~Wang, N.~Ashrafi, S.~Ashrafi, S.~Talwar, S.~Sajuyigbe, M.~Tur, A.~F.
  Molisch, and A.~E. Willner, ``Multipath effects in millimetre-wave wireless
  communication using orbital angular momentum multiplexing,'' \emph{Sci.
  Rep.}, vol.~6, p. 33482, 2016.

\bibitem{Chen2018A}
R.~Chen, W.~Yang, H.~Xu, and J.~Li, ``A {2-D} {FFT}-based transceiver
  architecture for {OAM-OFDM} systems with {UCA} antennas,'' \emph{{IEEE}
  Trans. Veh. Technol.}, vol.~67, no.~6, pp. 5481--5485, 2018.

\bibitem{Ashkin1986Observation}
A.~Ashkin, J.~M. Dziedzic, J.~E. Bjorkholm, and S.~Chu, ``Observation of a
  single-beam gradient force optical trap for dielectric particles,''
  \emph{Opt. Lett.}, vol.~11, no.~5, pp. 288--290, 1986.

\bibitem{He1995Direct}
H.~He, M.~E.~J. Friese, N.~R. Heckenberg, and H.~Rubinsztein-Dunlop, ``Direct
  observation of transfer of angular momentum to absorptive particles from a
  laser beam with a phase singularity,'' \emph{Phys. Rev. Lett.}, vol.~75,
  no.~5, pp. 826--829, 1995.

\bibitem{Simpson1997Mechanical}
N.~B. Simpson, K.~Dholakia, L.~Allen, and M.~J. Padgett, ``Mechanical
  equivalence of spin and orbital angular momentum of light: an optical
  spanner,'' \emph{Opt. Lett.}, vol.~22, no.~1, pp. 52--54, 1997.

\bibitem{Friese1996Optical}
M.~E.~J. Friese, J.~Enger, H.~Rubinsztein-Dunlop, and N.~R. Heckenberg,
  ``Optical angular-momentum transfer to trapped absorbing particles,''
  \emph{Phys. Rev. A}, vol.~54, no.~2, pp. 1593--1596, 1996.

\bibitem{O2002Intrinsic}
A.~T. O'Neil, I.~Macvicar, L.~Allen, and M.~J. Padgett, ``Intrinsic and
  extrinsic nature of the orbital angular momentum of a light beam,''
  \emph{Phys. Rev. Lett.}, vol.~88, no.~5, p. 053601, 2002.

\bibitem{Garc2003Observation}
V.~Garc\'es-Ch\'avez, D.~Mcgloin, M.~J. Padgett, W.~Dultz, H.~Schmitzer, and
  K.~Dholakia, ``Observation of the transfer of the local angular momentum
  density of a multiringed light beam to an optically trapped particle,''
  \emph{Phys. Rev. Lett.}, vol.~91, no.~9, p. 093602, 2003.

\bibitem{Courtney2013Dexterous}
C.~R.~P. Courtney, B.~W. Drinkwater, C.~E.~M. Demore, S.~Cochran, A.~Grinenko,
  and P.~D. Wilcox, ``Dexterous manipulation of microparticles using
  {Bessel}-function acoustic pressure fields,'' \emph{Appl. Phys. Lett.}, vol.
  102, no.~12, p. 123508, 2013.

\bibitem{Courtney2014Independent}
C.~R.~P. Courtney, C.~E.~M. Demore, H.~Wu, A.~Grinenko, P.~D. Wilcox,
  S.~Cochran, and B.~W. Drinkwater, ``Independent trapping and manipulation of
  microparticles using dexterous acoustic tweezers,'' \emph{Appl. Phys. Lett.},
  vol. 104, no.~15, p. 154103, 2014.

\bibitem{Anh2012Acoustic}
A.~Anh\"auser, R.~Wunenburger, and E.~Brasselet, ``Acoustic rotational
  manipulation using orbital angular momentum transfer,'' \emph{Phys. Rev.
  Lett.}, vol. 109, no.~3, p. 034301, 2012.

\bibitem{Hell1994Breaking}
S.~W. Hell and J.~Wichmann, ``Breaking the diffraction resolution limit by
  stimulated emission: stimulated-emission-depletion fluorescence microscopy,''
  \emph{Opt. Lett.}, vol.~19, no.~11, pp. 780--782, 1994.

\bibitem{Li2013Beating}
L.~Li and F.~Li, ``Beating the rayleigh limit: Orbital-angular-momentum-based
  super-resolution diffraction tomography,'' \emph{Phys. Rev. E}, vol.~88,
  no.~3, p. 033205, 2013.

\bibitem{Maurer2011What}
C.~Maurer, A.~Jesacher, S.~Bernet, and M.~Ritsch-Marte, ``What spatial light
  modulators can do for optical microscopy,'' \emph{Laser Photonics Rev.},
  vol.~5, no.~1, pp. 81--101, 2011.

\bibitem{Swartzlander2001Peering}
G.~A. Swartzlander, ``Peering into darkness with a vortex spatial filter,''
  \emph{Opt. Lett.}, vol.~26, no.~8, pp. 497--499, 2001.

\bibitem{F2005Spiral}
S.~F\"{u}rhapter, A.~Jesacher, S.~Bernet, and M.~Ritsch-Marte, ``Spiral phase
  contrast imaging in microscopy,'' \emph{Opt. Express}, vol.~13, no.~3, pp.
  689--694, 2005.

\bibitem{Wang2015Gradual}
J.~Wang, W.~Zhang, Q.~Qi, S.~Zheng, and L.~Chen, ``Gradual edge enhancement in
  spiral phase contrast imaging with fractional vortex filters,'' \emph{Sci.
  Rep.}, vol.~5, p. 15826, 2015.

\bibitem{Swartzlander2008Astronomical}
G.~A. Swartzlander, E.~L. Ford, R.~S. Abdul-Malik, L.~M. Close, M.~A. Peters,
  D.~M. Palacios, and D.~W. Wilson, ``Astronomical demonstration of an optical
  vortex coronagraph,'' \emph{Opt. Express}, vol.~16, no.~14, pp.
  10\,200--10\,207, 2008.

\bibitem{F2005Spiralinterferometry}
S.~F\"{u}rhapter, A.~Jesacher, S.~Bernet, and M.~Ritsch-Marte, ``Spiral
  interferometry,'' \emph{Opt. Lett.}, vol.~30, no.~15, pp. 1953--1955, 2005.

\bibitem{Guo2013Electromagnetic}
G.~Guo, W.~Hu, and X.~Du, ``Electromagnetic vortex based radar target
  imaging,'' \emph{J. Nat. Univ. Defense Technol.}, vol.~35, no.~6, pp. 71--76,
  2013.

\bibitem{Liu2015Orbital}
K.~Liu, Y.~Cheng, Z.~Yang, H.~Wang, Y.~Qin, and X.~Li,
  ``Orbital-angular-momentum-based electromagnetic vortex imaging,''
  \emph{{IEEE} Antennas Wireless Propag. Lett.}, vol.~14, pp. 711--714, 2015.

\bibitem{Yuan2016Electromagnetic}
T.~Yuan, H.~Wang, Y.~Qin, and Y.~Cheng, ``Electromagnetic vortex imaging using
  uniform concentric circular arrays,'' \emph{{IEEE} Antennas Wireless Propag.
  Lett.}, vol.~15, pp. 1024--1027, 2016.

\bibitem{Zhang2017Large}
C.~Zhang and D.~Chen, ``Large-scale orbital angular momentum radar pulse
  generation with rotational antenna,'' \emph{{IEEE} Antennas Wireless Propag.
  Lett.}, vol.~16, pp. 2316--2319, 2017.

\bibitem{Liu2017Super}
K.~Liu, Y.~Cheng, Y.~Gao, X.~Li, Y.~Qin, and H.~Wang, ``Super-resolution radar
  imaging based on experimental {OAM} beams,'' \emph{Appl. Phys. Lett.}, vol.
  110, no.~16, p. 164102, 2017.

\bibitem{Bu2018Implementation}
X.~Bu, Z.~Zhang, L.~Chen, X.~Liang, H.~Tang, and X.~Wang, ``Implementation of
  vortex electromagnetic waves high-resolution synthetic aperture radar
  imaging,'' \emph{{IEEE} Antennas Wireless Propag. Lett.}, vol.~17, no.~5, pp.
  764--767, 2018.

\bibitem{Liu2016Generation}
K.~Liu, Y.~Cheng, X.~Li, Y.~Qin, H.~Wang, and Y.~Jiang, ``Generation of orbital
  angular momentum beams for electromagnetic vortex imaging,'' \emph{{IEEE}
  Antennas Wireless Propag. Lett.}, vol.~15, pp. 1873--1876, 2016.

\bibitem{Lin2016Super}
M.~Lin, Y.~Gao, P.~Liu, and J.~Liu, ``Super-resolution orbital angular momentum
  based radar targets detection,'' \emph{Electron. Lett.}, vol.~52, no.~13, pp.
  1168--1170, 2016.

\bibitem{Lin2016Improved}
------, ``Improved {OAM}-based radar targets detection using uniform concentric
  circular arrays,'' \emph{Int. J. Antennas Propag.}, vol. 2016, pp. 1--8,
  2016.

\bibitem{Chen2018Orbital}
R.~Chen, W.-X. Long, Y.~Gao, and J.~Li, ``Orbital angular momentum-based
  two-dimensional super-resolution targets imaging,'' in \emph{Proc. IEEE
  Global Conf. Signal Inf. Process.}, 2018.

\bibitem{Yuan2017Beam}
T.~Yuan, Y.~Cheng, H.~Wang, and Y.~Qin, ``Beam steering for electromagnetic
  vortex imaging using uniform circular arrays,'' \emph{{IEEE} Antennas
  Wireless Propag. Lett.}, vol.~16, pp. 704--707, 2017.

\bibitem{Yuan2016Orbital}
T.~Yuan, H.~Liu, Y.~Cheng, Y.~Qin, and H.~Wang,
  ``Orbital-angular-momentum-based electromagnetic vortex imaging by
  least-squares method,'' in \emph{Proc. IEEE Int. Geosci. Remote Sens. Symp.},
  2016, pp. 6645--6648.

\end{thebibliography}
\begin{IEEEbiography}[{\includegraphics[width=1in,height=1.25in,clip,keepaspectratio]{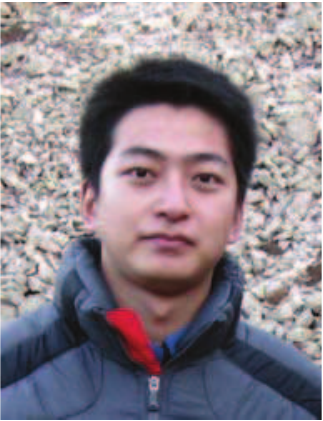}}]{Rui Chen}
(S'08-M'11) received the B.S., M.S. and Ph.D. degrees in Communications and Information Systems from Xidian University, Xi’an, China, in 2005, 2007 and 2011, respectively. From 2014 to 2015, he was a visiting scholar at Columbia University in the City of New York. He is currently an associate professor and Ph.D. supervisor in the school of Telecommunications Engineering at Xidian University. He has published about 50 papers in international journals and conferences and held 10 patents. He is an Associate Editor for International Journal of Electronics, Communications, and Measurement Engineering (IGI Global). His research interests include broadband wireless communication systems, array signal processing and intelligent transportation systems.
\end{IEEEbiography}

\begin{IEEEbiography}[{\includegraphics[width=1in,height=1.25in,clip,keepaspectratio]{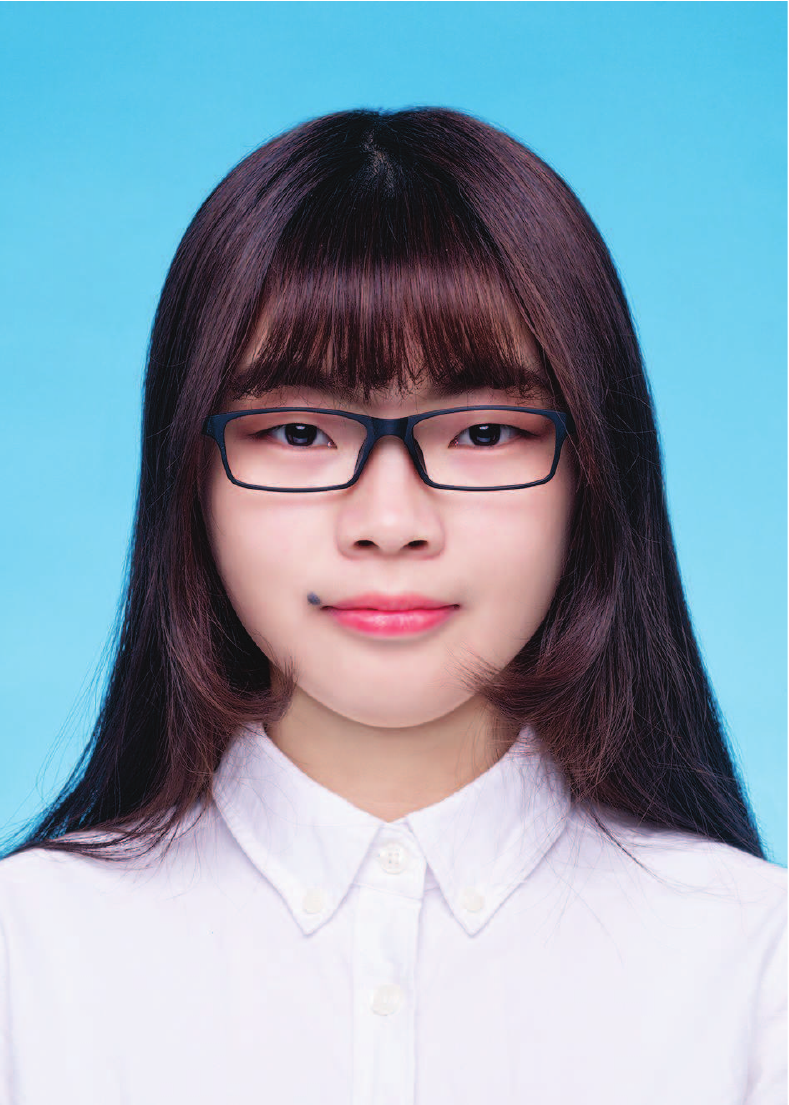}}]{Hong Zhou}
received the B.S. degree (with Highest Hons.) in Communications Engineering from Hefei University of Technology, Hefei, China in 2018. She is currently pursuing a M.S. degree in Communications and Information Systems at Xidian University. Her research interests include orbital angular momentum (OAM) communication systems and array signal processing.
\end{IEEEbiography}

\begin{IEEEbiography}[{\includegraphics[width=1in,height=1.25in,clip,keepaspectratio]{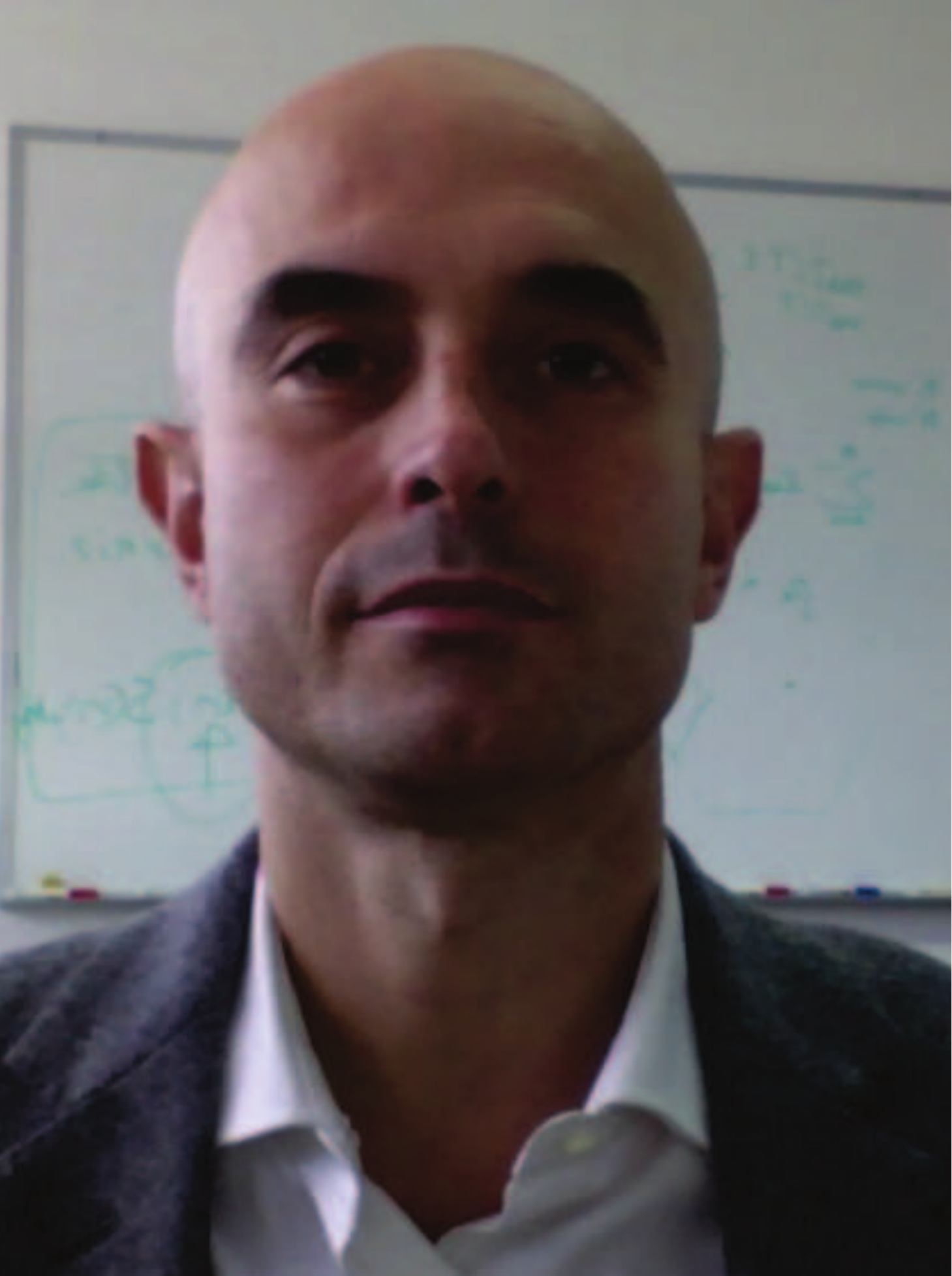}}]{Marco Moretti}
(M'03) received the degree in electronic engineering from the University of Florence, Florence, Italy, in 1995, and the Ph.D. degree from the Delft University of Technology, Delft, The Netherlands, in 2000. From 2000 to 2003, he was a Senior Researcher with Marconi Mobile. He is currently an Associate Professor with the University of Pisa, Pisa, Italy. His research interests include resource allocation for multicarrier systems, synchronization, and channel estimation. He is currently an Associate Editor of the IEEE TRANSACTIONS ON SIGNAL PROCESSING.
\end{IEEEbiography}

\begin{IEEEbiography}[{\includegraphics[width=1in,height=1.25in,clip,keepaspectratio]{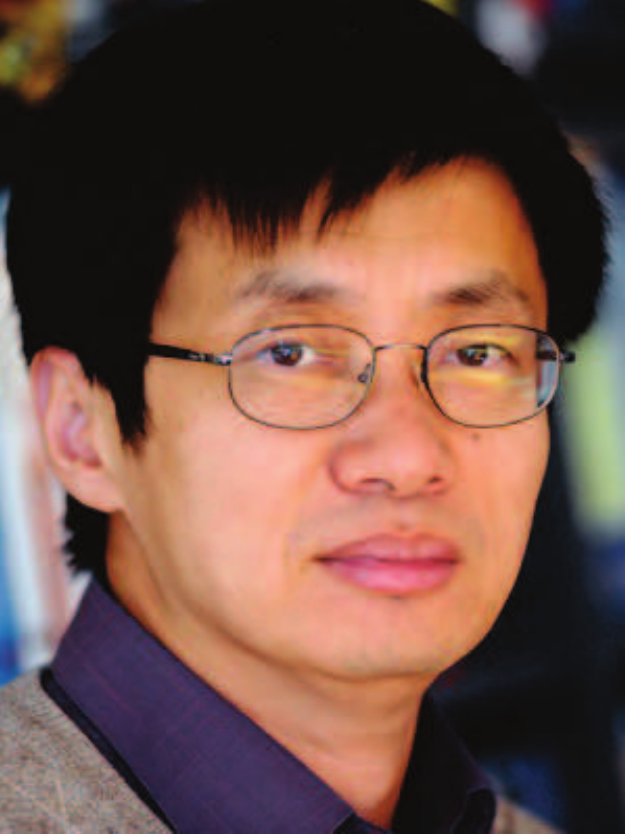}}]{Xiaodong Wang}
(S'98-M'98-SM'04-F'08) received the Ph.D. degree in electrical engineering from Princeton University. He is currently a Professor of electrical engineering with Columbia University, New York, NY, USA. He has authored the book Wireless Communication Systems: Advanced Techniques for Signal Reception (Prentice-Hall, 2003). His research interests include computing, signal processing, and communications, and has published extensively in these areas. His current research interests include wireless communications, statistical signal processing, and genomic signal processing. He was a recipient of the 1999 NSF CAREER Award, the 2001 IEEE Communications Society and Information Theory Society Joint Paper Award, and the 2011 IEEE Communication Society Award for Outstanding Paper on New Communication Topics. He was an Associate Editor for the IEEE TRANSACTIONS ON COMMUNICATIONS, the IEEE TRANSACTIONS ON WIRELESS COMMUNICATIONS, the IEEE TRANSACTIONS ON SIGNAL PROCESSING, and the IEEE TRANSACTIONS ON INFORMATION THEORY. He is listed as an ISI highly cited author.
\end{IEEEbiography}

\begin{IEEEbiography}[{\includegraphics[width=1in,height=1.25in,clip,keepaspectratio]{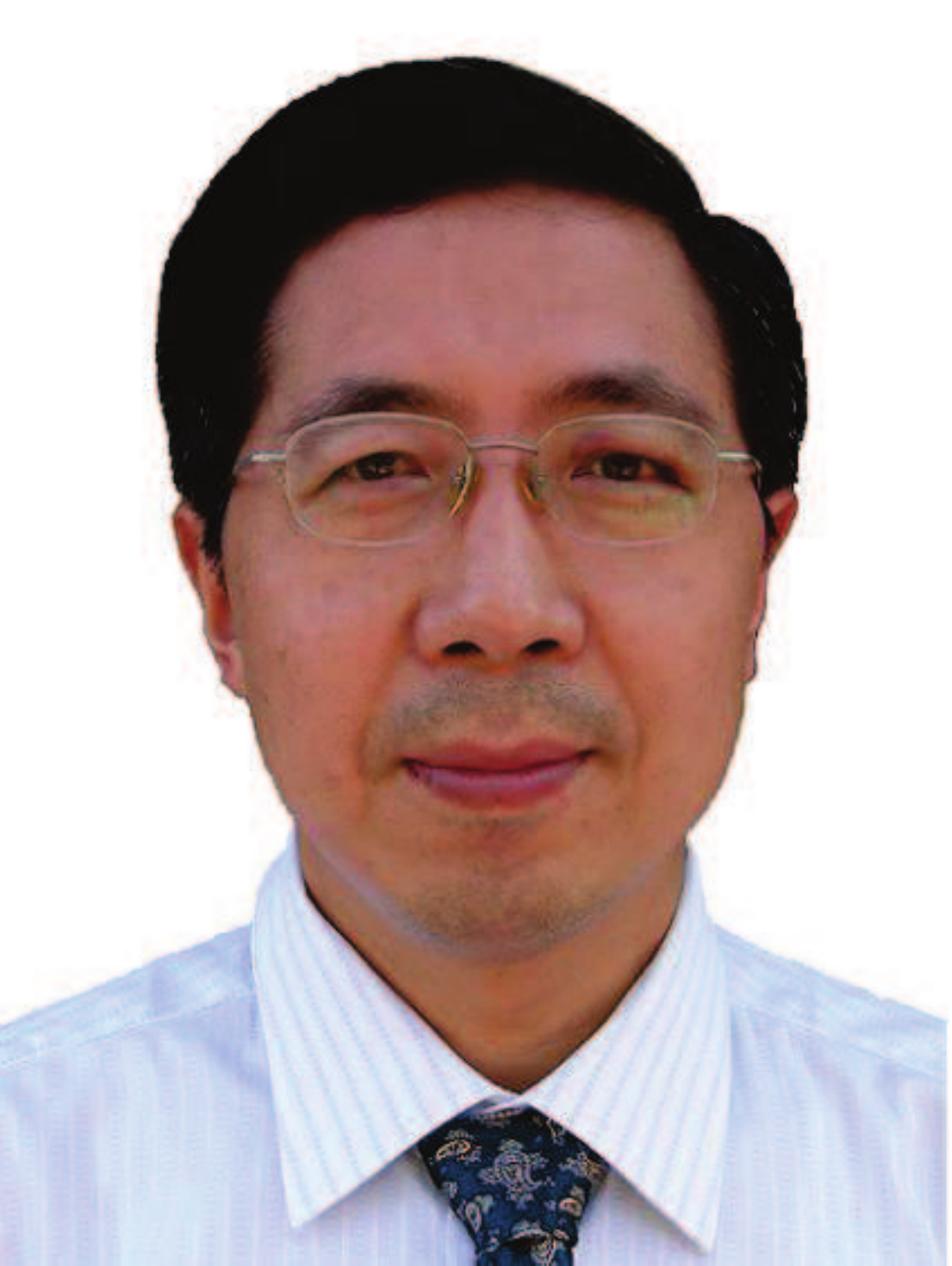}}]{Jiandong Li}
(SM'05) received the B.E., M.S., and Ph.D. degrees in communications engineering from
Xidian University, Xi’an, China, in 1982, 1985, and 1991, respectively. He was a Visiting Professor with the Department of Electrical and Computer Engineering, Cornell University, from 2002 to 2003. He has been a faculty member of the School of Telecommunications Engineering, Xidian University, since 1985, where he is currently a Professor and the Vice Director of the Academic Committee, State Key Laboratory of Integrated Service Networks. His major research interests include wireless communication theory, cognitive radio, and signal processing. He was recognized as a Distinguished Young Researcher by NSFC and a Changjiang Scholar by the Ministry of Education, China, respectively. He served as the General Vice Chair of ChinaCom 2009 and the TPC Chair of the IEEE ICCC 2013.
\end{IEEEbiography}

\end{document}